\documentclass[useAMS,usenatbib]{mn2e}
\usepackage{graphicx}
\usepackage{amsmath}
\usepackage{grffile}
\usepackage{amssymb}
\usepackage{lscape}
\usepackage{array}
\title[The XMM-LSS Class 1 cluster sample over the extended 11 deg$^2$]{The XMM-LSS survey: the Class 1 cluster sample over the extended 11 deg$^2$ and its spatial distribution}
\author[N. Clerc et al. ]{N. Clerc$^{1}$\thanks{E-mail:
nclerc@mpe.mpg.de},
C. Adami$^{2}$,
M. Lieu$^{3}$,
B. Maughan$^{4}$,
F. Pacaud$^{5}$,
M. Pierre$^{6}$,
\newauthor T. Sadibekova$^{6}$,
G.~P. Smith$^{3}$,
P. Valageas$^{7,8}$,
B. Altieri$^{9}$,
C. Benoist$^{10}$,
\newauthor S. Maurogordato$^{10}$,
J.~P. Willis$^{11}$
\\
$^{1}$Max Planck Institut f\"ur Extraterrestrische Physik, Postfach 1312, 85741 Garching bei M\"unchen, Germany.\\
$^{2}$Aix-Marseille Universit\'e, CNRS, LAM (Laboratoire d'Astrophysique de Marseille) UMR 7326, 13388, Marseille, France.\\
$^{3}$School of Physics and Astronomy, University of Birmingham, Edgbaston, Birmingham, B15 2TT, UK\\
$^{4}$H.~H. Wills Physics Laboratory, University of Bristol, Tyndall Ave, Bristol BS8 1TL, UK\\
$^{5}$Argelander-Institut f\"ur Astronomie, University of Bonn, Auf dem H\"ugel 71, 53121 Bonn, Germany.\\
$^{6}$Laboratoire AIM, CEA/DSM/IRFU/SAp, CEA Saclay, 91191 Gif-sur-Yvette, France.\\
$^{7}$Institut de Physique Th\'eorique, CEA Saclay, 91191 Gif-sur-Yvette, cedex, France.\\
$^{8}$CNRS, URA 2306, 91191 Gif-sur-Yvette, cedex, France.\\
$^{9}$ESAC, Villafranca del Castillo, 28692 Madrid, Spain.\\
$^{10}$Laboratoire J.~L. Lagrange OCA-CNRS-UNSA, BP4429, Nice Cedex 04, France.\\
$^{11}$Department of Physics and Astronomy, University of Victoria, 3800 Finnerty Road, Victoria, BC, Canada\\
}
\begin{document}

\date{Accepted 2014 August 8. Received 2014 July 8; in original form 2014 March 24.}

\pagerange{\pageref{firstpage}--\pageref{lastpage}} \pubyear{2002}

\maketitle

\label{firstpage}

\begin{abstract}
This paper presents 52 X-ray bright galaxy clusters selected within the 11~deg$^2$ XMM-LSS survey. 51 of them have spectroscopic redshifts ($0.05<z<1.06$), one is identified at $z_{\rm phot}=1.9$, and all together make the high-purity "Class 1" (C1) cluster sample of the XMM-LSS, the highest density sample of X-ray selected clusters with a monitored selection function.
Their X-ray fluxes, averaged gas temperatures (median $T_X=2$~keV), luminosities (median $L_{X,500}=5\times10^{43}$~ergs/s) and total mass estimates (median $5\times10^{13} h^{-1} M_{\odot}$) are measured, adapting to the specific signal-to-noise regime of XMM-LSS observations.
Particular care is taken in deriving the sample selection function by means of realistic simulations reproducing the main characteristics of XMM observations.
The redshift distribution of clusters shows a deficit of sources when compared to the cosmological expectations, regardless of whether WMAP-9 or Planck-2013 CMB parameters are assumed. This lack of sources is particularly noticeable at $0.4 \lesssim z \lesssim 0.9$. However, after quantifying uncertainties due to small number statistics and sample variance we are not able to put firm (i.e.~$>3 \sigma$) constraints on the presence of a large void in the cluster distribution.
We work out alternative hypotheses and demonstrate that a negative redshift evolution in the normalization of the $L_{X}-T_X$ relation (with respect to a self-similar evolution) is a plausible explanation for the observed deficit. We confirm this evolutionary trend by directly studying how C1 clusters populate the $L_{X}-T_X-z$ space, properly accounting for selection biases.
We also point out that a systematically evolving, unresolved, central component in clusters and groups (AGN contamination or cool core) can impact the classification as extended sources and be partly responsible for the observed redshift distribution.
We provide in a table the catalogue of 52 clusters together with their measured properties.
\end{abstract}

\begin{keywords}
cosmology: observations -- catalogues -- galaxies: clusters: general -- X-rays: galaxies: clusters.
\end{keywords}


\section{Introduction}

Studying the spatial distribution of galaxy clusters in a volume of cosmological size enables the cartography of large-scale structure in the Universe through its most massive building blocks. The number counts distribution of clusters is therefore an excellent test for cosmological models and the growth of structure \citep[e.g.][]{borgani01, henry09, vikhlinin09, mantz10,rozo10,pierre11,benson13}.

From an observational point of view, galaxy clusters are advantageously high signal astrophysical sources: as expected from simple scaling arguments \citep[e.g.][]{kaiser86}, the most massive virialized objects are also the largest in size and their observable properties scale up with mass. This is especially true in X-ray wavelengths because of the large amount of X-ray photons emitted by the hot ($T\sim10^7$~K) intra-cluster baryonic gas trapped in their deep potential wells. With typical X-ray luminosities of $10^{43-45}$~ergs/s, clusters can be detected up to large cosmological distances in a systematic and controlled way.
On the other hand, galaxy clusters are rare objects and robust analyses of their spatial distribution (e.g.~redshift distribution, 2-point correlation function...) and their ensemble properties (e.g.~scaling relations, mass distribution...) require medium to large area surveys in order to accumulate statistical power.
These characteristics motivated the assembly of large samples of galaxy clusters detected in X-rays. In particular, studies based on the \emph{ROSAT} all-sky survey \citep{truemper93} such as REFLEX \citep{boehringer01}, NORAS \citep{boehringer00}, the ROSAT North Ecliptic Pole Survey \citep{henry01}, the 400d \citep{burenin07} delivered solid cosmological results based on galaxy clusters \citep[e.g.][]{schuecker03,vikhlinin09,mantz10}.
During the last decade, the very sensitive XMM-Newton \citep{fassbender11,willis13} and Chandra \citep[e.g.][]{tozzi13} observatories revealed the presence of their characteristic emission beyond redshifts of 1, and even up to $z\sim2$ \citep{gobat11}, demonstrating the ability to construct deep samples of cosmological interest.

The XMM-LSS survey \citep{pierre04} is unique in this respect: it covers a contiguous sky area of 11~deg$^2$ with XMM pointed observations, reaching a sensitivity of $10^{-14}$~ergs/s/cm$^2$ for extended sources in the [0.5-2]~keV band. Thanks to the wide and complementary multi-wavelength coverage (from radio to $\gamma$-rays) and dedicated follow-up effort, it constitutes a relevant field for studies of galaxy clusters and groups. The complete X-ray source catalogue is published in \citet{chiappetti13} along with the optical associations. The XMM-XXL survey is currently extending its area to 50~deg$^2$ following a similar strategy as for the detection and characterization of galaxy clusters, with international support and expertise (Pierre et al. 2014, in prep.)
The sample used in this work is unique in terms of X-ray and spectroscopic redshifts completenesses:
(i)~spectroscopic redshifts (from observations of cluster galaxy members) enable to position clusters in 3D space and accurate derivation of their physical properties: gas temperature, luminosity, physical size;
(ii)~a trade-off between completeness, purity and assessment of selection effects has been carefully designed: \citet{pacaud06} indeed demonstrated the existence of an uncontaminated sample of extended sources detected on XMM-LSS images called "C1", 29 of them were detected in the first 5~deg$^2$ of the XMM-LSS survey \citep{pacaud07}.

In this study, we focus on the redshift distribution of the complete set of 52 XMM-LSS C1 clusters and its cosmological modelling. Results from the catalogue of sources detected in the Planck survey by the Sunyaev-Zeldovich effect indicated a deficit of clusters at all redshifts when compared to expectations from the Planck CMB cosmological model \citep{planckszcosmo}. Tension could be alleviated by modifying the mass-observable relation.
Studies based on galaxy clusters indeed appeal for a simultaneous modelling of the cosmological halo mass function, the mass-observable relations (scaling laws and their evolutions with redshift) and modelling of selection effects \citep[e.g.][]{pacaud07,pratt09,vikhlinin09,mantz10,allen11,clerc12b,planckszcosmo}.
Therefore, part of our results concerns the $L_X-T$ relation of C1 clusters and its evolution. This latter point is particularly debated in current studies of clusters detected in XMM data. \citet{pacaud07} pointed out the importance of selection biases in such studies. \citet{reichert11} found a negative trend (relative to self-similar expectations) by analysing an heterogeneous sample of objects. Both the XMM Cluster Survey \citep[XCS,][]{hilton12} and the XMM CLuster Archive Super Survey \citep[X-CLASS,][]{clerc12b} indicated a negative evolution in the normalization of the relation. However, differences in the selection and analyses methods make such comparisons difficult. 

This paper is organized as follows. In Section~\ref{sect:construction_sample} we describe the dataset and our choice of sample for this study. An in-depth characterization of the survey selection function is presented in Section~\ref{sect:selfunc}. Derivation of cluster properties is detailed in Section~\ref{sect:cluster_properties}. The spatial distribution and the luminosity-temperature relation of clusters in the sample are shown in Section~\ref{sect:results}, and we discuss further the modelling of the redshift distribution in Section~\ref{sect:discussion}. Section~\ref{sect:conclusions} summarizes our findings.

Throughout this paper, we assume $\Omega_m=0.3$, $\Omega_{\Lambda} = 0.7$ and $h=0.7$ with $H_0=100 h$~km.s$^{-1}$, except otherwise stated. In particular, our discussion of the redshift distribution (Sect.~\ref{sect:results} and~\ref{sect:discussion}) alternates between WMAP-9 \citep{hinshaw13} and Planck CMB \citep{planckcmb} cosmologies.
In all that follows, $M_{\delta} (=M_{\delta c})$ is the mass within a sphere of radius $R_{\delta}(=R_{\delta c})$, inside which the mass density is $\delta$ times the critical density of the Universe at the considered redshift. Transformations between different values of $\delta$ will assume a NFW profile \citep{nfw}.


\section[]{Construction of the sample}
	\label{sect:construction_sample}

The complete XMM-LSS 11~deg$^2$ source catalogue is presented in full extent by \citet{chiappetti13}. In this section, we briefly recall the main characteristics of the 11~deg$^2$ XMM-LSS survey and our procedure for detecting sources, with particular emphasis on the confirmation of C1 galaxy clusters.

	\subsection[]{The XMM dataset}
	
\begin{figure}
	\includegraphics[width=84mm]{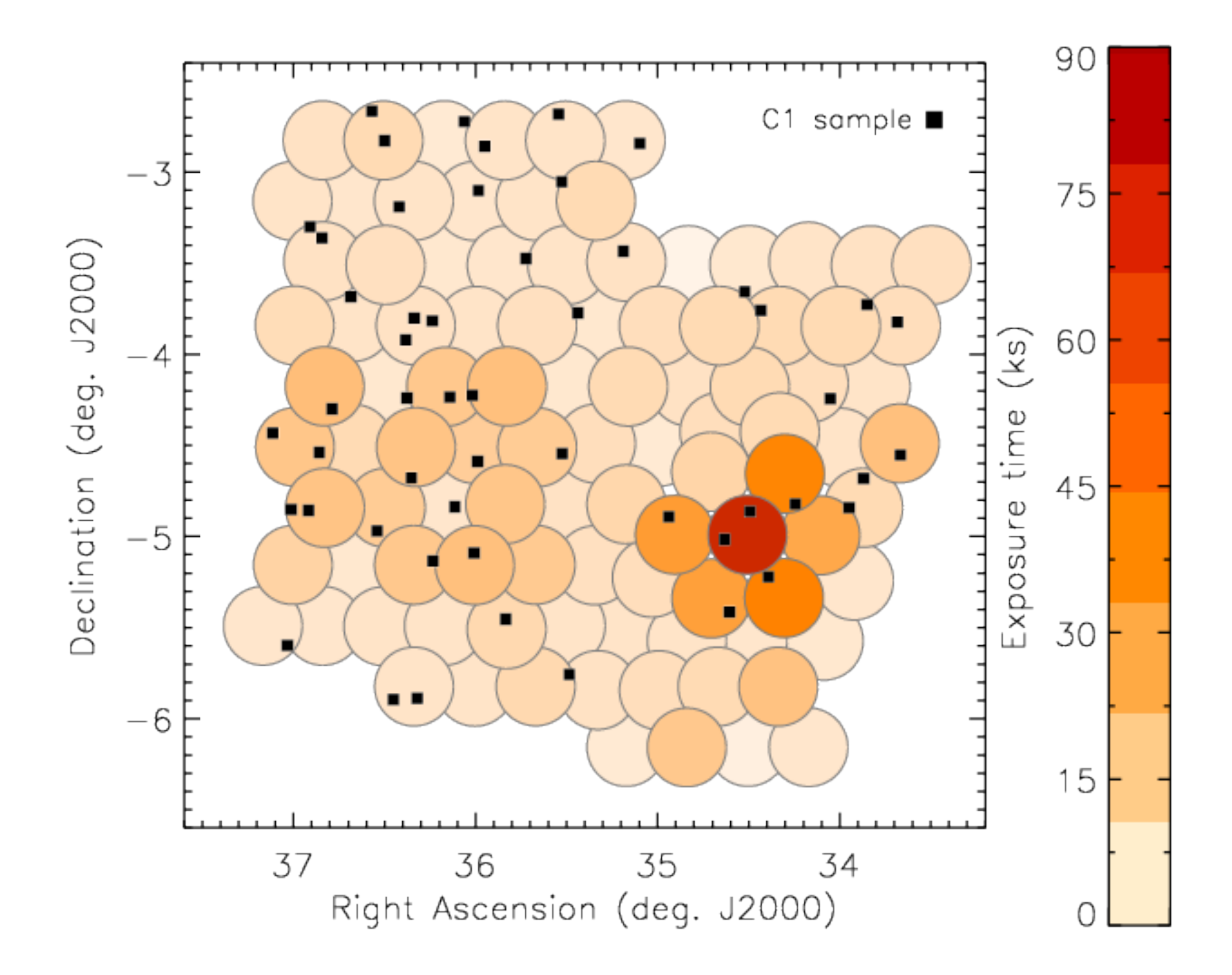}
	\caption{Layout of the 98 XMM observations constituting the XMM-LSS 11 deg$^2$ survey. Positions of C1 clusters presented in this work are overlaid as black squares. The on-axis, clean, exposure time of each pointing is shown by the colour scale. The size of each XMM observation is on scale and corresponds to the $13\arcmin$  radius circular area analysed around each pointing centre.}
	\label{fig:xmmlss_layout} 
\end{figure}

The XMM-LSS survey is located at ${\rm R.A.} = 02^h22^m$ and $\delta=-04\degr 30\arcmin$ and consists of 98 XMM pointings on a $\sim 3.5 \degr \times 3.5 \degr$ contiguous footprint (Fig.~\ref{fig:xmmlss_layout}). It represents a sub-area of the larger XMM-XXL survey \citep[][]{pierre11} for which a full description of the observation strategy and quality is given in Pierre et al. (2014, in prep.)
Each of those pointings corresponds to a single observation with the EPIC detectors (MOS1, MOS2 and PN) in full frame imaging mode, spanning a field of view of roughly $30 \arcmin$  diameter each.
	
The centre of each observation is defined by the exposure-weighted location of the optical axes of all three XMM telescopes. The work presented in this study relies on X-ray data collected by each pointing up to an off-axis radius of 13\arcmin.
The total geometrical area of the survey amounts to 11.1\,deg$^2$, 3.0\,deg$^2$ (27\%) of that area consists in overlaps between adjacent pointings (Fig.~\ref{fig:xmmlss_layout}). The remaining area (8.1\,deg$^2$) is covered by one unique pointing.
For each individual observation, an event list is created and filtered from solar proton flares by automatic inspection of the high-energy light-curves. High count-rate periods associated to particle flares are flagged and removed for each of the three detectors separately.
Cleaned exposure times after flares removal amount to $\sim 10$~ks for the majority of the pointings (Fig.~\ref{fig:xmmlss_layout}), although spatial fluctuations in the survey depth do exist and will be modeled in the selection function (Sect.~\ref{sect:selfunc}).

	\subsection[]{Source detection}
Our source detection procedure builds upon the algorithm described in \citet{pacaud06, pacaud07} with several revisions as detailed in \citet{chiappetti13}. These revisions mainly consist in a transcription of the algorithm code from IDL to Python and a correction in the relative astrometry.

Individual detector images are created in the [0.5-2]~keV band from cleaned event lists and binned in pixels of $2.5\arcsec$ width.
Sources are detected on each pointing image separately: in a first step, all three EPIC images are co-added, then filtered using the {\sc mr\_filter} multi-resolution algorithm \citep{starck1998}. Such a filtering adequately accounts for the Poisson nature of the noise in the background and source areas. A {\sc SExtractor} \citep{bertin96} pass over this image provides an initial detection list and a first guess for their positions.
In a second step, each source from this list is characterized thanks to a maximum likelihood fitting algorithm ({\sc XAmin}, \citealt{pacaud06}) specifically developed to assess the extension of sources in XMM-LSS data.
Among the output parameters of interest are the \emph{detection likelihood} quantifying the significance of the detection relative to a case without source and the \emph{extension likelihood}, comparing the case of an extended source and a point-like source. These likelihood values are defined (and corrected) so as to match a 2-parameter $\chi^2$ statistics, hence convertible into equivalent probabilities following $P=\exp(-L/2)$.

	\subsection[]{The C1 selection}
	\label{source_sel}
	
Following the methodology presented in \citet{pacaud06}, we define a C1 (Class 1) sample of sources by filtering on {\sc XAmin} output parameters, namely the \emph{extension} ($> 5\arcsec$), the \emph{extension likelihood} ($> 33$) and the \emph{detection likelihood} ($> 32$).
This set of parameters defines a subsample of extended sources. This subsample is almost unoccupied by point-like and spurious sources (1-3\% contamination, see Sect.~\ref{sect:selfunc}). It was originally defined by analysing extensive simulations of typical XMM-LSS pointings showing an exposure time of 10~ks and a nominal background level from \citet{readponman03}.
In order to provide a more accurate description of sensitivity fluctuations across the survey area, we extended this set of simulations to a wider range of exposures and background levels (see Sect.~\ref{sect:selfunc}).
Importantly, they show that C1 criteria remain stable against these variations, in the sense that contamination by point-like and spurious sources remain at a their low level, ensuring high purity.
However, it is worth noting that a same physical extended object would be detected with different output parameters depending on its spatial location in the survey. In other terms, the sample completeness slightly varies across the survey.

A total of 58 sources fulfilling C1 criteria were detected in the XMM-LSS survey thanks to the newest processing of each individual pointing.
Of them, 5 are duplicates detected on a neighboring pointing, 3 are actually nearby galaxies emitting X-rays as revealed by the inspection of optical/X-ray overlays and 1 is identified as a substructure in the X-ray emission of XLSSC~50, hence not considered as a distinct source.

All 32 sources presented in \citet{adami11} (C1 sample, their Table~2) were detected by our new processing. However, 6 of them were downgraded and no longer classified as C1: 3 were classified as C2\footnote{Lower-purity sample defined by a decrease in the \emph{extension likelihood} from 33 to 15.} and 3 were deblended as two distinct, non-C1 sources (XLSSC~12, 28 and~74~; see images in App.~\ref{app:clusters} of this paper and Fig.~C1 of \citealt{pacaud07}). In all three cases, such deblending arose because of point sources present close to the cluster centre. It is a direct consequence of the multi-scale approach implemented in our detection algorithm.
On the other hand, 5 sources within the 6~deg$^2$ area explored by \citet{adami11} were promoted as C1 sources with this new processing, 4 of them previously classified as C2.
We attribute these differences to changes in the {\sc XAmin} pipeline version, the event list processing and/or the XMM-SAS version and calibration data. Such changes can be viewed as additional "noise" in the images on top of the usual Poisson and background noise. However they are not handled by the noise model of our pipeline and thus induce variations in the final source list near the selection thresholds.
The maximal list of C1-classified sources (either by one or the other processing runs) contains 54 entries. We manually discarded one detection whose characterization as an extended source is doubtful, since it suffers from severe projection effects due to a bright X-ray emitting star and a large X-ray cluster (XLSSC~061 in Table~\ref{table_c1_catalogue} and Fig.~\ref{fig:c1_gallery}) in the foreground.

	\subsection[]{Cluster validation and redshifts}

Our procedure for confirming and validating C1 sources as galaxy clusters involves optical spectroscopic observations and is fully described in \citet{adami11}.
Spectroscopic follow-up campaigns were dedicated to the confirmation of C1 clusters, making use of different observing facilities in order to cover the range of redshifts encountered in the sample. Selection of spectroscopic targets was based on $ugriz$ optical imaging from the CFHT-LS survey, choosing in priority bright red-sequence galaxies in the vicinity of the cluster X-ray emission.
Each spectrum was reduced and its corresponding redshift measured by several independent persons. A final redshift value and associated quality flag was assigned by a moderator. We refer to \citet{adami11} for the meaning of these flags in terms of redshift reliability.
Additional galaxy redshifts from the VVDS deep and ultradeep surveys \citep{lefevre05} and the Subaru Deep Survey \citep{ueda08} as well as redshifts collected from the NED\footnote{{http://ned.ipac.caltech.edu/}, NASA/IPAC Extragalactic Database.} were added to the sample. Spectroscopic data is stored in {\sc Cesam}\footnote{{http://www.lam.fr/cesam/}, Centre de donn\'ees astrophysiques de Marseille.} and will be publicly released in the end of the XMM-XXL survey.

Cluster redshift validation was first based on identification of groups of galaxies sharing similar radial velocities along a line of sight by using the gapper method \citep[e.g.][]{biviano97}. We assigned membership in each putative group using a physical radius of 500~kpc around the X-ray position, computed using a cosmological angular distance at the mean group redshift. We used catalogues of galaxies with photometric redshifts derived from the CFHT-LS Wide imaging survey and inspected galaxy density maps in photometric redshifts slices centered around each group. A clear overlap between the X-ray isophotes and the density map at the (true) redshift of the source is expected. In this case, the nature of the X-ray source is confirmed as a galaxy cluster and its redshift validated ("C1 confirmed" clusters).
One source out of the 53 selected C1 candidates could not have any related spectroscopic observation and falls outside of the CFHT-LS footprint, preventing the derivation of a photometric redshift. Noting that it was classified as a C2 source by our former pipeline, it is discarded from this analysis.

Table~\ref{table_c1_catalogue} lists the C1 cluster catalogue including all 52 sources validated as bona-fide C1 clusters for this work. It represents a superset of \citet{pacaud07} and \citet{adami11} C1 samples and a subset of the full XMM-LSS source catalogue \citep{chiappetti13}.
All but one (98\%) have spectroscopically validated redshifts. We note that XLSSC~035 and XLSSC~048 are not "C1 confirmed clusters" according to the definition above, because of their current low number of securely identified spectroscopic members \citep[see][for a discussion]{adami11}.
The remaining cluster (XLSSU J021744.1-034536) is located at $z_{\rm phot}=1.9$ on the basis of a photometric redshift analysis involving deep, near-infrared, imaging data \citep{willis13} and included in the present sample since it clearly is detected as a C1~source in X-rays. A discussion of the errors on cluster spectroscopic redshifts can be found in \citet{adami11} and we choose to be conservative in quoting values up to the second decimal in Table~\ref{table_c1_catalogue}.
Three-color images and X-ray overlays non already published in \citet{pacaud07} are presented in App.~\ref{app:clusters}.

\begin{table*}
	\centering
	\caption{\label{table_c1_catalogue} The XMM-LSS C1 sample of galaxy clusters. ($^{a}$: XLSSU J021744.1-034536). Quoted uncertainties reflect 68\% confidence intervals limits. {\bf (1)} Cluster redshift, in brackets is the number of cluster members with spectroscopic redshifts and "L" stands for "literature" (Abell~329). ($^{b}$: photometric redshift from \citealt{willis13}, $^{c}$: see note in Table~2 of \citealt{adami11}). {\bf (2)} Absorbed flux in units $10^{-14}$~ergs/s/cm$^2$ measured in the [0.5-2]~keV band in a circular aperture of 0.5~Mpc at the cluster redshift. {\bf (3)} As computed from a $M_{500}-T$ relation (see text). {\bf (4)} Bolometric luminosity within $R_{500}$, units $10^{43}$~ergs/s. {\bf (5)} Rest-frame [0.5-2]~keV luminosity, units $10^{43}$~ergs/s. {\bf (6)-(7)} Mass estimates, units $10^{13} h^{-1} M_{\odot}$ from two different methods (Sect.~\ref{mass_method1} and~\ref{mass_method2}). {\bf (8)} 1: \citet{pacaud07}, 2: \citet{berge08}, 3: \citet{adami11}, 4: \citet{willis13}, 5: \citet{abell89}.
}
\begin{tabular}{cccccccccccccc}
\hline
{\sc xlssc}	&	R.A.	&	Dec	&	$z$ [$N_z$]	&	$R_{\rm spec}$	&	$F_{14}^{[0.5-2]}$	&	$T_X$	&	$R_{500}$	&	$L_{500}^{bol}$	&	$L_{500}^{[0.5-2]}$	&	$M_{500}$	&	$M_{500}$	&	Ref.	\\
	&	J2000	&	J2000	&		&	(\arcsec)	&	0.5 Mpc	&	(keV)	&	(Mpc)	&	(ergs/s)	&	(ergs/s)	&	M1	&	M2	&	\\
	&		&		&	{\bf (1)	}&		&	{\bf (2)}	&		&	{\bf (3)}	&	{\bf (4)}	&	{\bf (5)}	&	{\bf (6)	}&	{\bf (7)	}&	{\bf (8)}	\\
\hline
060	&	33.668	&	-4.552	&	0.14	 [L]&	360	&	$116.0 \pm 1.0$	&	$5.2 \pm 0.1$	&	1.088	&	$29.0 \pm 0.3$	&	$8.4 \pm 0.1$	&	35	&	13	&	5\\
076	&	33.682	&	-3.823	&	0.75	 [6]&	72	&	$2.1 \pm 0.3$	&	$1.3 \pm 0.1$	&	0.350	&	$10.3 \pm 1.3$	&	$4.9 \pm 0.6$	&	2.0	&	6.0	&	-\\
072	&	33.850	&	-3.726	&	1.00	 [7]&	54	&	$4.0 \pm 0.4$	&	$3.5_{-0.6}^{+1.0}$	&	0.526	&	$56.3 \pm 5.2$	&	$19.7 \pm 1.8$	&	5.6	&	10	&	4	\\
056	&	33.870	&	-4.681	&	0.35 [6]	&	135	&	$10.7 \pm 0.5$	&	$3.5_{-0.7}^{+0.9}$	&	0.783	&	$12.6 \pm 0.8$	&	$4.4 \pm 0.3$	&	13	&	7.8	&	-\\
078	&	33.948	&	-4.842	&	0.95 [3]	&	36	&	$1.2 \pm 0.2$	&	$3.3_{-0.8}^{+1.3}$	&	0.524	&	$15.1 \pm 2.2$	&	$5.4 \pm 0.8$	&	8.8	&	5.3	&	4	\\
057	&	34.051	&	-4.242	&	0.15 [16]	&	90	&	$18.3 \pm 0.9$	&	$2.0_{-0.2}^{+0.4}$	&	0.645	&	$3.1 \pm 0.2$	&	$1.3 \pm 0.1$	&	4.8	&	4.3	&	5\\
065	&	34.245	&	-4.821	&	0.43	 [3]&	55	&	$1.0 \pm 0.2$	&	$2.7_{-0.7}^{+1.2}$	&	0.645	&	$1.6 \pm 0.4$	&	$0.6 \pm 0.1$	&	8.9	&	2.7	&	3	\\
059	&	34.391	&	-5.223	&	0.65 [8]&	75	&	$1.4 \pm 0.2$	&	$2.7_{-0.5}^{+0.6}$	&	0.563	&	$6.5 \pm 0.7$	&	$2.5 \pm 0.3$	&	6.0	&	4.5	&	3	\\
-$^{a}$	&	34.433	&	-3.760	&	1.9$^{b}$	&	45	&	$1.3 \pm 0.2$	&	$6.6_{-2.2}^{+5.4}$	&	0.465	&	$91.6 \pm 14.8$	&	$23.5 \pm 3.8$	&	13	&	7.3	&	4	\\
079	&	34.494	&	-4.863	&	0.19	 [7]&	50	&	$0.6 \pm 0.2$	&	$2.2 \pm 0.4$	&	0.651	&	$0.12 \pm 0.06$	&	$0.05 \pm 0.03$	&	7.1	&	0.9	&	-\\
077	&	34.522	&	-3.656	&	0.20	 [4]&	80	&	$4.5 \pm 0.8$	&	$1.6_{-0.3}^{+1.2}$	&	0.541	&	$1.4 \pm 0.2$	&	$0.6 \pm 0.1$	&	4.1	&	2.7	&	-\\
080	&	34.605	&	-5.415	&	0.65	 [5]&	65	&	$1.2 \pm 0.1$	&	$1.5_{-0.2}^{+0.5}$	&	0.409	&	$4.1 \pm 0.6$	&	$1.9 \pm 0.3$	&	1.7	&	4.0	&	-\\
064	&	34.633	&	-5.016	&	0.88	 [3]&	72	&	$1.4 \pm 0.1$	&	$2.0_{-0.2}^{+0.4}$	&	0.424	&	$12.1 \pm 0.9$	&	$5.1 \pm 0.4$	&	4.4	&	5.5	&	3	\\
058	&	34.938	&	-4.891	&	0.33	 [9]&	60	&	$1.8 \pm 0.2$	&	$2.4_{-0.5}^{+0.7}$	&	0.631	&	$1.8 \pm 0.2$	&	$0.7 \pm 0.1$	&	6.1	&	2.8	&	3	\\
039	&	35.098	&	-2.841	&	0.23	 [4]&	108	&	$1.9 \pm 0.4$	&	$1.0 \pm 0.1$	&	0.417	&	$0.6 \pm 0.1$	&	$0.3 \pm 0.1$	&	2.1	&	2.0	&	1,3	\\
023	&	35.189	&	-3.433	&	0.33	 [3]&	63	&	$4.2 \pm 0.4$	&	$1.9 \pm 0.2$	&	0.565	&	$3.9 \pm 0.4$	&	$1.6 \pm 0.2$	&	4.1	&	4.3	&	1,3	\\
006	&	35.438	&	-3.772	&	0.43	 [16]&	117	&	$24.7 \pm 0.8$	&	$5.6_{-0.5}^{+0.7}$	&	0.964	&	$67.6 \pm 2.3$	&	$18.9 \pm 0.6$	&	24	&	16	&	1,3	\\
061	&	35.485	&	-5.757	&	0.26 [10]	&	78	&	$5.9 \pm 0.7$	&	$1.8_{-0.2}^{+0.6}$	&	0.570	&	$2.9 \pm 0.4$	&	$1.3 \pm 0.2$	&	5.2	&	3.9	&	-\\
040	&	35.523	&	-4.546	&	0.32	 [16]&	42	&	$2.1 \pm 0.3$	&	$3.9_{-1.4}^{+3.0}$	&	0.834	&	$1.8 \pm 0.4$	&	$0.6 \pm 0.1$	&	16	&	3.1	&	1,3	\\
036	&	35.527	&	-3.054	&	0.49	 [3]&	54	&	$10.7 \pm 0.6$	&	$3.8_{-0.5}^{+0.6}$	&	0.748	&	$30.6 \pm 1.8$	&	$10.3 \pm 0.6$	&	12	&	11	&	1,3	\\
047	&	35.544	&	-2.680	&	0.79 [14]	&	60	&	$1.6 \pm 0.3$	&	$3.2_{-1.1}^{+2.2}$	&	0.567	&	$13.7 \pm 2.2$	&	$5.0 \pm 0.8$	&	7.9	&	5.9	&	1,3	\\
048	&	35.722	&	-3.474	&	1.00 [2$^{c}$]	&	27	&	$1.3 \pm 0.2$	&	$3.0_{-1.0}^{+1.9}$	&	0.482	&	$17.1 \pm 3.1$	&	$6.3 \pm 1.2$	&	6.9	&	5.4	&	1,3	\\
075	&	35.834	&	-5.454	&	0.21	 [9]&	63	&	$2.5 \pm 0.4$	&	$1.09_{-0.03}^{+0.04}$	&	0.442	&	$0.7 \pm 0.1$	&	$0.36 \pm 0.05$	&	2.5	&	2.3	&	-\\
035	&	35.950	&	-2.858	&	0.07	 [1$^{c}$]&		60	&	$6.1 \pm 1.2$	&	$1.1 \pm 0.1$	&	0.468	&	$0.16 \pm 0.03$	&	$0.08 \pm 0.01$	&	1.3	&	1.1	&	1,3	\\
028	&	35.985	&	-3.100	&	0.30	 [8]&		45	&	$2.0 \pm 0.4$	&	$1.2_{-0.1}^{+0.2}$	&	0.437	&	$1.2 \pm 0.2$	&	$0.6 \pm 0.1$	&	2.4	&	2.7	&	1,3	\\
049	&	35.989	&	-4.588	&	0.50	 [4]&	35	&	$2.0 \pm 0.2$	&	$2.9_{-0.7}^{+1.3}$	&	0.642	&	$5.5 \pm 0.7$	&	$2.0 \pm 0.2$	&	5.7	&	4.4	&	1,3	\\
018	&	36.008	&	-5.090	&	0.32	 [9]&	90	&	$1.7 \pm 0.2$	&	$1.5 \pm 0.2$	&	0.492	&	$1.4 \pm 0.2$	&	$0.6 \pm 0.1$	&	3.7	&	2.5	&	1,3	\\
029	&	36.017	&	-4.225	&	1.05 [5]	&	50	&	$3.3 \pm 0.2$	&	$4.5_{-1.0}^{+1.1}$	&	0.593	&	$56.1 \pm 4.1$	&	$17.4 \pm 1.3$	&	9.4	&	11	&	1,4	\\
062	&	36.060	&	-2.721	&	0.06	 [4]&	60	&	$8.2 \pm 2.8$	&	$0.8 \pm 0.1$	&	0.392	&	$0.18 \pm 0.04$	&	$0.10 \pm 0.02$	&	1.0	&	1.1	&	-\\
053	&	36.114	&	-4.836	&	0.50	 [7]&	60	&	$2.7 \pm 0.3$	&	$6.7_{-2.8}^{+8.3}$	&	1.022	&	$8.7 \pm 1.7$	&	$2.2 \pm 0.4$	&	32	&	6.0	&	2	\\
044	&	36.141	&	-4.234	&	0.26	 [17]&	50	&	$2.7 \pm 0.3$	&	$1.2 \pm 0.1$	&	0.458	&	$1.2 \pm 0.1$	&	$0.6 \pm 0.1$	&	2.0	&	2.5	&	1,3	\\
021	&	36.234	&	-5.134	&	0.08	 [7]&	27	&	$4.2 \pm 0.7$	&	$0.76_{-0.05}^{+0.04}$	&	0.387	&	$0.16 \pm 0.02$	&	$0.09 \pm 0.01$	&	1.3	&	1.0	&	1,3	\\
001	&	36.238	&	-3.816	&	0.61	 [17]&	90	&	$7.9 \pm 0.4$	&	$3.3_{-0.4}^{+0.6}$	&	0.644	&	$36.9 \pm 2.0$	&	$13.2 \pm 0.7$	&	8.6	&	11	&	1,3	\\
054	&	36.320	&	-5.888	&	0.05	 [25]&	117	&	$47.5 \pm 2.0$	&	$1.8 \pm 0.1$	&	0.631	&	$0.86 \pm 0.04$	&	$0.37 \pm 0.02$	&	3.6	&	2.4	&	-\\
008	&	36.337	&	-3.801	&	0.30	 [11]&	63	&	$1.9 \pm 0.3$	&	$1.3_{-0.1}^{+0.3}$	&	0.461	&	$1.2 \pm 0.2$	&	$0.6 \pm 0.1$	&	2.8	&	2.5	&	1,3	\\
025	&	36.353	&	-4.679	&	0.27	 [10]&	63	&	$8.9 \pm 0.5$	&	$2.1_{-0.1}^{+0.2}$	&	0.611	&	$5.1 \pm 0.3$	&	$2.1 \pm 0.1$	&	4.8	&	5.3	&	1,3	\\
041	&	36.377	&	-4.239	&	0.14	 [6]&	81	&	$21.8 \pm 1.1$	&	$1.6 \pm 0.1$	&	0.573	&	$2.8 \pm 0.2$	&	$1.2 \pm 0.1$	&	3.6	&	4.2	&	1,3	\\
002	&	36.384	&	-3.920	&	0.77	 [8]&	36	&	$2.7 \pm 0.3$	&	$2.6_{-0.5}^{+0.6}$	&	0.518	&	$19.3 \pm 2.0$	&	$7.5 \pm 0.8$	&	6.9	&	7.4	&	1,3	\\
050	&	36.419	&	-3.189	&	0.14	 [13]&	86	&	$55.4 \pm 1.1$	&	$3.3_{-0.2}^{+0.3}$	&	0.838	&	$9.3 \pm 0.2$	&	$3.3 \pm 0.1$	&	15	&	7.6	&	1,3	\\
055	&	36.452	&	-5.895	&	0.23	 [13]&	117	&	$14.0 \pm 0.7$	&	$3.0_{-0.5}^{+0.6}$	&	0.758	&	$6.6 \pm 0.4$	&	$2.5 \pm 0.2$	&	8.5	&	5.9	&	-\\
051	&	36.498	&	-2.826	&	0.28	 [6]&	99	&	$1.1 \pm 0.3$	&	$1.4_{-0.1}^{+0.2}$	&	0.487	&	$0.6 \pm 0.2$	&	$0.3 \pm 0.1$	&	3.4	&	1.9	&	1,3	\\
011	&	36.541	&	-4.968	&	0.05	 [8]&	90	&	$12.3 \pm 1.3$	&	$0.78 \pm 0.05$	&	0.399	&	$0.16 \pm 0.01$	&	$0.09 \pm 0.01$	&	1.0	&	1.1	&	1,3	\\
052	&	36.568	&	-2.665	&	0.06	 [5]&	72	&	$12.2 \pm 1.6$	&	$0.68_{-0.03}^{+0.02}$	&	0.369	&	$0.19 \pm 0.02$	&	$0.10 \pm 0.01$	&	1.4	&	1.1	&	1,3	\\
009	&	36.686	&	-3.684	&	0.33 [8]	&	54	&	$2.2 \pm 0.5$	&	$0.8_{-0.1}^{+0.2}$	&	0.345	&	$1.5 \pm 0.3$	&	$0.8 \pm 0.1$	&	1.2	&	2.7	&	3	\\
005	&	36.788	&	-4.300	&	1.06	 [4]&	63	&	$1.0 \pm 0.1$	&	$2.7_{-0.7}^{+1.2}$	&	0.442	&	$14.9 \pm 2.0$	&	$5.8 \pm 0.8$	&	5.8	&	5.3	&	1,3	\\
010	&	36.843	&	-3.362	&	0.33 [5]	&	54	&	$5.8 \pm 0.5$	&	$2.5 \pm 0.4$	&	0.648	&	$5.3 \pm 0.6$	&	$2.1 \pm 0.2$	&	6.9	&	5.2	&	1,3	\\
013	&	36.858	&	-4.538	&	0.31	 [19]&	54	&	$2.4 \pm 0.3$	&	$1.2_{-0.2}^{+0.1}$	&	0.434	&	$1.5 \pm 0.2$	&	$0.7 \pm 0.1$	&	1.9	&	2.8	&	1,3	\\
003	&	36.909	&	-3.299	&	0.84	 [9]&	54	&	$3.8 \pm 0.4$	&	$3.2_{-0.7}^{+1.0}$	&	0.556	&	$34.5 \pm 3.6$	&	$12.5 \pm 1.3$	&	6.6	&	9.5	&	1,3	\\
022	&	36.916	&	-4.857	&	0.29	 [15]&	63	&	$9.8 \pm 0.3$	&	$2.0 \pm 0.1$	&	0.584	&	$6.8 \pm 0.2$	&	$2.9 \pm 0.1$	&	5.2	&	6.1	&	1,3	\\
027	&	37.014	&	-4.851	&	0.29	 [6]&	60	&	$6.4 \pm 0.4$	&	$2.9_{-0.5}^{+0.6}$	&	0.720	&	$5.2 \pm 0.4$	&	$2.0 \pm 0.1$	&	10	&	5.1	&	1,3	\\
074	&	37.034	&	-5.597	&	0.19	 [7]&	72	&	$4.2 \pm 0.8$	&	$1.1 \pm 0.1$	&	0.445	&	$0.9 \pm 0.2$	&	$0.5 \pm 0.1$	&	2.4	&	2.2	&	-\\
012	&	37.114	&	-4.432	&	0.43	 [5]&	117	&	$3.2 \pm 0.3$	&	$1.6_{-0.2}^{+0.1}$	&	0.474	&	$4.9 \pm 0.4$	&	$2.2 \pm 0.2$	&	2.7	&	5.0	&	3	\\
\hline
\end{tabular}
\end{table*}

\section[]{Characterization of the survey}
\label{sect:selfunc}
In order to obtain statistically relevant results from the present cluster sample, we derive a survey selection function. It is straightforwardly associated to the detection process described in previous section and our procedure follows closely \citet{pacaud06, elyiv12, clerc12b}.
	 
	\subsection[]{Simulations and observational selection function}

We first compute a selection function for each individual pointing. To this purpose, we designed simulations of realistic XMM observations containing point-sources representative of the observed AGN logN-logS \citep{moretti03} and extended sources. The latter are represented as $\beta$-models with $\beta=2/3$, with varying core radii and fluxes, placed at random positions on the detectors.
Our simulations span a variety of detector exposure times ($T_{exp}=$ 3, 7, 10, 20, 40\,ks) and particle background levels ($b=$ 0.25, 0.5, 1, 2, 4 times the nominal value of \citealt{readponman03}).
For each value of the exposure time ($T_{exp}$), of the background value ($b$), cluster extent (input core-radius $ext$) and input count-rate ($cr$), 2500 to 3500 fake XMM pointings were produced, each of them containing 4 to 8 extended sources. All simulated observations went through the same detection and characterization pipeline as used for real data.
As discussed earlier, our simulations show that contamination of a C1 selected sample by point-like and spurious sources is very low, ranging from 1\% to 3\% depending on exposure time and background level. This contamination rate results from simulations where the point-source population is spatially uncorrelated from the extended sources. We will discuss in Sect.~\ref{sect:discussion} an extreme case, in which point-sources systematically populate the centre of clusters in simulations.
The observational selection function is the probability of detecting and characterizing an extended source as a C1. Output detection lists were correlated in position with input source lists, and the selection function $P_{C1,{[\theta_1,\theta_2]}}(T_{exp},b,cr,ext)$ was derived by selecting only sources satisfying the C1 criteria (see Sect.~\ref{sect:construction_sample}) and detected within a given a range of off-axis values $[\theta_1,\theta_2]$.

	\subsection[]{Selection function of the actual dataset}
The next step in characterizing the survey selection function consists of linking simulations to real observations.
Each actual XMM pointing is assigned its proper selection function by interpolating along the $(T_{exp},b)$ grid used for simulations.
We estimated background levels ($b$) in real observations by matching the average local background level around each detected point-like source to the value seen in simulated pointings, similarly to \citet{elyiv12,clerc12b}.

\begin{figure}
	\includegraphics[width=84mm]{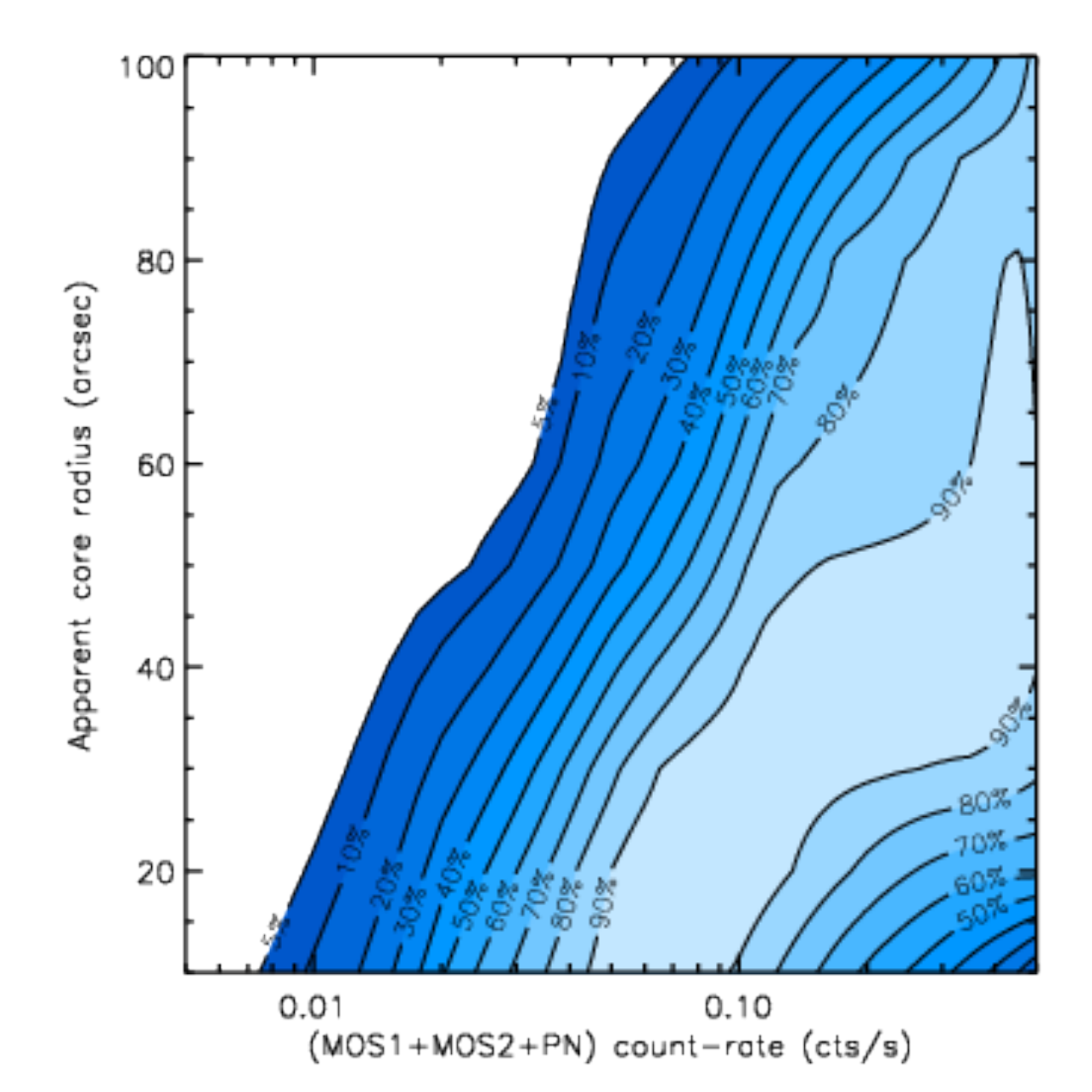}
	\caption{The average XMM-LSS 11~deg$^2$ C1 selection function in X-ray observables domain, as derived from simulations of realistic XMM-LSS observations. Contours represent the probability of detecting and classifying as C1 an extended source with a surface brightness profile following a $\beta$-model ($\beta=2/3$) of given core radius and given total flux (or count-rate). Pointing to pointing differences in sensitivity and pointings overlaps are taken into account and weighted according to their area on sky.}
	\label{fig:survey_selfunc} 
\end{figure}

The 8.1~deg$^2$ consisting of non-overlapping area (i.e. unique to each pointing) was treated by averaging all 98 individual pointing selection functions calculated up to 10$\arcmin$ off-axis radius (i.e. $[\theta_1,\theta_2] = [0,10\arcmin]$).
The remaining 3~deg$^2$ mainly (97\%) consist of area shared by exactly two adjacent pointings, whose centers are spaced by approximately  20-25~arcmin. Since cluster detection is performed independently on each pointing, we estimated the selection function of each overlap between pointing $A$ and pointing $B$ by:
\begin{equation}
	\label{equ_proba_overlap}
	P(AB,{\rm overlap}) = P(A)+P(B)-P(A)P(B)
\end{equation}
where $P(A)$ and $P(B)$ are the respective selection functions of each pointing, considering a $[\theta_1,\theta_2] = [10\arcmin,13\arcmin]$ range.
Fig.~\ref{fig:survey_selfunc} shows the averaged selection function of the 11~deg$^2$ area. It takes into account each unique (non-overlapping) patch of the survey as well as overlaps, weighted by their respective geometrical area.
The noticeable sharp decrease in probability for high-countrate, small-size, sources corresponds to a morphological misclassification into point-like sources by the automated algorithm.

\begin{figure}
	\includegraphics[width=84mm]{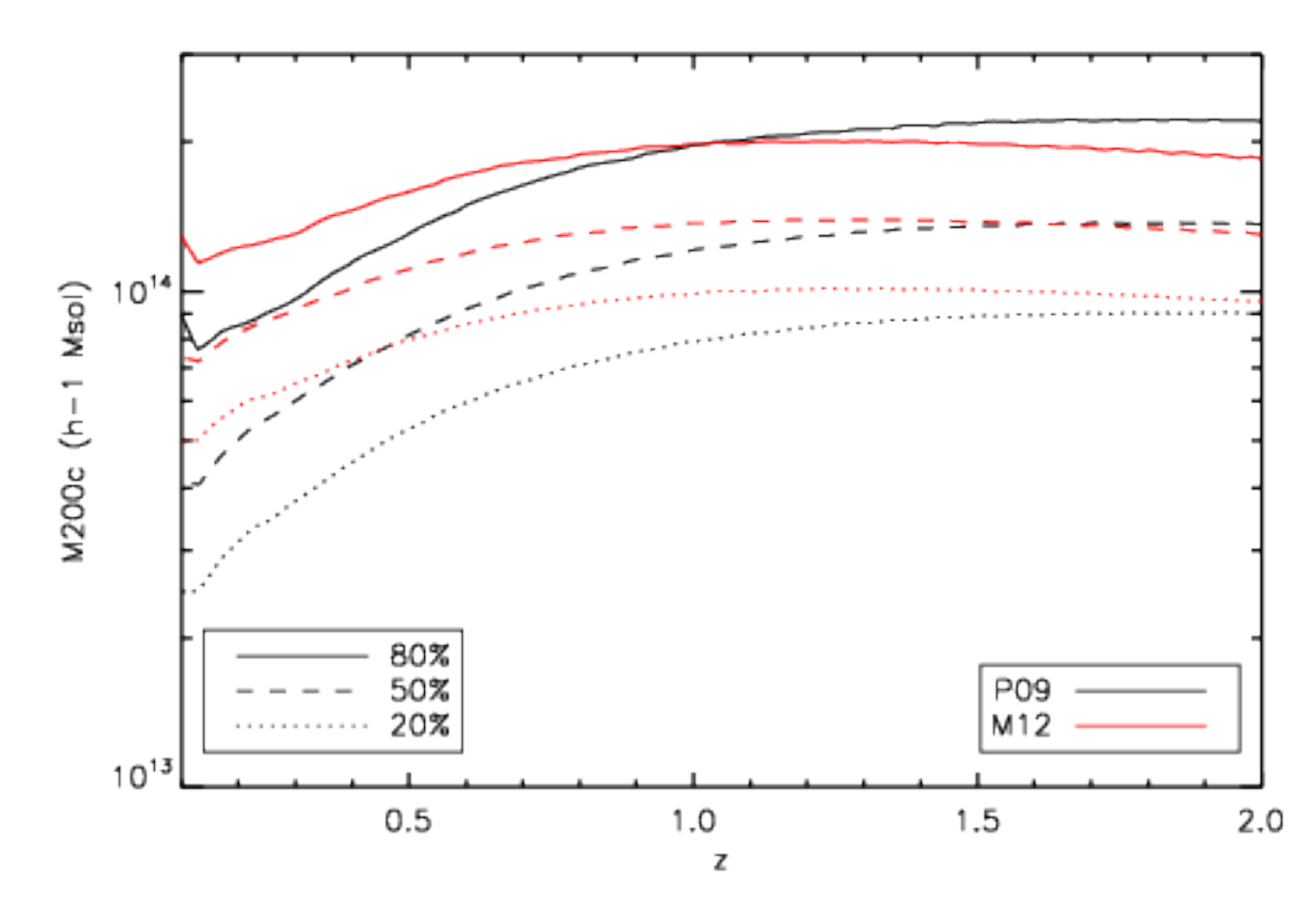}
	\caption{The XMM-LSS 11~deg$^2$ C1 selection function in mass-redshift plane, as derived from the observational function shown on Fig.~\ref{fig:survey_selfunc}. Planck CMB cosmology is assumed, conversion from mass to temperature follows \citet{arnaudpointecouteau05} and two different luminosity-temperature relations are tested: M12 \citep[]['ALL' sample]{maughan12} and P09 \citep[]['ALL' sample]{pratt09}, both following self-similar evolution.}
	\label{fig:survey_selfunc_mz} 
\end{figure}

Such a formulation of the survey selection function only depends on observational quantities (apparent size and flux). Assuming a cosmological model and a set of mass-observable relations, it can be rewritten in terms of a limiting mass as a function of redshift. We provide an example in Fig.~\ref{fig:survey_selfunc_mz}, assuming Planck~2013 cosmology and illustrate the changes due to the choice of model, specifically the luminosity-temperature relation.

\section[]{Characterization of individual cluster properties}
	\label{sect:cluster_properties}
		
	\subsection[]{X-ray spectral properties}
	\label{sect:spectral_fit}
An X-ray spectrum was extracted around each cluster position in a circular aperture. Similarly to \citet{pacaud07}, a background annulus is chosen so that it does not contain emission from the cluster. The spectral extraction radius ($R_{\rm spec}$) is optimized on the basis of the signal-to-noise estimated from the cluster surface-brightness profile.
Background-subtracted spectra are fitted with {\sc XSpec} v.12.8.0 \citep{arnaudxspec} using a single-temperature APEC plasma model (v.2.0.1) and assuming a galactic hydrogen density column given by \citet{kalberla2005}. Metallicity abundances were fixed at 0.3 times the solar value, except for XLSSC~60\footnote{Its spectrum contains enough photons to enable a simultaneous fit of temperature and metallicity and we find an abundance value of $0.29\pm 0.06$~$Z_{\odot}$ within $R_{\rm spec}$.}.
The median temperature measured in the sample is 2.1~keV, with a typical uncertainty of $\sim 15$\%.

A comparison with previously measured values for the 29 clusters in common with \citet{pacaud07} is presented in Appendix~\ref{app:compa_pacaud}, with an attempt to disentangle between the different causes of discrepancies. We reach the conclusion that a change in APEC models slightly impacts the temperatures for the coolest systems, while other results agree well within the error bars: changes in the X-ray processing, {\sc XSpec} version and plasma models only create scatter around the one-to-one relation.

Assuming the $M_{500}-T_X$ relation of \citet{sun09} (converted into a $R_{500c}-T_X$ relation, see their Table~6 for the "Tier 1-2+clusters" sample), we assign a value of $R_{500c}$ to our clusters:

\begin{equation}
\label{eq_r500_sun09}
R_{500} ({\rm Mpc}) = 0.600 \, h^{-1} \, \Big( \frac{T}{3 {\rm keV}} \Big) ^{0.55} \, E(z)^{-1}
\end{equation}
where $E(z) = H(z)/H_0$ is the normalized Hubble constant.

	\subsection[]{Flux and luminosities measurements}
	\label{sect:flux_mes}
	
		\subsubsection[]{Net count-rates and physical fluxes}
X-ray cluster fluxes were measured in the [0.5-2]~keV band directly on images created from cleaned event lists. Two methods were tested and compared: (i)~modelling the surface brightness radial profile, and (ii)~integrating the source flux in growing circular apertures ("growth curve analysis" \citealt[][]{boehringer04, suhada12}), as applied in \citet{adami11, clerc12b}. We detail here our procedures.

\begin{figure}
	\includegraphics[width=84mm]{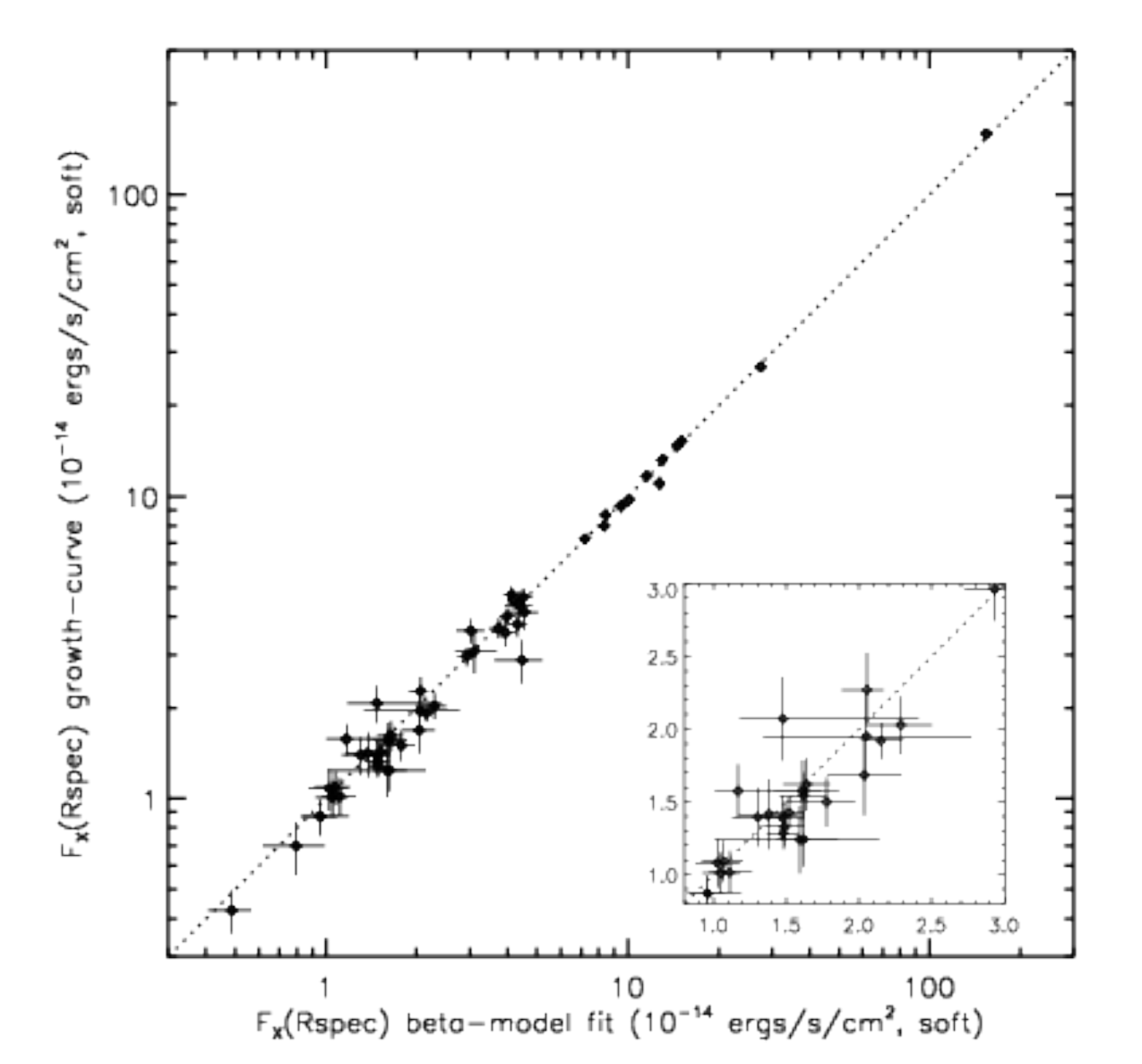}
	\caption{Comparison of [0.5-2]~keV absorbed fluxes derived with two different methods for all clusters in the sample. Measurements reported on the x-axis were obtained by fitting a $\beta$-model to the X-ray surface brightness profile. The y-axis corresponds to an aperture photometry measurement (growth curve analysis). This comparison is performed within a radius $R_{\rm spec}$, defined as an optimal signal-to-noise extraction region (see text). The inset zooms over a crowded area in the plot.}
	\label{fig:comp_flux} 
\end{figure}

(i) The first method follows \citet{pacaud07} by assuming a one-dimensional $\beta$-model \citep{cavalierefusco} and three free parameters: angular core-radius, $\beta$ and normalization. A local background level is estimated by means of a double-component model (vignetted and unvignetted) adjusted in a source-free area over each of the three XMM EPIC detectors.
A local, one-dimensional, analytic PSF model \citep[as in][]{arnaud2002} is convolved to the model $\beta$-profile and accounts for the telescopes spatial resolution. Model and data profiles are binned to ensure a minimal 3-$\sigma$ signal-to-noise ratio in each bin. $\chi^2$ statistics provide a best-fit value and confidence levels on a $\beta$-core radius grid. The normalization is derived from the number of counts collected in the fit area.
Given the generally low signal-to-noise ratios of C1 clusters and the high background levels in XMM images, $\chi^2$ contours are degenerate in the two-dimensional $\beta$-core radius parameter space \citep[e.g.][]{alshino10}. We ranked $\chi^2$ contours and surface brightness profiles according to their level of degeneracy and found that objects with more than 300 net counts in the [0.5-2]~keV band provide well-behaved $\chi^2$ surfaces. For those 30 clusters, the 3-parameter model is then considered as a good description of its surface brightness profile. For the remaining 22 objects, we instead forced $\beta=2/3$ and only derived a best-fit core-radius associated to this particular choice of prior.
The surface-brightness model corresponding to the set of best-fit parameters can then be integrated up to any given radius and provides the cluster flux within the corresponding aperture (e.g.~$R_{\rm spec}$ in Fig.~\ref{fig:comp_flux}).

(ii) The second method does not assume any model as for the cluster surface brightness profile. Cluster emissivity is integrated in circular annuli of growing sizes around the cluster centroid. Background subtraction is controlled via an annulus whose size is adjusted by hand, and a double-component model (vignetted+unvignetted) is fitted over the pixels inside the annulus. This background model is then transported at the source location. Surrounding sources detected by the detection pipeline are masked out and any removed area is accounted for by assuming circular symmetry of the count-rate profile.
Since this method involves manual intervention, we checked for its robustness by comparing the results of two independent measurers (T.~Sadibekova, N.~Clerc). In all 22 cases that underwent this comparison we found results in agreement within 1-$\sigma$ error bars.

Finally, conversion factors from count-rates to physical, galactic-absorbed, fluxes were computed using {\sc XSpec} and the best-fit APEC model derived from X-ray spectral analysis (Sect.~\ref{sect:spectral_fit}).
Fig.~\ref{fig:comp_flux} displays a comparison of fluxes measured by methods (i) and (ii). It shows that both agree within their 1-$\sigma$ error bars.
This comparison is shown within the radius $R_{\rm spec}$, specifically chosen for maximizing the signal-to-noise ratio of each cluster emissivity (see Sect.~\ref{sect:spectral_fit} and numerical values in Table~\ref{table_c1_catalogue}). Hence, discrepancies due to $\beta$-model extrapolation or background removal uncertainties are kept at their lowest level.
Finally, we show the good agreement between these calculations and previously published values \citep{pacaud07} in Appendix~\ref{app:compa_pacaud}.

		\subsubsection[]{Bolometric and rest-frame band luminosities}
	\label{sect:lum_comput}

We derived X-ray bolometric and [0.5-2]~keV (rest-frame) luminosities by combining physical fluxes measured in the [0.5-2]~keV (observer frame) band with the best-fit spectral model found in Sect.~\ref{sect:spectral_fit}. We quote in Table~\ref{table_c1_catalogue} luminosities measured within $R_{500}$ as estimated for each cluster (Eq.~\ref{eq_r500_sun09}). The median bolometric (resp. soft-band) luminosity of our sample is $4.9 \times 10^{43}$~ergs/s (resp. $2.0 \times 10^{43}$~ergs/s) within $R_{500}$. Individual uncertainties are dominated by count-rate measurement uncertainties (including background removal) and their median level is $11$\%.

	\subsection[]{Mass estimates}
	\label{sect:mass_estimates}
	
	The ultimate quantity of interest describing galaxy clusters is their individual, total, mass as derived from X-ray data. Given the low signal-to-noise ratios associated to each cluster, we choose to only provide rough estimates and no error bars. However, we illustrate results obtained by two different methods, each one depending on different observables.
	
		\subsubsection[]{\label{mass_method1}Method 1: isothermal, hydrostatic equilibrium}
Similarly to \citet{pacaud07} we can estimate the mass within $R_{500}$ by using: 
\begin{enumerate}
\item a mass-temperature relation \citep[][see Eq.~\ref{eq_r500_sun09}]{sun09} in order to obtain a value of $R_{500}$,
\item the best-fit $\beta$-model profile injected into the equation of hydrostatic equilibrium under the assumption of isothermality of the intra-cluster medium \citep[e.g.][]{ettori2000}, which reads:
\end{enumerate}
\begin{equation}
\label{equ_mass_beta}
M_{500} (M_{\odot}) = 1.11 \times 10^{14} \beta \, R_c \, T \, \frac{x_{500}^3}{1+x_{500}^2},
\end{equation}
where $\beta$ and $R_c$~(Mpc) are the best-fit $\beta$-model parameters found in Sect.~\ref{sect:flux_mes}, $T$ is expressed in~keV and $x_{500}=R_c/R_{500}$.

Mass estimates obtained by this method are listed in Table~\ref{table_c1_catalogue} under column label "M1". In principle, we note that assuming a $R_{500}-T$ relation is unnecessary and redundant since the equation linking $R_{500}$ to $M_{500}$ is univocal. We discuss our choice further in Appendix~\ref{app:m500stabil}.

		\subsubsection[]{\label{mass_method2}Method 2: luminosity-mass relation}

The second method starts from growth-curve flux measurements. They provide luminosities integrated within a cylindrical aperture of any given size. We make use of a $L_X-M_{500}$ relation from \citet[][]{sun12} and iteratively find the value of $R_{500}$ that leads to a converged set of $L_{X,500}$ and $M_{500}$ values. This approach is similar to the one presented by \citet[][]{suhada12} for the XMM-SPT cluster sample. However, in contrast to the XMM-SPT clusters, temperature measurements are available in this work. They are used to model the cluster X-ray spectrum (using APEC v.2.0.2) and convert from instrumental count-rates to physical quantities.
We checked that this iterative analysis returns values of $L_{X,500}$ consistent with the computation described in Sect.~\ref{sect:lum_comput} within their respective error bars, although the underlying scaling relations used in deriving the radius $R_{500}$ differ.
Values obtained with this method are quoted in Table~\ref{table_c1_catalogue} under column label "M2".

		\subsubsection[]{Comparison of values}
\begin{figure}
	\includegraphics[width=84mm]{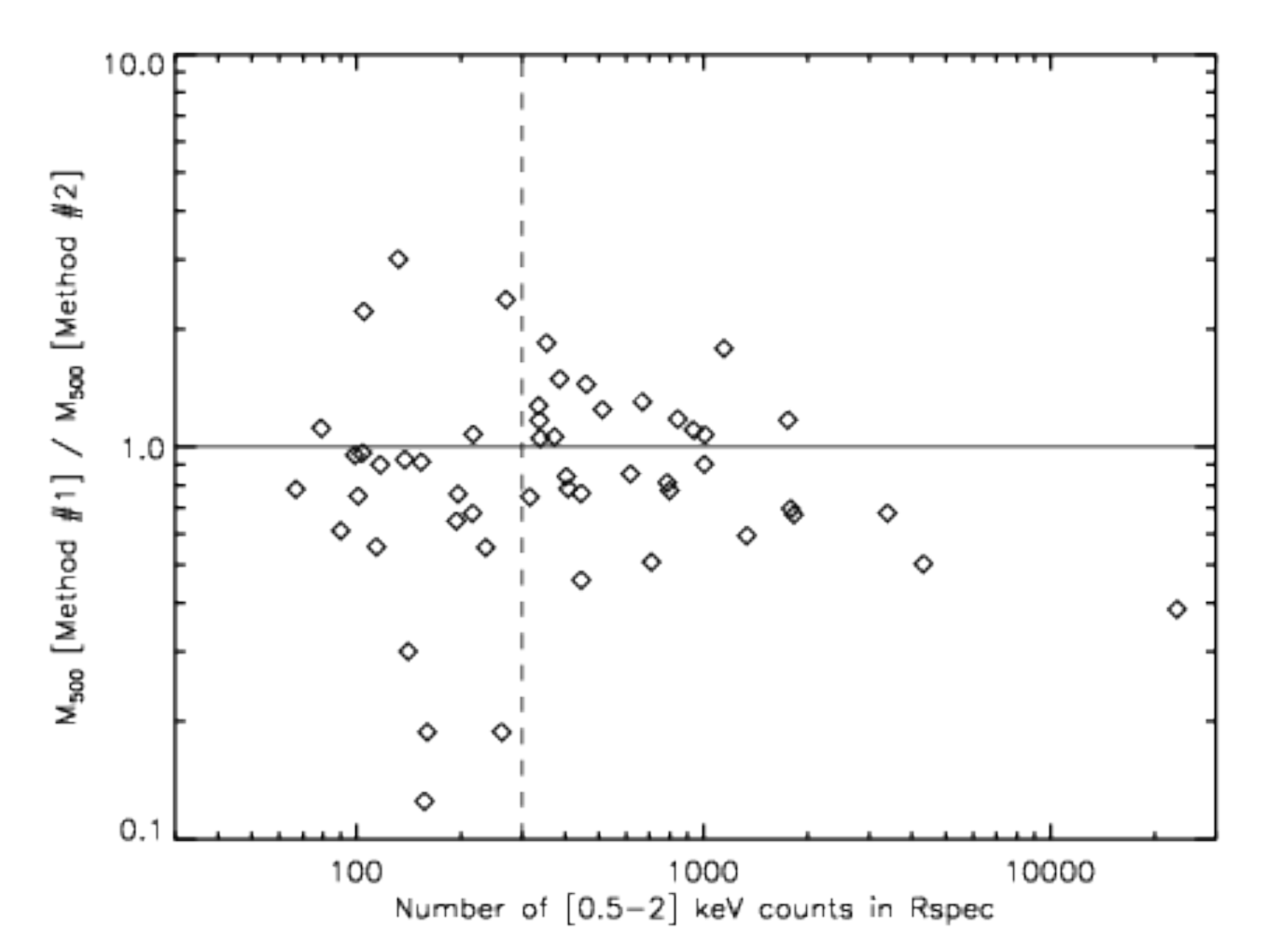}
	\caption{Comparison of mass estimates for the 52 clusters in this sample. These are obtained with two independent method, both based on X-ray data: \emph{Method~1} assumes hydrostatic equilibrium and a gas distribution following a $\beta$-model. \emph{Method~2} relies on a given luminosity-mass relation \citep{sun12}.
	Clusters located right of the dashed line have their surface brightness profile well-described by a 2-parameter $\beta$-model, while clusters on the left were imposed $\beta=2/3$ (see text).}
	\label{fig:comp_masses} 
\end{figure}

Figure~\ref{fig:comp_masses} compares mass estimates obtained from these two methods. Although the scatter is high and differences up to a factor of $\sim 5$ are clearly apparent, the broad agreement is encouraging and confirms the validity of our X-ray analysis. Indeed, both methods rely on distinct X-ray observables. Method~1 does not require knowledge of the total luminosity (or flux) of the cluster, while this is a key ingredient of Method~2. Conversely, Method~2 requires knowledge of the gas temperature only for converting from count-rates to luminosities (which includes a $k$-correction), while it directly enters Method~1 (see Eq.~\ref{equ_mass_beta}). The correlation between these values thus reflects the intrinsic correlation between mass, luminosity and temperature in the intra-cluster medium.
Based on these results, we find that the median mass ($M_{500c}$) of clusters in the present sample is $\sim 5 \times 10^{13}$~$h^{-1} \, M_{\odot}$.
		
\section[]{Results}
\label{sect:results}

	\subsection[]{Redshift distribution}
	\label{sect:redshift_distribution}

\begin{figure}
	\includegraphics[width=84mm]{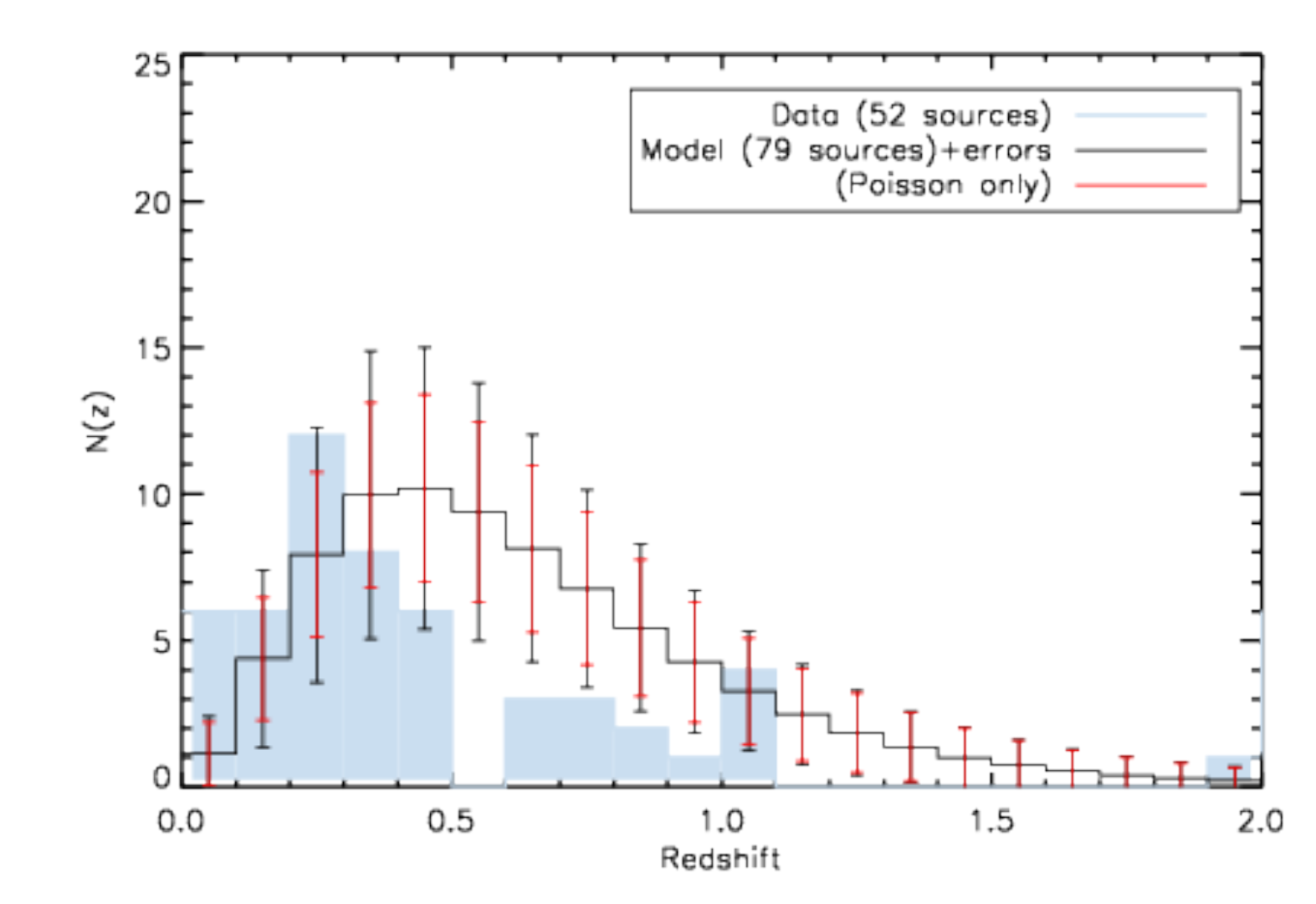}
	\caption{The XMM-LSS 11~deg$^2$ C1 redshift distribution in $\Delta z=0.1$ bins (shaded histogram, 52 sources total). The line histogram is predicted from an ab-initio modelling including WMAP-9yr cosmological model \citep{hinshaw13}, \citet{tinker08} mass function and our selection function (Fig.~\ref{fig:survey_selfunc}), after converting halo masses into X-ray observables thanks to \citet{arnaudpointecouteau05} and \citet{maughan12} scaling relations. This model predicts $79 \pm 12$ sources. 1-$\sigma$ uncertainties are computed analytically and account for both shot-noise (red bars) and sample variance (see text and \citealt{valageas11}).}
	\label{fig:dndz_wmap9} 
\end{figure}

Fig.~\ref{fig:dndz_wmap9} shows the redshift distribution of all 52 sources entering this analysis (plain blue histogram). Typical uncertainties on cluster redshift are much smaller than the bin size of $\Delta z=0.1$ used for this plot (of the order few $\times 10^{-3}$, \citealt{adami11}).
A model distribution is superimposed, whose derivation follows the steps presented in \citet{pacaud07, pierre11} \citep[see also][]{clerc12a}. In brief, a model halo distribution \citep{tinker08} is projected onto the X-ray observable space (+redshift) using standard scaling relations \citep{arnaudpointecouteau05, maughan12} and XMM detector responses. Evolution of these scaling relations follows a self-similar model. Our observational selection (shown on Fig.~\ref{fig:survey_selfunc}) filters out undetectable clusters. To this purpose it assumes a fixed physical core-radius of 180\,kpc for all clusters. The resulting distribution is further integrated over all quantities but redshift and provides the expected $dn/dz$ distribution.
This quantity is summed in finite redshift bins and uncertainties on the histogram are computed following the formalism described in \citet{valageas11}. They account for both shot-noise (i.e. $\sqrt{n_i}$) and sample variance in each bin. These uncertainties are almost uncorrelated (e.g.~$C_{ij,i \neq j} \ll \sqrt{C_{ii} C_{jj}}$).
Given the number of assumptions involved here, Sect.~\ref{sect:discussion} will discuss further the hypotheses entering the derivation of this model histogram.

	\subsection[]{$L_X-T$ scaling relation and evolution}
	\label{sect:lt_evolution}
	
	The luminosity-temperature relation ($L_X-T$) is one of the main studied scaling laws of X-ray clusters \citep[see e.g.][]{arnaudevrard99, pratt09, mittal11, maughan12, takey13}. It reflects the history of heating and cooling of the intra-cluster gas \citep[see e.g.][]{voit05} and relates two major cluster properties ultimately linked to the total cluster mass \citep{kaiser86}.
Taking advantage of the wide redshift range spanned by our sample, we address the evolution in the normalization of this scaling law assuming various local relations selected in the literature. We pay particular attention to the role of selection effects in this analysis.
	
		\subsubsection{$L_X-T$ relation in the sample}
\begin{figure*}
	\includegraphics[width=\linewidth]{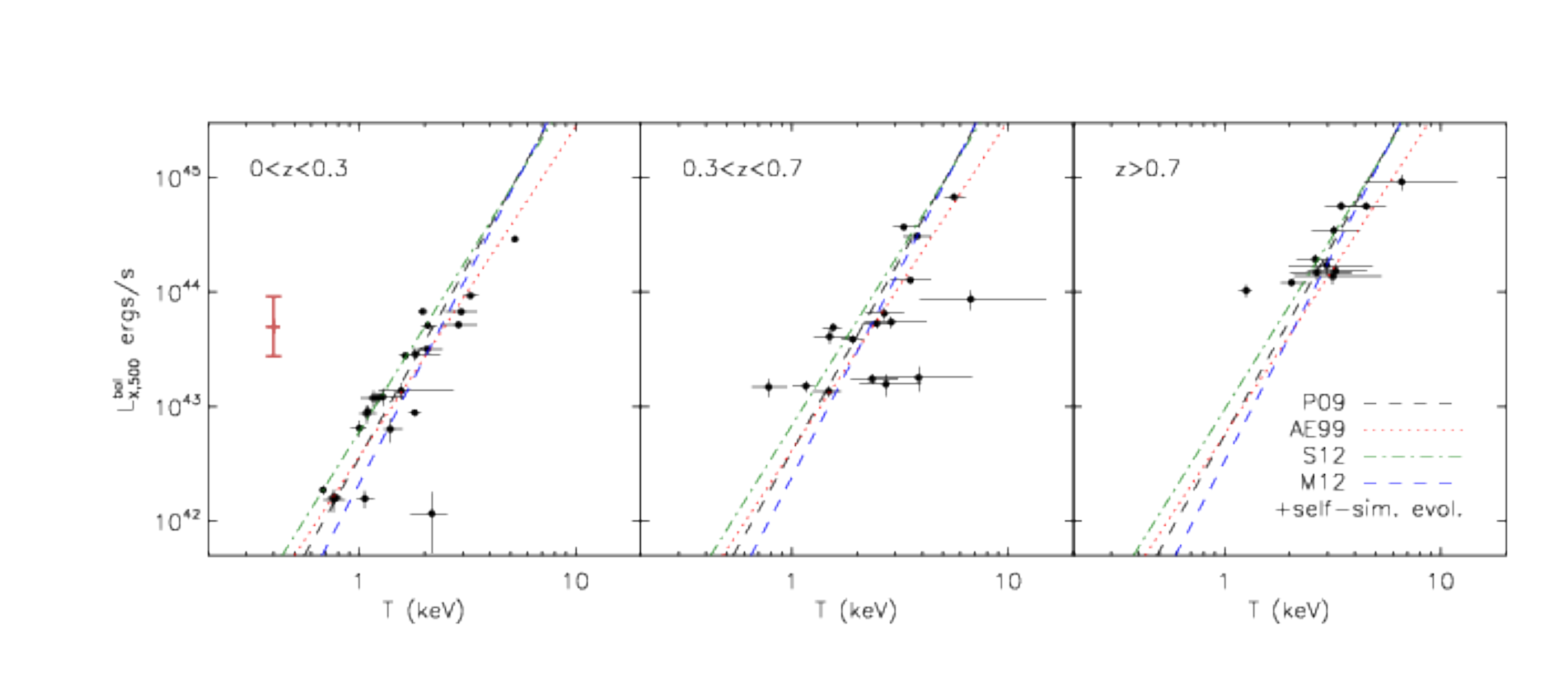}
	\caption{The luminosity-temperature relation of our cluster sample. Luminosities are bolometric and measured within $R_{500}$ as inferred from a mass-temperature relation (see text). Temperatures are measured within an aperture maximizing the signal-to-noise of the spectral fit. Several local scaling relations are overplotted for comparison, each being evolved following a self-similar evolution $L_X(T,z) \propto E(z)$: P09 \citep{pratt09}, AE99 \citep{arnaudevrard99}, S12 \citep{sun12} and M12 \citep{maughan12}. A typical $\sigma_{\ln L|T}=0.6$ scatter is illustrated by the red thick error bar.  Scaling laws are displayed at the median redshift of each panel subsample ($z=0.20,0.43,0.96$).}
	\label{fig:lx-t_selfsimilar} 
\end{figure*}

Figure~\ref{fig:lx-t_selfsimilar} shows the relationship between the bolometric luminosity and the temperature measured for the 52 clusters in this sample. The X-ray luminosity correlates well with gas temperature in all three redshift slices displayed on Fig.~\ref{fig:lx-t_selfsimilar} as expected from basic scaling arguments \citep{kaiser86}. We plot on the same figure a selection of recent scaling relations \citep{arnaudevrard99, pratt09, maughan12, sun12}.
Although these scaling relations were derived with samples of clusters spanning relatively wide redshift ranges, the numerical values corresponding to these relations were published for the local Universe ($z=0$). Hence in the following we will refer to them as "local" scaling relations.
Each of them is evolved self-similarly to the median redshift in each panel of Fig.~\ref{fig:lx-t_selfsimilar} ($z=0.20,0.43,0.96$), i.e. $L_X(T,z) = L(T,0)\times E(z)$. The slope of these extrapolated relations appears compatible with our data points, as is the normalization.
However, this visually good agreement has to be checked against selection effects and we describe our findings in the next paragraph.

		\subsubsection{Evolution of the $L_X-T$ normalization and impact of selection biases}

Since many objects in the sample are close to the detection threshold, it is necessary to correctly account for selection effects before any attempt to interpret the data point distribution in Fig.~\ref{fig:lx-t_selfsimilar} \citep[e.g.][]{pacaud07, mantz10, allen11, reichert11}.
A typical misinterpretation can arise from Malmquist bias: intrinsically brighter objects are favorably present in the sample, which translates into an average luminosity at a given temperature and redshift higher than the true expected luminosity. Such a bias is increasingly important as the intrinsic scatter in the studied scaling law is high. This is the case for the $L_X-T$ relation \citep[of the order of $\sigma_{\ln L|T} \sim 0.6$, e.g.][]{pratt09}.
We follow the same approach as in~\citet{pacaud07} for studying the evolution in the normalization of the relation from our sample. Namely, we model at each redshift the $L_X-T$ relation assuming a local ($z=0$) relation and a normalization following $E(z) \, (1+z)^{\alpha}$. The exponent $\alpha$ is a free parameter in the analysis. The scatter $\sigma_{\ln L|T}$ is kept at its $z=0$ value. The resulting distribution is folded with the survey selection function (Fig.~\ref{fig:survey_selfunc}) after passing it through the XMM instrumental response. The likelihood of each cluster being drawn from this particular model is computed taking into account uncertainties on the temperature\footnote{In truth, the full C-statistic given by {\sc XSpec} from the spectral fitting and converted into a probability distribution as a function of $T_X$.} and neglecting uncertainties on luminosities.

\begin{figure*}
	\includegraphics[width=\linewidth]{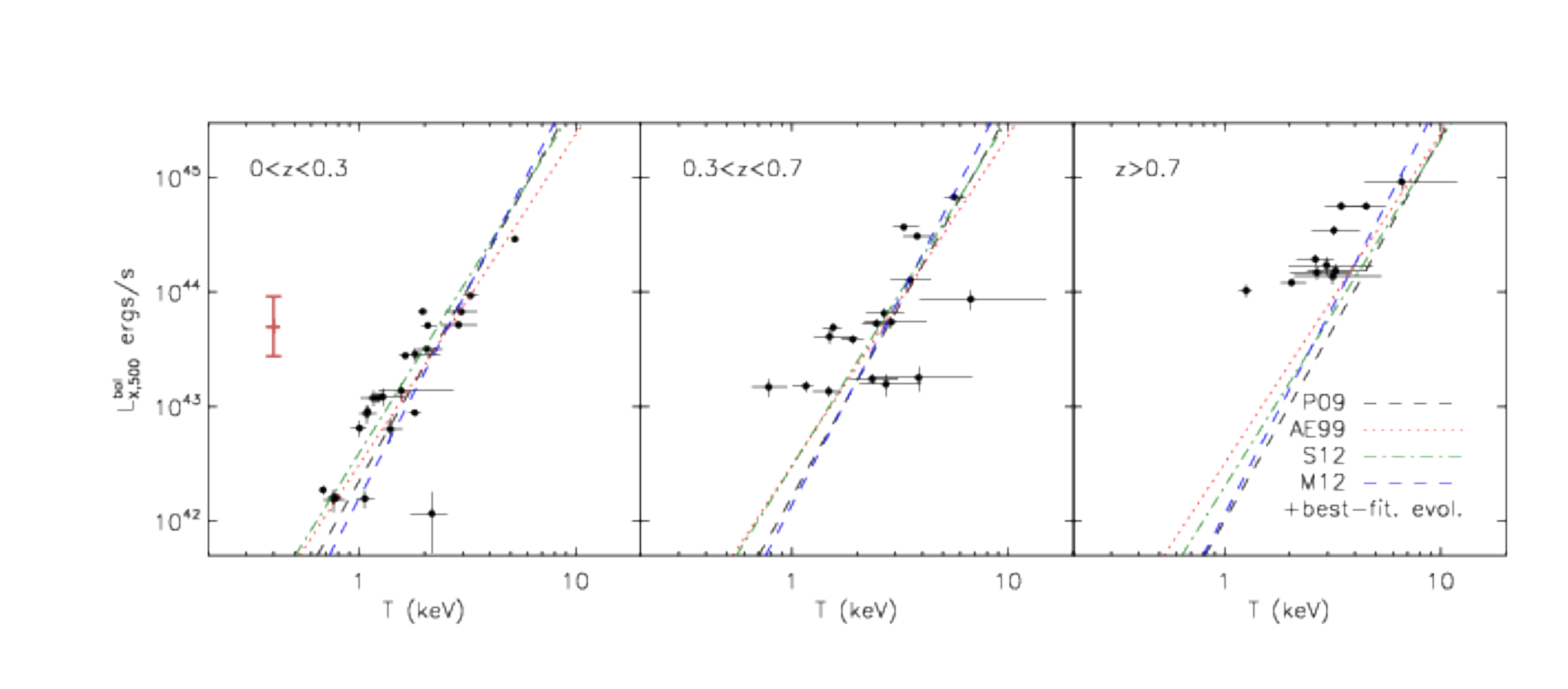}
	\caption{Same data points as Fig.~\ref{fig:lx-t_selfsimilar}. The evolution of local scaling relations now takes into account selection effects -- in particular Malmquist bias -- and corresponds to our best-fit, 1-parameter, model $L_X(T,z) \propto E(z).(1+z)^{\alpha}$. The corresponding best-fit values of $\alpha$ are listed in Table~\ref{table_lx-t_fitvalues} (column "corrected").}	\label{fig:dndz_lx-t_bestfit} 
\end{figure*}

Repeating this procedure for a range of values in $\alpha$ enables the derivation of 68\% confidence intervals for $\alpha$, as quoted in Table~\ref{table_lx-t_fitvalues} (under column "corrected").
The resulting value depends on the assumed local ($z=0$) scaling relation because of the differences in their slopes, normalizations and scatters. Fig.~\ref{fig:dndz_lx-t_bestfit} visually illustrates our results. The disagreement between data points and best-fit scaling relations is only apparent and due to the impact of selection effects. This is reflected in the numbers shown in Table~\ref{table_lx-t_fitvalues} (column "uncorrected") where the same procedure is applied, but the selection function is artificially neglected (i.e. we assume that $P_{C1}=1$).
In the latter case, the normalization is found to be positive or mildly negative (with respect to self-similar evolution, $\alpha=0$), while a proper account for selection effects clearly hints toward a negative evolution. This result is in agreement with the findings of \citet[][]{reichert11, hilton12, clerc12b}, although these studies differ in their treatment of selection effects and modelling.

We note here that self-similar evolution does not necessarily imply a $\propto E(z)$ scaling of the $L_X-T$ normalization \citep{maughan14}. This actually depends on the assumed slope of the $M_{gas}-M$ scaling relation and \citet{maughan14} find a $E(z)^{0.4}$ behavior instead. The present significance of the negative evolution is thus reduced when expressed relative to this particular scaling.

\begin{table}
	\centering
\caption{\label{table_lx-t_fitvalues}The evolution of the $L_{X,500}^{bol}-T$ relation measured from the sample of clusters presented in this work (see Fig.~\ref{fig:dndz_lx-t_bestfit}). Different local ($z=0$) scaling relations are assumed: P09 \citep{pratt09}, AE99 \citep{arnaudevrard99}, S12 \citep{sun12}, M12 \citep{maughan12}. Their normalization is evolved following $E(z).(1+z)^{\alpha}$. "Corrected" ("Uncorrected") refers to the best-fit value of $\alpha$ found with (without) accounting for selection biases.}
		\begin{tabular}{@{}lcc@{}}
\hline
						&	Best-fit $\alpha$	&	Best-fit $\alpha$	\\
Local scaling relation	&	Corrected		&	Uncorrected			\\
\hline
P09 "ALL, $L_1-T_1$, BCES Ortho."		&	$-2.5 \pm 0.4$	&	$-0.6 \pm 0.4$\\
AE99	 (+ $\sigma_{\ln L|T}=0.6$)			&	$-0.9 \pm 0.4$	&	$0.6 \pm 0.3$\\
S12 "$L_1-T_{500}$ S+R"					&	$-2.3 \pm 0.4$	&	$-1.1 \pm 0.3$\\
M12 "all, $[0-1] R_{500}$" (M12)			&	$-1.6 \pm 0.4$	&	$0.1 \pm 0.4$\\
\hline
		\end{tabular}
\end{table}

	\subsection[]{Tridimensional distribution of clusters}

\begin{figure*}
	\begin{tabular}{cc}
		\includegraphics[width=84mm]{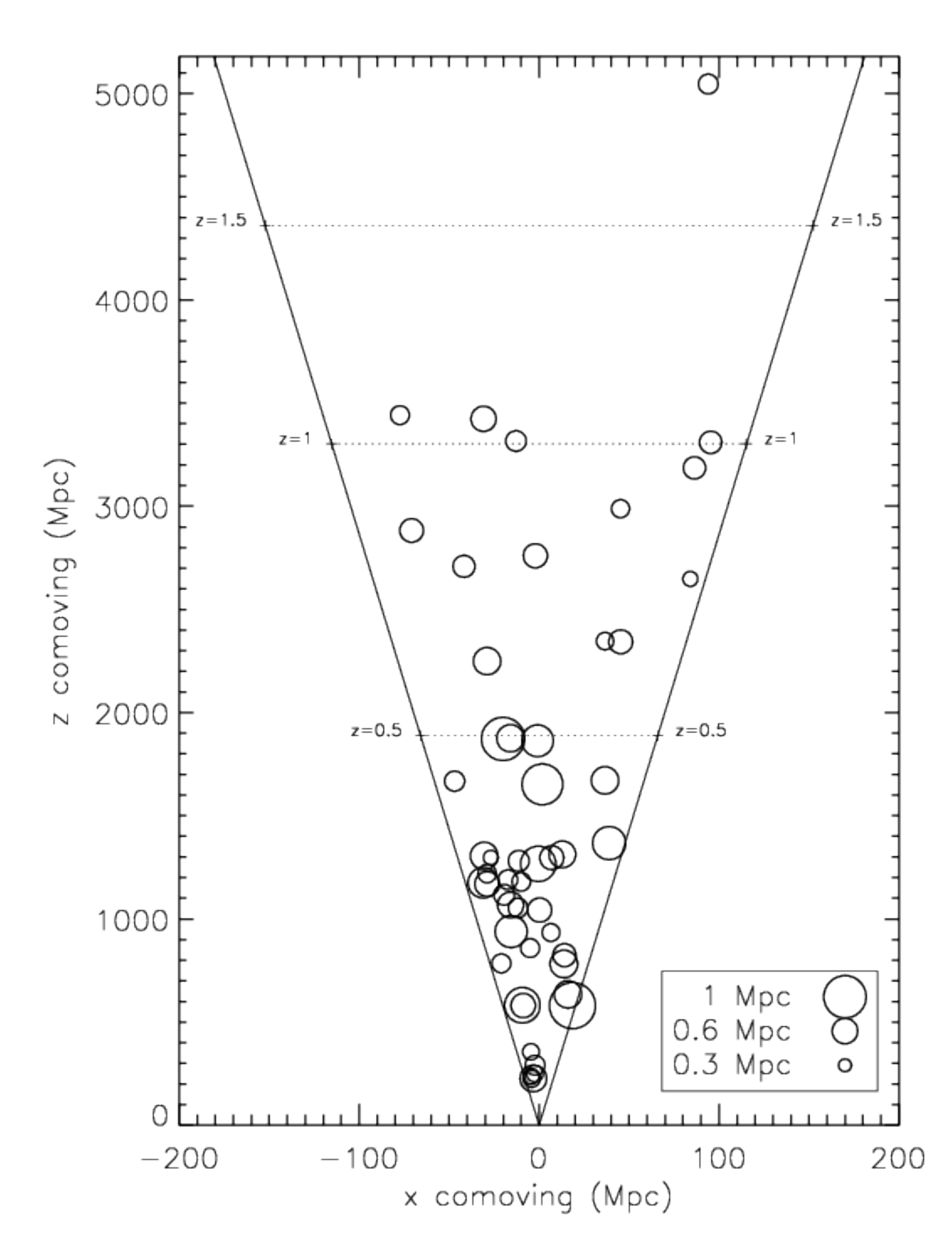} &
		\includegraphics[width=84mm]{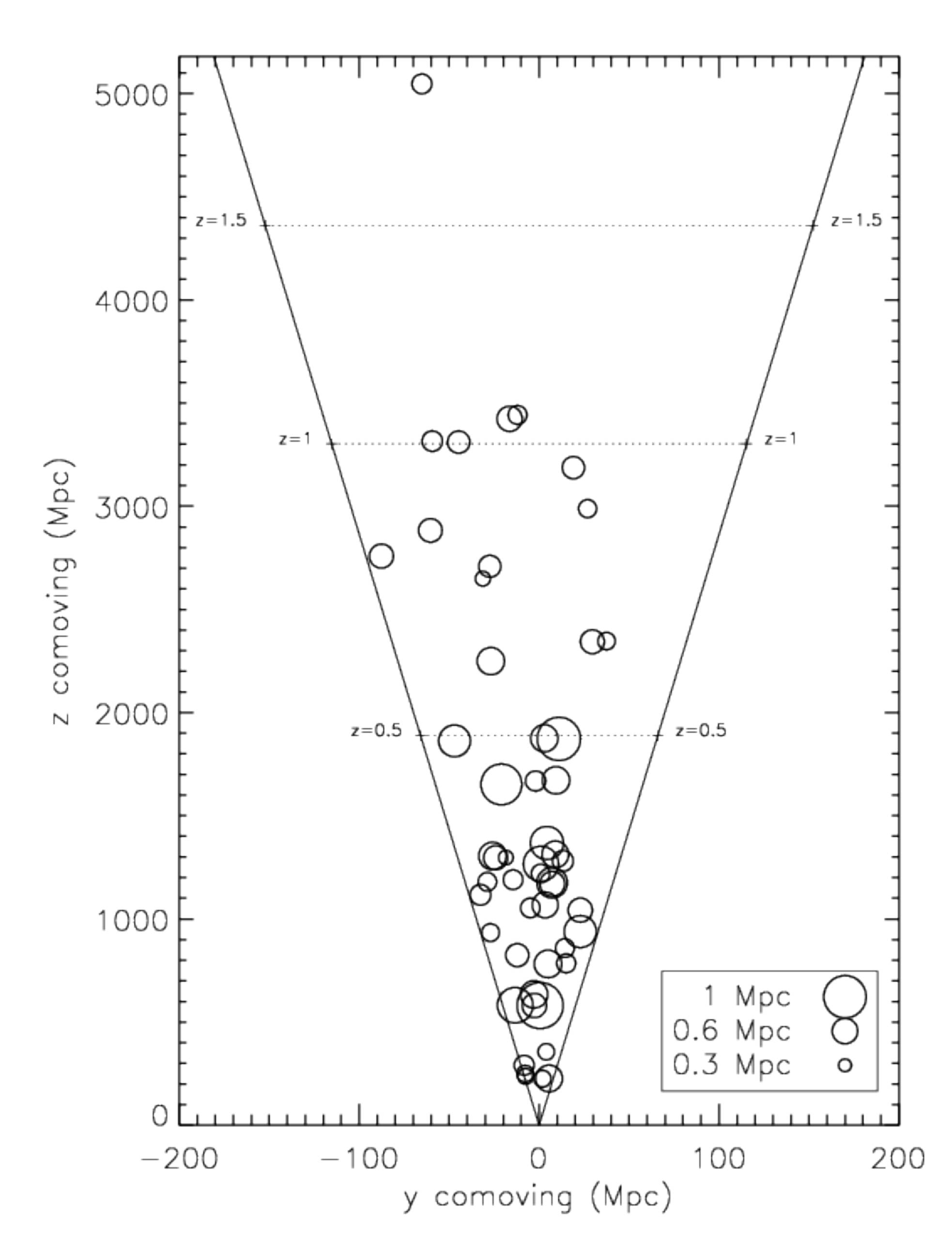} \\
	\end{tabular}
	\caption{The tridimensional distribution of XMM-LSS C1 clusters viewed from two perpendicular directions, both orthogonal to the line-of-sight ($z$-axis). The observer is located at $(x,y,z)=(0,0,0)$ in this comoving coordinate system. Each cluster is represented by a symbol whose size is proportional to $R_{500}$ as inferred from X-ray temperature measurements (see Eq.~\ref{eq_r500_sun09}). The approximate XMM-LSS survey boundaries are materialized by two solid lines.}
	\label{fig:tridimensional} 
\end{figure*}

The sample presented in this work is unique in several respects: it is drawn from an homogeneous, contiguous, X-ray survey, it is 98\% complete in terms of spectroscopic redshift availability, spans a wide range of redshifts and masses and the sample selection function is well understood. This is summarized on Fig~\ref{fig:tridimensional}, where each C1 cluster from Table~\ref{table_c1_catalogue} is drawn at its location in a comoving coordinates frame (the observer is located at the origin and the line of sight is the $z$-axis). We note that the number of objects (52) presented in this study is too low to enable robust, quantitative, interpretations of the 3-dimensional distribution of clusters in the volume (e.g.~correlation function analysis). However, this is an open window for the on-going XMM-XXL survey~\citep[][Pierre et al. in prep.]{pierre11}, which multiplies the surveyed area by a factor 5 (in two separate areas).

\begin{figure*}
	\includegraphics[width=110mm]{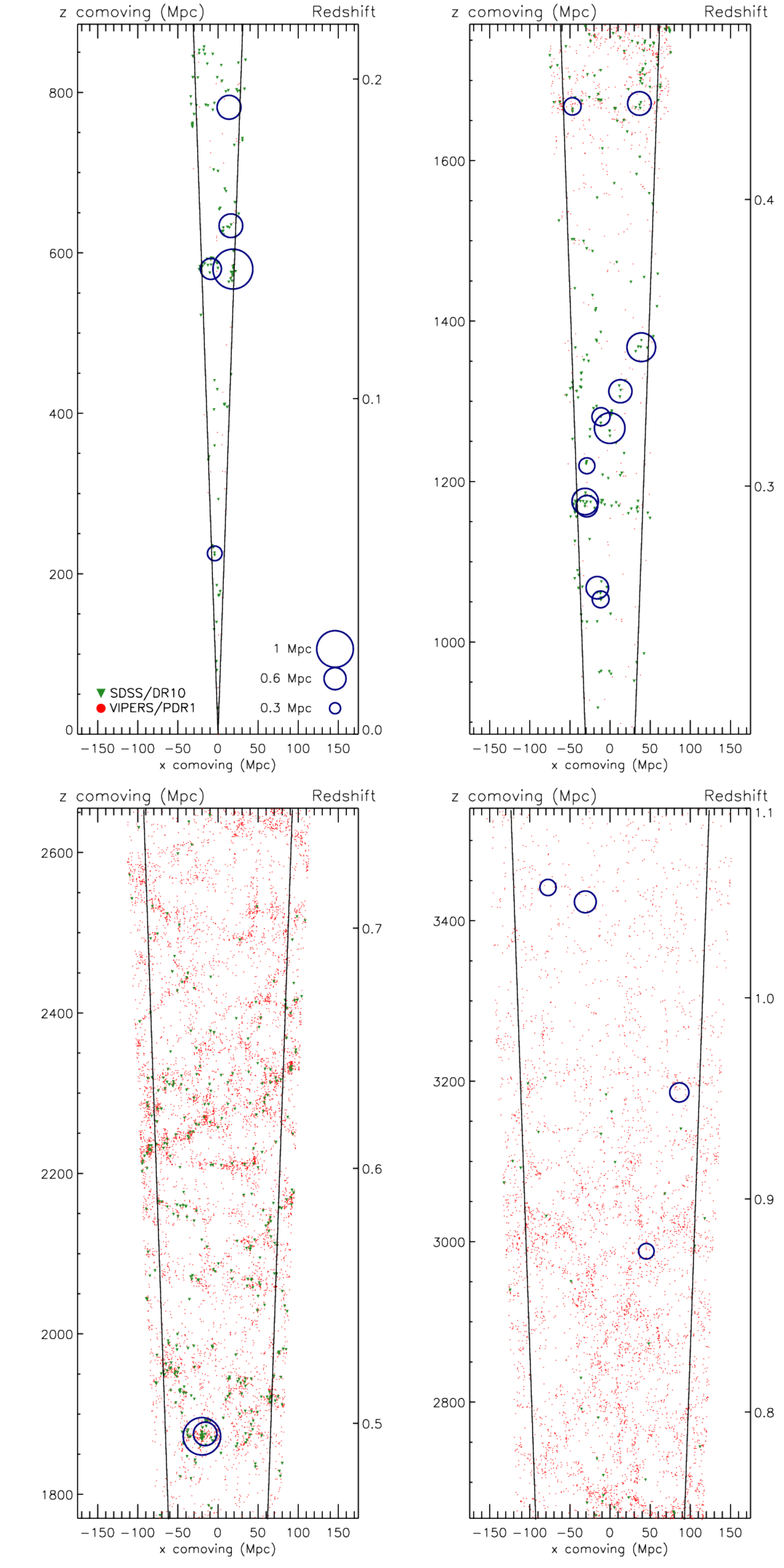}
	\caption{The tridimensional distribution of XMM-LSS C1 clusters (blue circles, sizes proportional to $R_{500}$) in the context of galaxy spectroscopic surveys. Green triangles stand for galaxies in the SDSS DR10 release \citep{ahn13}. Red points stand for VIPERS objects with flags between 2.X and 9.X \citep{garilli13}. Only objects with $33<{\rm R.A}<38$~deg and $-5.1<\delta<-4.1$~deg are represented (i.e. in the common sky overlap of all three surveys).}
	\label{fig:vipers} 
\end{figure*}

Figure~\ref{fig:vipers} shows the location of galaxies detected in two large spectroscopic samples: BOSS-DR10 \citep{ahn13} and VIPERS-PDR1 \citep{garilli13,guzzo13}. The declination range was shrunk to only display objects in the common sky area. Moreover, radial selection effects differ from one survey to the other, in particular, VIPERS galaxies are preferentially selected in the $0.5 \lesssim z \lesssim 1$ redshift range. As we will describe in the next section, the most striking visual result in this figure is the absence of C1 X-ray clusters in this redshift domain, while clustering seems apparent in the VIPERS dataset.

\section[]{Discussion}
\label{sect:discussion}

We focus now on modelling the redshift distribution of clusters as presented in Sect.~\ref{sect:redshift_distribution} and Fig.~\ref{fig:dndz_wmap9}.
The first redshift bin ($0<z<0.1$) contains 6 groups (XLSSC~035, 062, 021, 054, 011, 052) of low mass and small size. It is thus likely that our assumption of a fixed core-radius value (180~kpc) fails in this redshift range. For example, assuming a core-radius size of half this value roughly doubles the number of predicted clusters in this bin. However, given the small volume of Universe involved, we refrain from deriving quantitative results from this bin. Hence, the following results rely on the 46 clusters with $z>0.1$.

	\subsection[]{A gap in the redshift distribution?}

The redshift distribution model of Fig.~\ref{fig:dndz_wmap9} presents a rough agreement with the observed redshift histogram of the sample.
Rebinning the histograms such that each model bin contains at least 5 objects, we compute a $\chi^2$ value of 18.0 using 10 bins.
Given that $P(\chi^2 \leq 18.0) = 96$\%, we conclude on a marginal agreement between data and this model.

\begin{figure}
	\includegraphics[width=84mm]{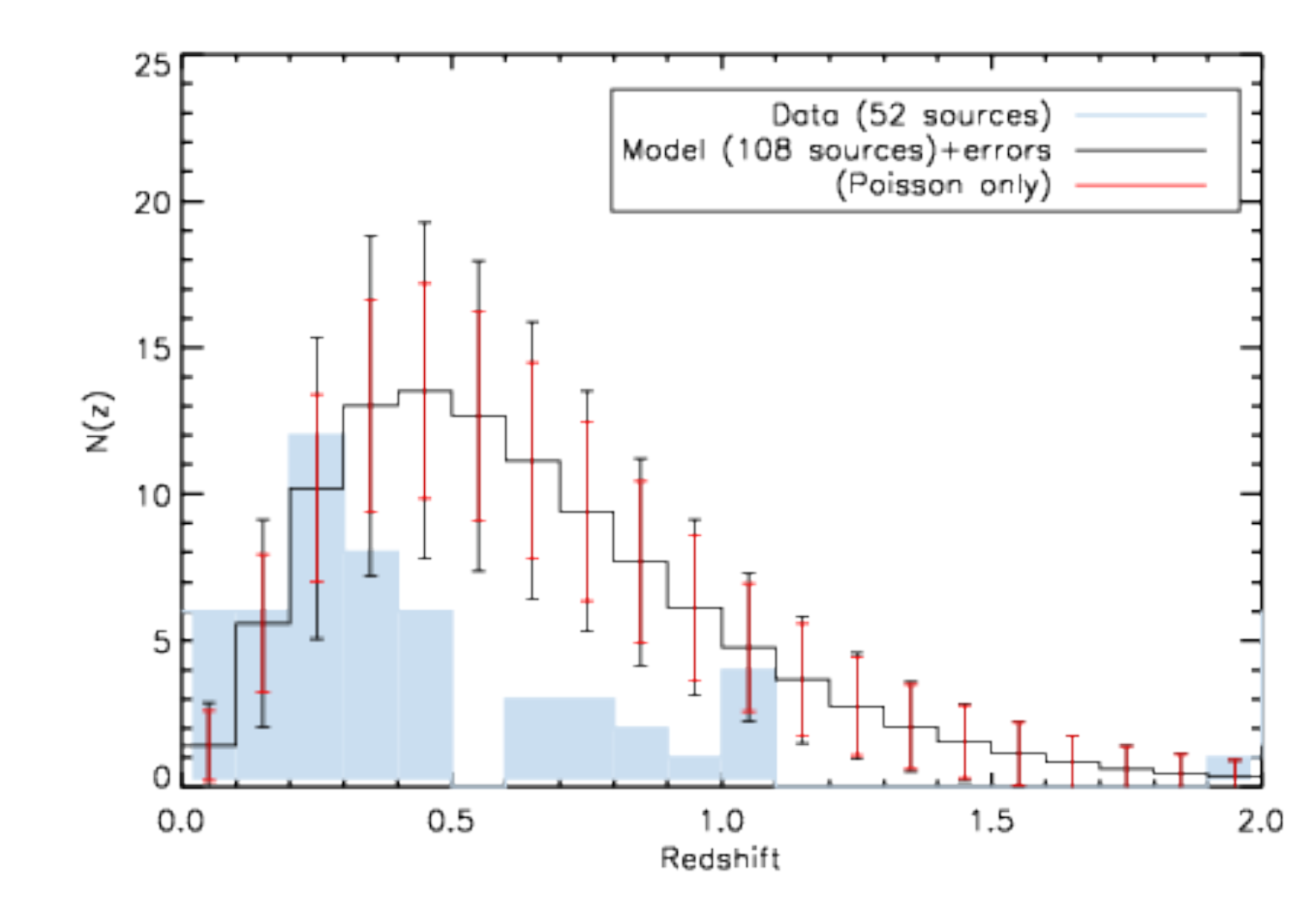}
	\caption{Same figure as Fig.~\ref{fig:dndz_wmap9}, with WMAP-9yr cosmology replaced by Planck~2013 CMB cosmology \citep{planckcmb} in the model derivation (plain line and errors). This model predicts $108 \pm 14$ C1 clusters in the $0 \leq z \leq 2$ range.}
	\label{fig:dndz_planck2013} 
\end{figure}

Updating the cosmological model to the best-fit model derived from Planck CMB results \citep{planckcmb}, we obtain a new curve, as shown in Fig.~\ref{fig:dndz_planck2013}. Because of higher values of $\sigma_8$ (0.834 instead of 0.821) and $\Omega_m$ (0.316 instead of 0.279), the total number of predicted clusters is higher in each $\Delta z$ bin. Taking into account uncertainties in each bin, we find a $\chi^2$ of 28.0 (in 12 bins). Since $P(\chi^2 \leq 28.0) = 99.7$\%, we reject the hypothesis that the observed redshift distribution derives from this model, with a false-alarm error probability of $\sim 0.3$\% (equivalent to a 3-$\sigma$ rejection).

Fig.~\ref{fig:dndz_planck2013} suggests that the strongest disagreement between data and model occurs in the $0.4 \lesssim z \lesssim 0.9$ range, in which too many clusters are expected. The data histogram alone suggests the presence of a "gap" in this range, reinforced by the fact that 4 C1 clusters are clearly detected at $z \sim 1$, ruling out a severe sensitivity effect.
Limiting to this "ad hoc" redshift range provides a $\chi^2$ of 15.4 in 5 bins, also leading to model rejection with false-alarm error probability  of 0.4\%.
Taking these results at face value let us postulate the existence of a substantial lack of massive structures around $z \sim 0.7$. This deficit would be rare enough to be marginally accounted for in our sample variance calculations.
However, in what follows we discuss several physical effects that can be held responsible for this observation and argue for a combination of observational and physical effects.

	\subsection[]{$L_X-T$ evolution and redshift histogram}

\begin{figure}
	\includegraphics[width=84mm]{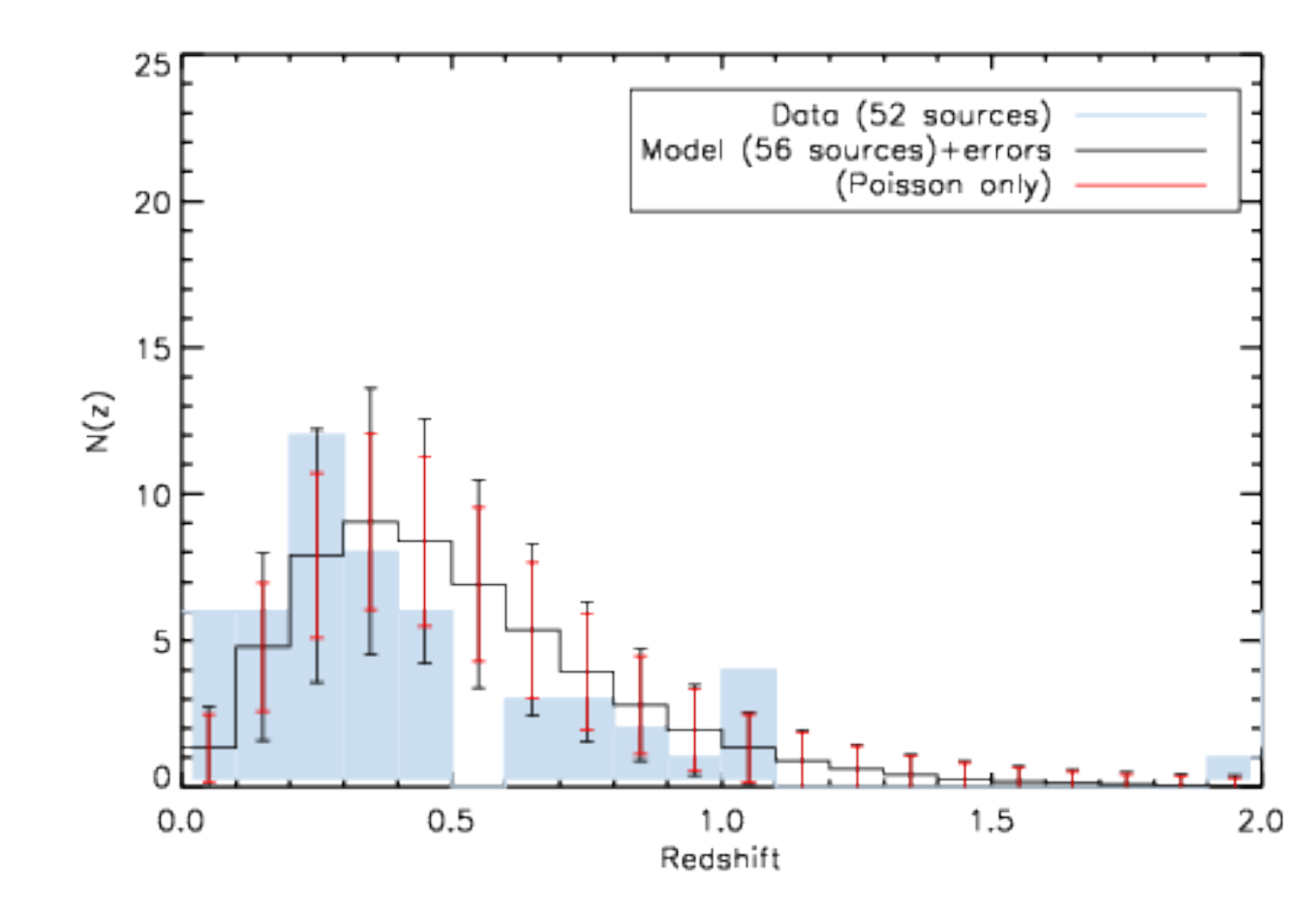} 
	\caption{Same figure as Fig.~\ref{fig:dndz_planck2013}, but assuming an evolution of the luminosity-temperature following our best-fit result, $\alpha=-1.6$ (Table~\ref{table_lx-t_fitvalues}, 4th row) instead of self-similar evolution ($\alpha=0$). The model predicts $56 \pm 10$ C1 clusters.}
	\label{fig:dndz_planck2013_zevol} 
\end{figure}

Model redshift distributions shown in Figs.~\ref{fig:dndz_wmap9} and~\ref{fig:dndz_planck2013} depend on the exact scaling relations used for converting cluster masses into observed properties. This is particularly true for the luminosity-temperature relation, which is involved in the computation of a cluster mass from its apparent flux.
As shown in Sect.~\ref{sect:lt_evolution}, our dataset prefers a negative evolution of the $L_X-T$, namely a normalization evolving as $E(z).(1+z)^{-1.6 \pm 0.4}$ instead of $E(z)$.
Fig~\ref{fig:dndz_planck2013_zevol} demonstrates the impact of such an evolution on the modeled redshift distribution, still under the assumption of a Planck~2013 CMB cosmology (i.e.~equivalent to Fig.~\ref{fig:dndz_planck2013}). With this evolution, clusters of a given mass become fainter with increasing redshift, thus we expect less detections at higher redshifts, simply due to a dimming of these objects.
In this case, the $\chi^2$ value is 6.1 (in 8 bins), sufficiently low to prevent the model rejection.
Hence, the additional assumption of a non self-similar $L_X-T$ reconciles the observed redshift histogram with our model and reduces the significance of the central gap.

We note that a complete analysis should simultaneously fit a cosmological model and a scaling relation model in order to properly account for Eddington bias \citep[e.g.][]{mantz10}. This kind of self-consistent analysis in the signal-to-noise regime of "XMMLSS-like" clusters is particularly well handled by the $z$-CR-HR method presented in \citet{clerc12a} and will be applied to the much larger sample of XMM-XXL clusters.

Finally, we stress that previous results were derived assuming a local $L_X-T$ from \citet{maughan12}. The other three local scaling laws considered earlier (Table~\ref{table_lx-t_fitvalues}) lead to even higher cluster densities, hence in stronger disagreement with this dataset (see also \citealt{clerc12b} for a discussion).

	\subsection[]{Central AGN contamination and redshift distribution}
	\label{sect:agn_cont}

Simplifying hypotheses entering the derivation of our selection function do not account for the true physical nature of galaxy clusters and possibly impact the model redshift distribution shown on Fig.~\ref{fig:dndz_wmap9} and~\ref{fig:dndz_planck2013}. We explore in this section the impact of an evolving AGN contamination fraction in the center of galaxy clusters by means of simple simulations and a demonstrative toy-model.

		\subsubsection[]{Simulations with point-source contamination}
		
The presence of a central AGN in galaxy cluster is a common source of concern for extended source detection algorithms. A central, blended, point-like source increases the total flux of a cluster in X-rays, thus increasing its detection probability. However, beyond a certain flux the detected source can no longer be classified as a "secure" extended source on morphological grounds only, given the sharply peaked profile of the blend.
We modified our set of XMM cluster simulations by adding a point-source with varying flux at the center of each $\beta$-model. We processed these simulations with the exact same methodology as for real data and derived the C1 detection probability of a cluster as a function of its count-rate, extent and central contamination.

\begin{figure*}
	\includegraphics[width=\linewidth]{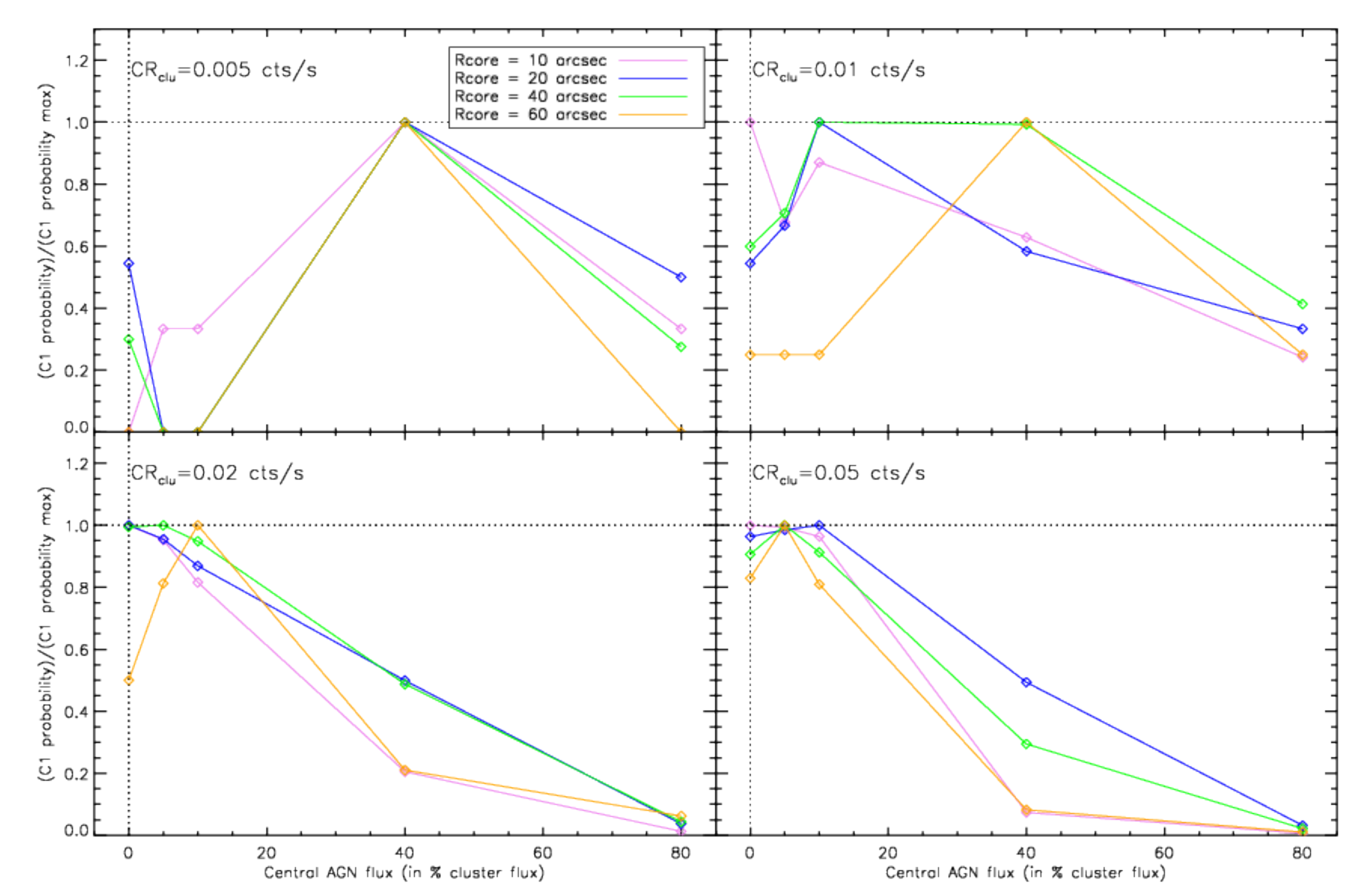}
	\caption{Impact of central AGN contamination on the C1 selection function. These curves show the change in C1 detection probability of a cluster as a function of a central AGN flux as compared to the probability at zero contamination level. They are derived from realistic simulations of $\beta$-model sources with core-radius $R_{core}$ and various count-rates on the XMM detectors ($CR_{\rm clu}$). Each curve is normalized to its maximal value and is computed for a typical 10~ks exposure with nominal detector background.}
	\label{fig:agn_contamination} 
\end{figure*}

Our raw results are displayed on Fig.~\ref{fig:agn_contamination}.
As expected, a very bright central AGN (e.g.~80\% contamination) causes a misclassification of the detected source and a decrease in the C1 detection probability.
On the other hand, a 10-20\% contamination in relatively bright and extended clusters (e.g.~$0.02-0.05$~cts/s and $40-60\arcsec$ core-radius) slightly enhances the detection probability ("flux boosting"), while bright, compact clusters are less affected at this level of contamination. The increase in detection with AGN contamination is conspicuous for faint (e.g.~$0.005-0.01$~cts/s) clusters, provided their morphology is not too compact and the AGN flux remains reasonable (below $\sim 50$\%).

Obviously, characterization of extended sources contaminated by a point is heavily dependent upon the surrounding background level in the X-ray images. We repeated this exercise for a higher background level and a different exposure time (20~ks). The general statements above remain unchanged, but they impact the selection function at slightly different locations in the extent-flux-contamination parameter space.

		\subsubsection[]{Impact of point-source contamination on $dn/dz$}

\begin{figure}
	\includegraphics[width=84mm]{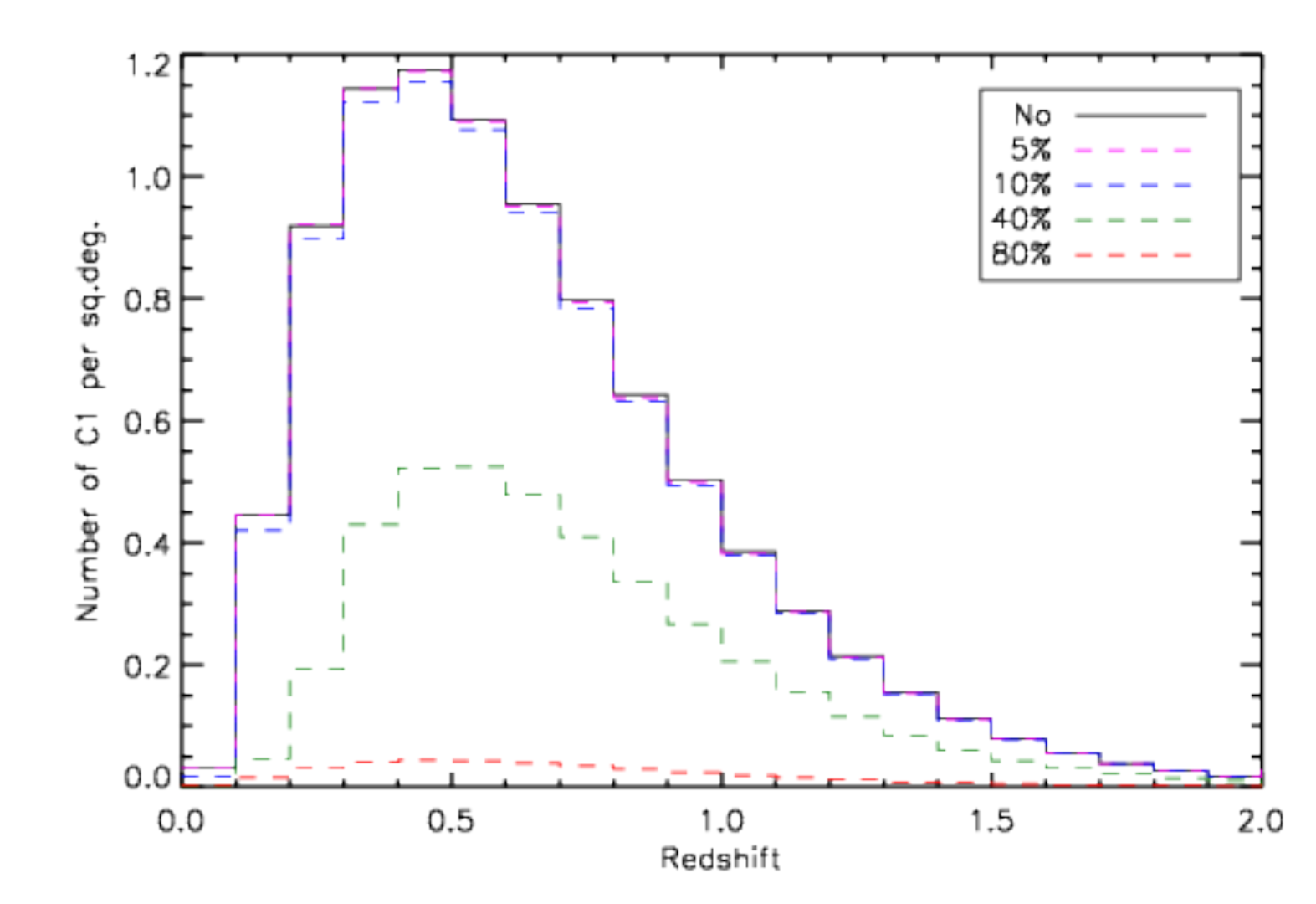}
	\caption{Impact of central AGN contamination on the modeled C1 redshift distribution. These curves are obtained by folding the selection functions of Fig.~\ref{fig:agn_contamination} into our model predicting the number density of C1 detections in a typical XMM pointing. The percentages correspond to different levels of central AGN contamination, assuming that it arises in all clusters at all redshifts.}
	\label{fig:dndz_agncont} 
\end{figure}

\begin{figure}
	\includegraphics[width=84mm]{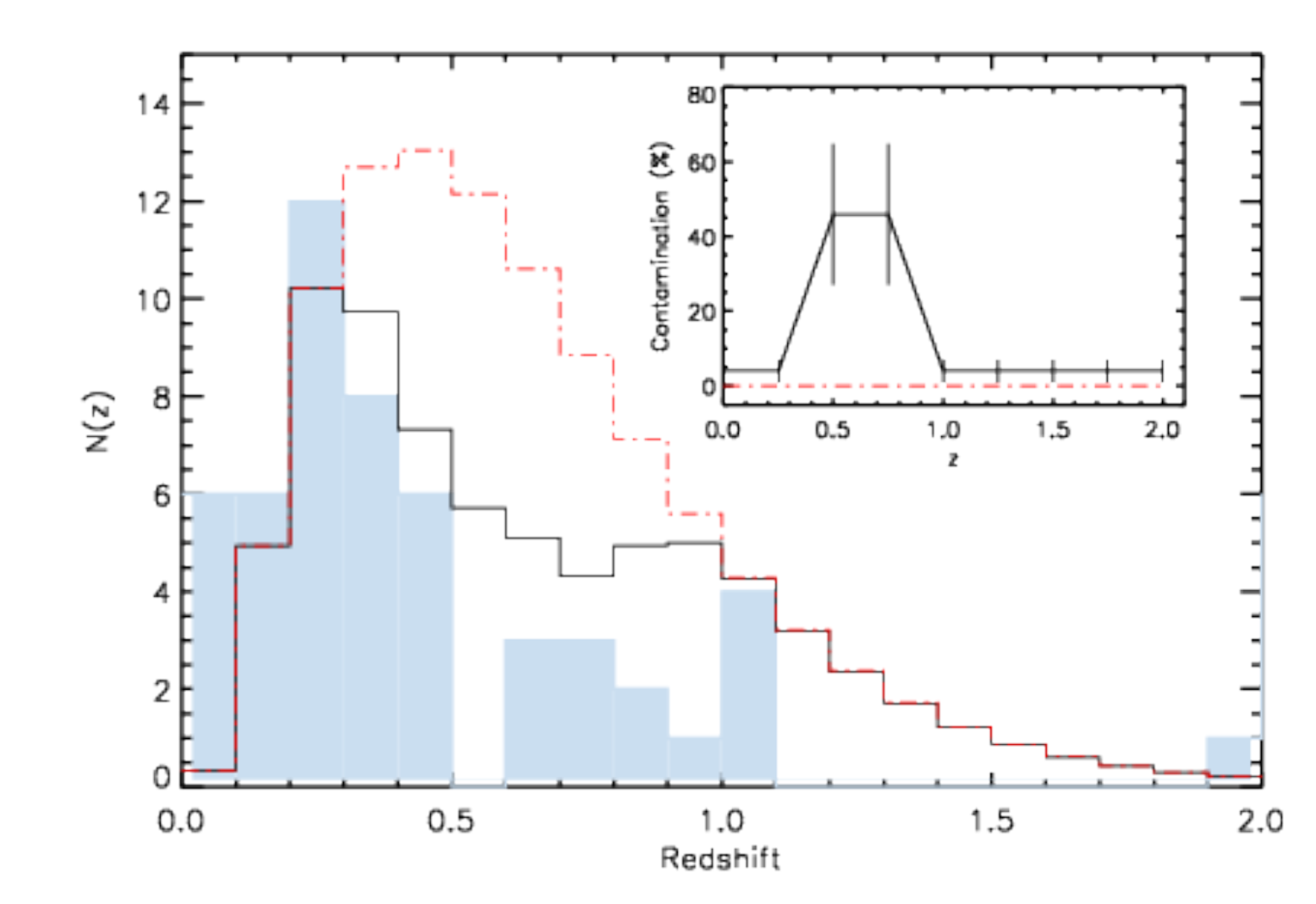}
	\caption{Impact of an evolving fraction of central AGN contamination onto the modeled cluster redshift distribution. Blue histogram corresponds to this sample. Red dashed line is identical to Fig.~\ref{fig:dndz_planck2013}, i.e.~corresponds to a model in which all clusters are AGN-free. The black line follows a similar model, with a supplementary assumption that clusters contain a central AGN. The AGN relative flux is evolving with redshift as shown in the inset.}
	\label{fig:dndz_agncont_toymodel} 
\end{figure}

We fold the modified selection functions of Fig.~\ref{fig:agn_contamination} into the model predicting C1 cluster redshift distribution. Results are shown in Fig.~\ref{fig:dndz_agncont}. Each separate histogram shows the expected density of clusters assuming that all of them are contaminated by a central AGN at a given level (from 0\% to 80\%)\footnote{Note this figure does not use the full 11~deg$^2$ selection function but the one computed for a 10~ks pointing with nominal background level.}
From this analysis, we conclude that high contamination rates dramatically reduce the number of C1 detections because of misclassifications of the sources. On the other hand, a slight increase in number of C1 is barely perceptible for light AGN contamination (5-10\%), lowering the impact of "flux boosting" on C1 detection.
Relying on this simple model, we conclude that mild ($>20$\%) contamination is a valid explanation for the lack of observed C1 in the $0.5<z<1$) redshift range, but we rule out the possibility that central point-source flux excess is the cause for the apparent increase at $z \sim 1$ in the observed redshift distribution of C1 clusters (Fig~\ref{fig:dndz_wmap9}).
We illustrate this finding more clearly on Fig.~\ref{fig:dndz_agncont_toymodel}. This shows a toy-model in which clusters around $z \sim 0.7$ are contaminated by a central AGN with flux ratios $0.5 \pm 0.25$ and almost uncontaminated in the remaining redshift intervals. This simple model creates a "gap" in the redshift distribution, simply because of selection effects.

\subsubsection[]{Towards a more realistic AGN population model in clusters}

The toy-model depicted in the previous section is an "ad hoc" illustration of the impact of selection effects on the redshift histogram of objects we classify as galaxy clusters. 
Recent findings in the field of X-ray surveys show that: i)~X-ray AGN preferentially live in $10^{13.5}$ M$_{\odot}$ haloes \citep[see e.g.][for a review]{cappelluti2012} and ii)~the density of low-luminosity AGN ($L_X \lesssim 10^{44}$~ergs/s) peaks at redshifts $\sim0.5-1$ \citep[e.g.][]{hasinger2005,ueda2014}. This suggests a likely enhancement of AGN flux in low-mass clusters at redshifts around $\sim 0.7$. Such an enhancement should depend both on the number density and emissivity of clusters and the AGN luminosity function across cosmic times. We propose here a simple model in order to estimate the typical flux contamination of clusters by AGN. A more detailed investigation is deferred to future work (Ramos Ceja et al., in prep.)

\begin{figure}
	\includegraphics[width=84mm]{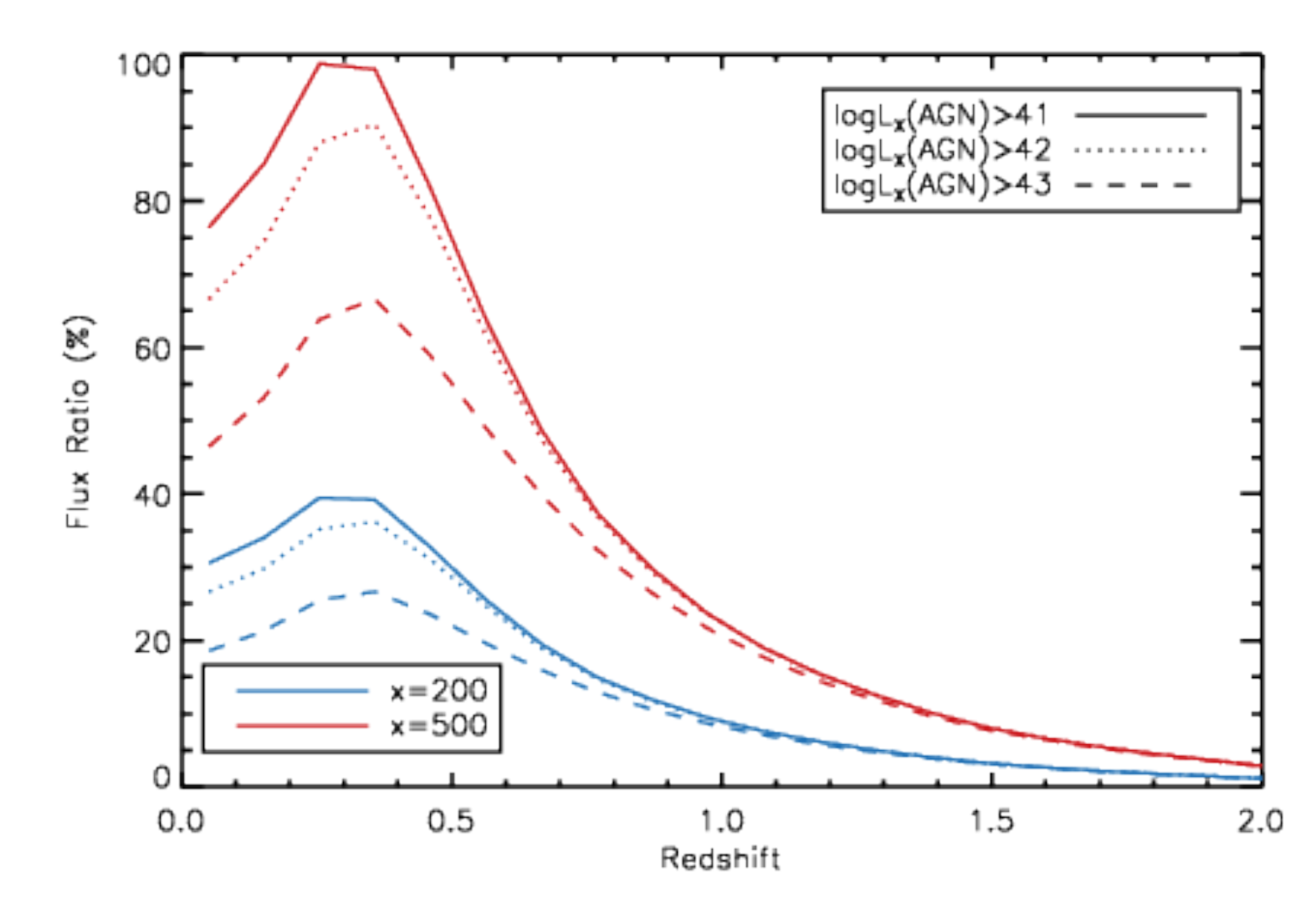}
	\caption{Outcome of a model for AGN contamination in low-mass clusters. This set of curves shows the typical ratio of AGN to cluster [0.5-2]~keV flux based upon luminosity functions from \citet{ueda2014} and our model for cluster evolution in the range of mass $10^{13}-10^{14} M_{\odot}$. We assume that AGN populate the inner 0.5~Mpc of clusters with a volume density $x=200$ or 500 times the field volume density. Different lines show different thresholds for the AGN [2-10]~keV luminosity.}
	\label{fig:ratio_agncluflux} 
\end{figure}

In a first step we calculate the [0.5-2]~keV emissivity of AGN per unit volume at different redshifts. This is done by linking the AGN X-ray luminosity function computed by \citet{ueda2014} to a bank of AGN spectral models \citep[adapted from][]{gilli2007}. Such spectra depend on the source redshift, the intrinsic power-law index $\Gamma$, the absorbing column density of obscuring material $N_H$ and the source luminosity in the [2-10]~keV band. Assuming $\Gamma=1.9$ \citep{gilli2007}, we integrate the distribution of AGN over values of $N_H$ ($10^{20}-10^{26}$ cm$^{-2}$) and $L_X$ ($10^{41}-10^{47}$~ergs/s) and we obtain the total [0.5-2]~keV flux emitted by AGN as a function of redshift. As already shown in \citet{ueda2014}, it peaks at $z \sim 0.7$.
In a second step, we derive the distribution of group/cluster fluxes using the cluster model described in Sect.~\ref{sect:redshift_distribution}, restricted to haloes of mass $10^{13} \leq M_{200b}/(h^{-1} M_{\odot}) \leq 10^{14}$.
We finally derive the typical AGN-to-cluster flux ratio in the [0.5-2]~keV band. We assume a typical cluster radius of 0.5~Mpc (hence a volume $\sim 0.5$ Mpc$^3$) and a linear scaling of AGN number density with matter density (i.e.~the number of AGN per unit volume is $x$ times higher in clusters than in the field, $x$ being of the order a few hundreds). Integrating this ratio weighted by the number of clusters at a given redshift provides the series of curves shown on Fig.~\ref{fig:ratio_agncluflux}.

This model suggests an enhancement of the flux ratio between AGN and cluster around $z \sim 0.4$, and a slight decrease at higher redshifts. The order of magnitude of the contamination level must reach $\sim 40-50$\% in order to account for the gap in the C1 redshift histogram (Fig.~\ref{fig:dndz_agncont_toymodel}): this is in rough agreement with the level calculated in our simple model (e.g.~for $x=500$ and $L_X>10^{43}$~ergs/s).
Although instructive, this model presents two main caveats, as it neglects interactions between AGNs and clusters. First, physical mechanisms activating cluster central AGN differ from those in the field \citep[e.g.][]{fabian2012}. Moreover, AGN in clusters may not be centrally concentrated and more likely to spread within the cluster volume \citep[e.g.][]{branchesi2007, haines2012}.

	\subsection[]{Surface brightness profiles and redshift distribution}

The predicted redshift distribution of XMM-LSS clusters discussed so far relies on the assumption of a particular surface brightness profile for the clusters. Indeed, the selection function as shown in Fig.~\ref{fig:survey_selfunc} is a function of the cluster angular core-radius. This core-radius is relative to a $\beta=2/3$ model, since the extended sources in simulations follow this surface brightness profile. It enters our cosmological model by assuming a physical value of $R_c=180$\,kpc.
Similarly to Section~\ref{sect:agn_cont}, we investigate the impact of such an hypothesis on the predicted redshift histogram.

\begin{figure}
	\includegraphics[width=84mm]{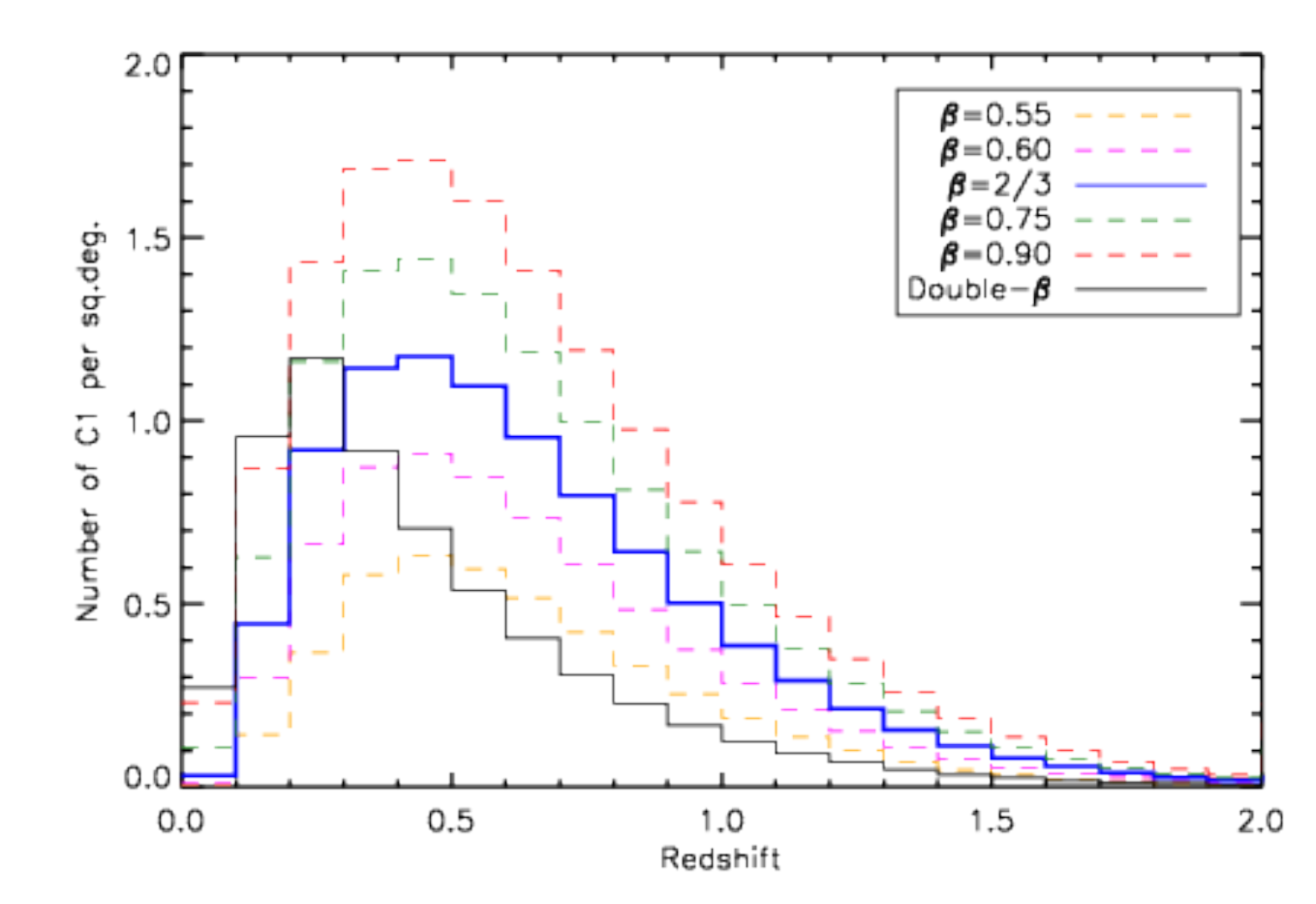}
	\caption{Impact of varying surface brightness (SB) profiles on the modeled C1 redshift distribution. These curves are obtained by folding a set of modified selection functions accounting for different SB profiles into our cosmological model.
Our reference single-$\beta=2/3$ SB model is represented by the thick blue line. Other single-$\beta$ models appear with dashed lines.
An additional SB model with two superimposed $\beta=2/3$ components simulates a population of cool-core clusters.
Each curve assumes that all clusters at all redshifts follow a SB model profile with a same physical size (180~kpc core-radius for single-$\beta$ models, 170 and 40~kpc core-radii for the double-component profile).}
	\label{fig:dndz_betavar} 
\end{figure}

In a first step, we relaxed the assumption on $\beta$ and created a new set of simulations with different values: $\beta= 0.55$, 0.60, 0.75 and 0.90. We analysed these images with our pipeline and derived the C1 detection probability as a function of core-radius, input count-rate and $\beta$.
Unsurprinsingly, changing the value of $\beta$ at fixed core-radius and flux changes the C1 detection probability. Higher values lead to sharply peaked profiles, hence increased detection rates. Very concentrated clusters ($R_c=10\arcsec$) with high $\beta$ values ($\gtrsim 0.75$) are misclassified as point-sources.
Folding the modified selection functions into the model predicting the C1 redshift distribution leads to results shown on Fig.~\ref{fig:dndz_betavar}. Each histogram represents the predicted density of C1 clusters assuming all of them follow a $\beta$-profile with $R_c=180$\,kpc and different $\beta$.
As precedently in the case of AGN contamination, these results can be interpreted in two complementary ways. First, a possible explanation for the lack of clusters in the current sample can be attributed to selection biases due to a (evolving) variation of surface brightness profiles in the cluster population. Secondly and conversely, statistical surface brightness analyses of a (X-ray selected) cluster sample must correctly take into account selection biases, since such a sample may be biased towards sharply peaked profiles (e.g.~high values of $\beta$ or small core-radius values.)

In order to estimate the impact of a centrally luminous, cool-core, cluster on our selection function, we followed \citet{eckert11} and modeled such objects with a "double-$\beta$" model.
Namely, a broad $\beta=2/3$ component with a physical core-radius of size 170~kpc hosts a smaller $\beta=2/3$ component with a size 40~kpc and a relative peak intensity 15 times higher than the larger one. These sources were inputs of our realistic XMM simulations. Running our detection algorithm provides a new selection function, which is subsequently folded into our cosmological model and provides an expected redshift distribution assuming that all galaxy clusters follow this particular profile. The result is also shown on Fig.~\ref{fig:dndz_betavar} and can be interpreted in a similar fashion as in the case of central point-source contamination: a cool-core slightly enhances the detection probability at lower redshifts ($z \lesssim 0.3$), thus increasing the number density of C1 clusters. At higher redshifts, such clusters appear more concentrated and are less likely to be classified as "secure extended source" (C1) by the detection algorithm.

Recently, \citet{mcdonald13} investigated the X-ray surface brightness evolution of 83 clusters detected by the South Pole Telescope through their Sunyaev-Zeldovich signature. Based on high-resolution {\it Chandra} data they found no change in the cooling properties of those clusters (central entropy and cooling time). However, they found an evolution in the distribution of surface brightness profiles, namely a deficit of cuspy, cool-core clusters at high redshifts ($z>0.75$) as compared to lower redshifts.
Our observed redshift distribution is compatible with such an evolution, as is visible on Fig.~\ref{fig:dndz_betavar} by comparing the plain black and plain thick blue lines. A large presence of cool-cores at lower redshifts would decrease the observed density of clusters in the $0.4 \lesssim z \lesssim 0.7$ range, while a lack of cool-cores at higher redshifts would enhance the number density above $z>0.8$, hence leading to a "gap" in the observed redshift distribution of C1 clusters.
Proper assessment of such important evolution trends will ideally be addressed by comparing multi-wavelength cluster selection, by targeted {\it Chandra} follow-up observations of the sample and by investigation of numerical simulations.

\section[]{Conclusions}
\label{sect:conclusions}

This work presents a detailed study of a complete sample of 52 galaxy clusters. All are detected in a X-ray contiguous survey with XMM-Newton and covering 11~deg$^2$ on sky.
We used an improved version of the well-qualified {\sc XAmin} algorithm \citep{pacaud06} to reduce event lists, produce images in the [0.5-2]~keV band and detect sources in the X-ray data. Relying on a set of extensive, realistic simulations, we defined a clean and uncontaminated "C1" sample of extended sources.
A series of follow-up optical observations confirmed the nature of these galaxy clusters. All but one are spectroscopically confirmed and their redshift is determined with high accuracy ($\Delta z < 0.01$).

We measured X-ray properties of each of those clusters in a homogeneous way. We demonstrated the good agreement between fluxes measured assuming a model surface-brightness profile and fluxes measured by aperture photometry.
X-ray spectra provided temperature measurements of the intra-cluster gas, which displays a typical 15\% relative uncertainty, as well as X-ray luminosities within $R_{500}$, radius estimated from scaling relations. 
We finally computed cluster total mass estimates with two different methods, one relying on the assumption of intra-cluster medium isothermality and hydrostatic equilibrium, the other on a scaling relation between luminosity and mass. We interpret their rough agreement as a reflection of the underlying X-ray scaling relations between mass, temperature and luminosity, and as a consequence of the consistency in our measurements.

We modeled the C1 cluster selection function across the wide survey area using a set of synthetic simulations and found a theoretical limiting mass ($M_{200c}$) of $1-2 \times 10^{14} h^{-1} M_{\odot}$ (80\% detection probability), depending on redshift and detailed assumptions of the mass-to-observable conversion.
Folding this selection function into a cosmological model allowed us to compare the observed redshift distribution of clusters in the sample to theoretical expectations.
Accounting for uncertainties due to small number statistics and cosmological sample variance, we find a marginal agreement between the predicted model and the observed distribution, and we note that results depend on the choice of model. In particular, assuming the Planck CMB cosmological model leads to predict too high a density of objects at $0.4<z<0.9$, as compared to the current sample.

We compared several bolometric luminosity-temperature relations extrapolated from the recent literature to our data points. Taking advantage of our knowledge of sample selection effects, we suggest a simple parametrization for the evolution of the normalization of the $L_X-T$ and confirm a negative trend with respect to a pure self-similar evolution. This result is observed in numerical simulations \citep[e.g.][]{short10}, when pre-heating of the intra-cluster medium occurs at the early stage of cluster formation. Interestingly, different assumptions for local ($z=0$) scaling relations lead to different results for their evolution (see numbers quoted in Table~\ref{table_lx-t_fitvalues}).
This is due to their different slopes and normalizations and the fact that our sample spans different loci of the $L_X-T$ plane as redshift increases. Indeed, selection effects move the high-redshift sample to higher temperatures and luminosities.
Importantly, we confirm that mistakenly neglecting selection effects substantially changes these conclusions and leads to a quasi self-similar evolution.

We concluded this study by investigating the reality of the apparent "gap" in the redshift distribution of clusters between $0.4 \lesssim z \lesssim 0.9$:
\begin{itemize}
\item Considering the negative evolution of the $L_X-T$ relation found earlier, we find a milder disagreement between the Planck~2013 cosmology model and our dataset. We note however that a complete analysis should self-consistently address the cosmological model and the evolution of scaling relations.
\item We explored to what extent the observed redshift distribution can be explained in terms of a selection bias due to central point-source contamination of clusters. Based on our simulations, we cannot attribute the increase of cluster density at $z=1$ to a detection enhancement ("flux boosting"). We propose instead an ad-hoc scenario in which the AGN contamination evolves with redshift, both in its occurrence and its strength and peaks at $0.5 \lesssim z \lesssim 1$. Interestingly, combining our cluster population model to the luminosity function of AGN in the Universe points towards a similar trend, although a more thorough modeling is needed (in paticular using numerical simulations).
\item We finally described the impact of various surface brightness profiles on the C1 selection function. Our results show that the enhanced presence of cuspy, cool-core clusters at low redshifts (e.g.~at $z<0.75$, as observed by \citealt{mcdonald13}) could also lead to an apparent gap in the observed C1 redshift distribution, since cool-core clusters in $0.4 \lesssim z < 0.75$ would be considered as less likely extended sources by detection algorithms.
\end{itemize}

Comparison of cluster catalogues selected in X-ray and other wavelengths (e.g.~optical and S-Z) will confirm these possible scenarios. So will do detailed studies with the {\it Chandra} observatory by assessing the point-source content and surface brightness shape evolution of those galaxy clusters.
Future results will therefore rely on the larger XMM-XXL survey (50~deg$^2$ at a similar depth), separated in two independent fields in order to beat sample variance. Echoing our discussion on selection effects and their impact on cosmological observables (here $dn/dz$), a large effort has been undertaken within the XMM-XXL team in order to develop and compare multi-wavelength cluster selections, to perform detailed follow-up of selected samples and analyses of clusters from numerical simulations.
The full-sky survey of eRosita \citep{predehl10,merloni12} starting early 2016 will bring the statistical power ($\sim 10^5$ X-ray galaxy clusters up to $z \gtrsim 1$) needed to break degeneracies between the multiple assumptions entering the cosmological analysis of X-ray cluster surveys and will offer an unprecedented tridimensional view on the large-scale structure as traced by its most massive constituents.

\section*{Acknowledgments}
We are thankful to the anonymous referee for his careful reading and his help in improving the quality of this manuscript.

The authors thank P.~Giles, T.~Ponman and J.~B. Melin for useful discussions on the X-ray properties of the clusters, J.~P. Le F\`evre for the cluster database management, C.~Libbrecht and A.~Gueguen for their contribution in the X-ray pipeline development. We thank Ian McCarthy for assistance with the calculation of cluster masses.

ML acknowledges a postgraduate studentship from the UK Science and Technology Facilities Council (STFC).  GPS acknowledges support from the Royal Society and STFC.

FP acknowledges support from the DLR Verbunforschung grant 50 OR 1117.

Based on observations obtained with XMM-Newton, an ESA science mission with instruments and contributions directly funded by ESA Member States and NASA.

Based on observations collected at TNG (La Palma, Spain), Magellan (Chile), and at ESO Telescopes at the La Silla and Paranal Observatories under programmes ID 072.A-0312, 074.A-0476, 076.A-0509, 070.A-0283, 072.A-0104, 074.A-0360, 089.A-0666, and 191.A-0268.

This research has made use of the XMM-LSS database, operated at CeSAM/LAM, Marseille, France.

Based on observations obtained with MegaPrime/MegaCam, a joint project of CFHT and CEA/IRFU, at the Canada-France-Hawaii Telescope (CFHT) which is operated by the National Research Council (NRC) of Canada, the Institut National des Science de l'Univers of the Centre National de la Recherche Scientifique (CNRS) of France, and the University of Hawaii. This work is based in part on data products produced at Terapix available at the Canadian Astronomy Data Centre as part of the Canada-France-Hawaii Telescope Legacy Survey, a collaborative project of NRC and CNRS. 

This paper uses data from the VIMOS Public Extragalactic Redshift Survey (VIPERS). VIPERS has been performed using the ESO Very Large Telescope, under the "Large Programme" 182.A-0886. The participating institutions and funding agencies are listed at {http://vipers.inaf.it}.

Funding for SDSS-III has been provided by the Alfred P. Sloan Foundation, the Participating Institutions, the National Science Foundation, and the U.S. Department of Energy Office of Science. The SDSS-III web site is {http://www.sdss3.org/}.
SDSS-III is managed by the Astrophysical Research Consortium for the Participating Institutions of the SDSS-III Collaboration including the University of Arizona, the Brazilian Participation Group, Brookhaven National Laboratory, Carnegie Mellon University, University of Florida, the French Participation Group, the German Participation Group, Harvard University, the Instituto de Astrofisica de Canarias, the Michigan State/Notre Dame/JINA Participation Group, Johns Hopkins University, Lawrence Berkeley National Laboratory, Max Planck Institute for Astrophysics, Max Planck Institute for Extraterrestrial Physics, New Mexico State University, New York University, Ohio State University, Pennsylvania State University, University of Portsmouth, Princeton University, the Spanish Participation Group, University of Tokyo, University of Utah, Vanderbilt University, University of Virginia, University of Washington, and Yale University.

\appendix

\section[]{Comparison with Pacaud et al. (2007) measurements}
\label{app:compa_pacaud}

As described in Sect.~\ref{sect:construction_sample}, 29 of the 52 clusters presented in Table~\ref{table_c1_catalogue} pertain to the sample published by \citet[][hereafter P07]{pacaud07}. They are issued from a similar datasets and were analysed through similar methodologies. However, several changes have occurred between these two analyses:
\begin{itemize}
\item the {\sc XAmin} software used to extract sources, leading to slight changes in the masking of sources and definition of the optimal extraction radius $R_{\rm spec}$. 
\item the event list processing and spectral extraction algorithms, in particular the {\sc XMM-SAS}\footnote{XMM Science Analysis Software} version.
\item the version of {\sc XSpec} (from v.11.3.2 to v.12.8.0).
\item the APEC model, specifically the ATOMDB\footnote{http://atomdb.org/} database models (from v.1.3.1 to v.2.0.1).
\end{itemize}

	\subsection{Updated cluster redshift}
XLSSC~35 has been updated to a new redshift value of $z=0.07$ instead of $z=0.17$ as quoted in P07. The presence of a giant elliptical at $z=0.069$ coincident with the X-ray peak argues in favor of this cluster redshift, although a superposition of two layers cannot be ruled out \citep[see discussion in][]{adami11}.

	\subsection{Temperature measurements}
	
\begin{figure}
	\includegraphics[width=84mm]{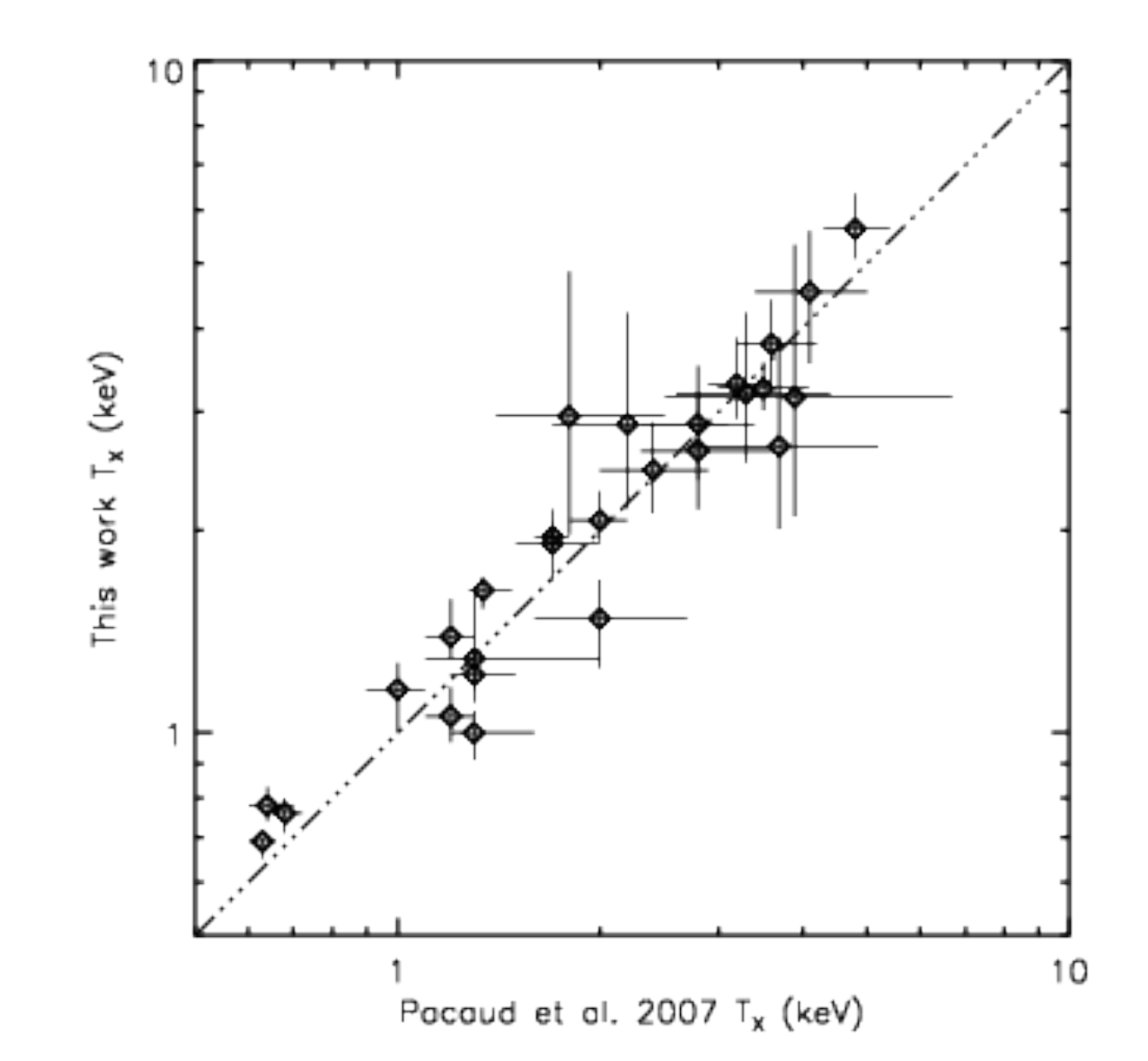}
	\caption{Comparison of temperature measurements the 29 clusters in common between \citet{pacaud07} and this work (Table~\ref{table_c1_catalogue}). Error bars delimitate 68\% confidence intervals. The dashed line shows the equality relation.}
	\label{fig:compa_tx_pacaud07} 
\end{figure}	

We first compare the X-ray spectral measurements, through the values of temperatures $T_X$. Fig.~\ref{fig:compa_tx_pacaud07} compares values listed in P07 to those published in this work. Taking into account measurement uncertainties, we conclude to a good agreement between these two series. In order to address the discrepancies, we defined a series of additional temperature measurements for all clusters by applying changes as listed in Table~\ref{table_tx_changes}.
We summarize our findings as follows, the numbers referring to this table:
\begin{enumerate}
\item (5) provided identical results as P07 (2), proving that there is no other source of bias than the one listed in the table.
\item (4)-(5) provided identical results, as did (3)-(7). Since only the {\sc XSpec} version was changed in these comparisons, we cannot attribute the discrepancies to the spectral fitting routines in {\sc XSpec}.
\item (6)-(7) led to the identification of a bias in the temperatures of cool systems ($T \lesssim 2$~keV), attributed to a change in the APEC plasma models. The newest APEC version (v.2.0.1) delivers higher temperatures as the v.1.3.1, the effect being more pronounced for the coolest systems (from 1$\sigma$ differences above 1~keV to 2-3$\sigma$ differences below 1~keV). As a straightforward consequence, this bias is also found in the (1)-(3) comparison.
\item (1)-(6) provided almost identical results, except for the presence of two outliers with large error bars. We attribute these differences to the changes in spectral extraction regions and the low signal-to-noise of the spectra involved.
\item (3)-(5) and (3)-(2) provided similar results, although with some scatter around the one-to-one relation. This is attributed to the change in event lists creation and spectral extraction routines.
\end{enumerate}
In brief, we attribute the scatter around the equality line in Fig.~\ref{fig:compa_tx_pacaud07} to changes in the event lists processing and our defintion of source extraction regions, while the small bias at low temperatures is attributed to a recent change in APEC models.

Note that the updated redshift value for XLSSC~35 leads to an updated value of the temperature that is consistent with the previously published value ($T_X=1.2 \pm 0.1$~keV in P07, $T_X=1.1 \pm 0.1$ in this work).

\begin{table}
	\centering
\caption{\label{table_tx_changes}List of X-ray temperature measurement experiments designed to address the discrepancies between previously published values \citep[][P07]{pacaud07} and this work. Each line corresponds to a series of measurements of the 29 clusters in common between the two samples. Line~(1) corresponds to values listed in Table~\ref{table_c1_catalogue} while line~(2) to those listed in Table~1 of P07, as compared in Fig.~\ref{fig:compa_tx_pacaud07}.}
		\begin{tabular}{@{}lcccc@{}}
\hline
	&	Extraction/	&	Event lists	&	{\sc Xspec} &	APEC		\\
	&	mask regions	&	version	&	version	& version	\\
\hline
\hline
(1)	this work	&	New	&	3.2	&	12.8.0	&	2.0.1	\\
(2) P07			&	\emph{Old}	&	\emph{2.1}	&	\emph{11.3.2}	&	\emph{1.3.1}	\\
\hline
(3)				&	Old	&	3.2	&	11.3.2	&	1.3.1	\\
(4)				&	Old	&	2.1	&	12.8.0	&	1.3.1	\\
(5)				&	Old	&	2.1	&	11.3.2	&	1.3.1	\\
(6)				&	Old	&	3.2	&	12.8.0	&	2.0.1	\\
(7)				&	Old	&	3.2	&	12.8.0	&	1.3.1	\\
\hline
		\end{tabular}
\end{table}

	\subsection{Flux measurements}

\begin{figure}
	\includegraphics[width=84mm]{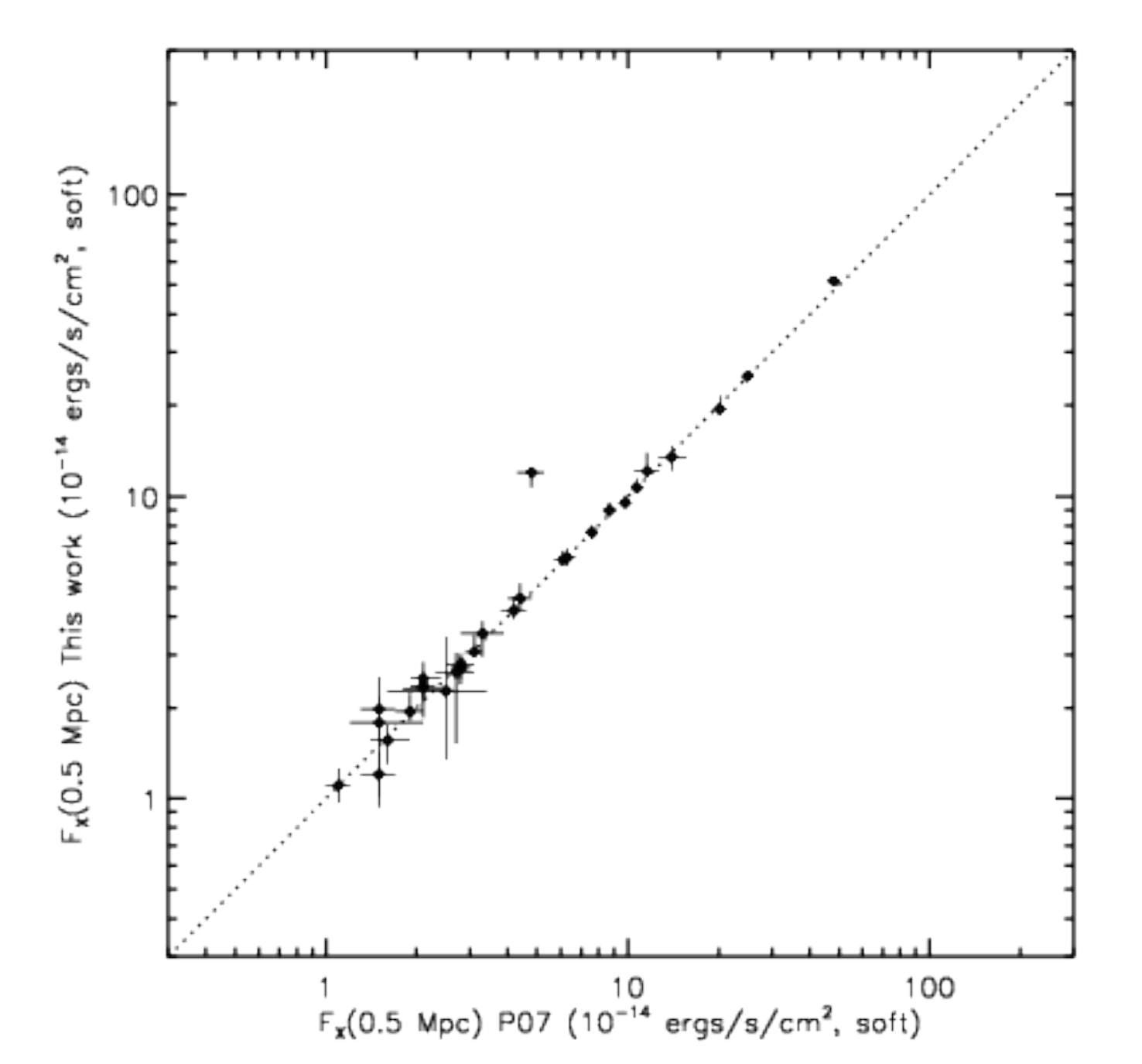}
	\caption{Comparison of [0.5-2]~keV flux measurements for the 29 clusters in common between \citet{pacaud07} and this work (Table~\ref{table_c1_catalogue}) in 0.5~Mpc apertures at the cluster redshift. Error bars delimitate 68\% confidence intervals. The dashed line shows the equality relation. The clear outlier corresponds to XLSSC~35 whose redshift has been updated to 0.07 (instead of 0.17 in P07).}
	\label{fig:compa_fx_pacaud07} 
\end{figure}

We compare on Fig.~\ref{fig:compa_fx_pacaud07} our flux measurements with those obtained in P07. The methodology adopted in P07 is very similar to our "method~(i)" in Sect.~\ref{sect:flux_mes}. Namely, it relies on fitting a surface brightness profile by means of $\beta$-models. For this reason, we do not compare fluxes measured with method~(ii) with those of P07.
As this comparison is shown in a radius of 0.5~Mpc at the cluster redshift, we corrected our angular apertures from the change of cosmological model. Note that XLSSC~35 was quoted with a redshift of 0.17 in P07, implying a smaller angular aperture than the one used for this work.
Overall, both flux measurements are in excellent agreement.

\section[]{Derivation of masses and assumption on $R_{500}$}
\label{app:m500stabil}

\begin{figure}
	\includegraphics[width=84mm]{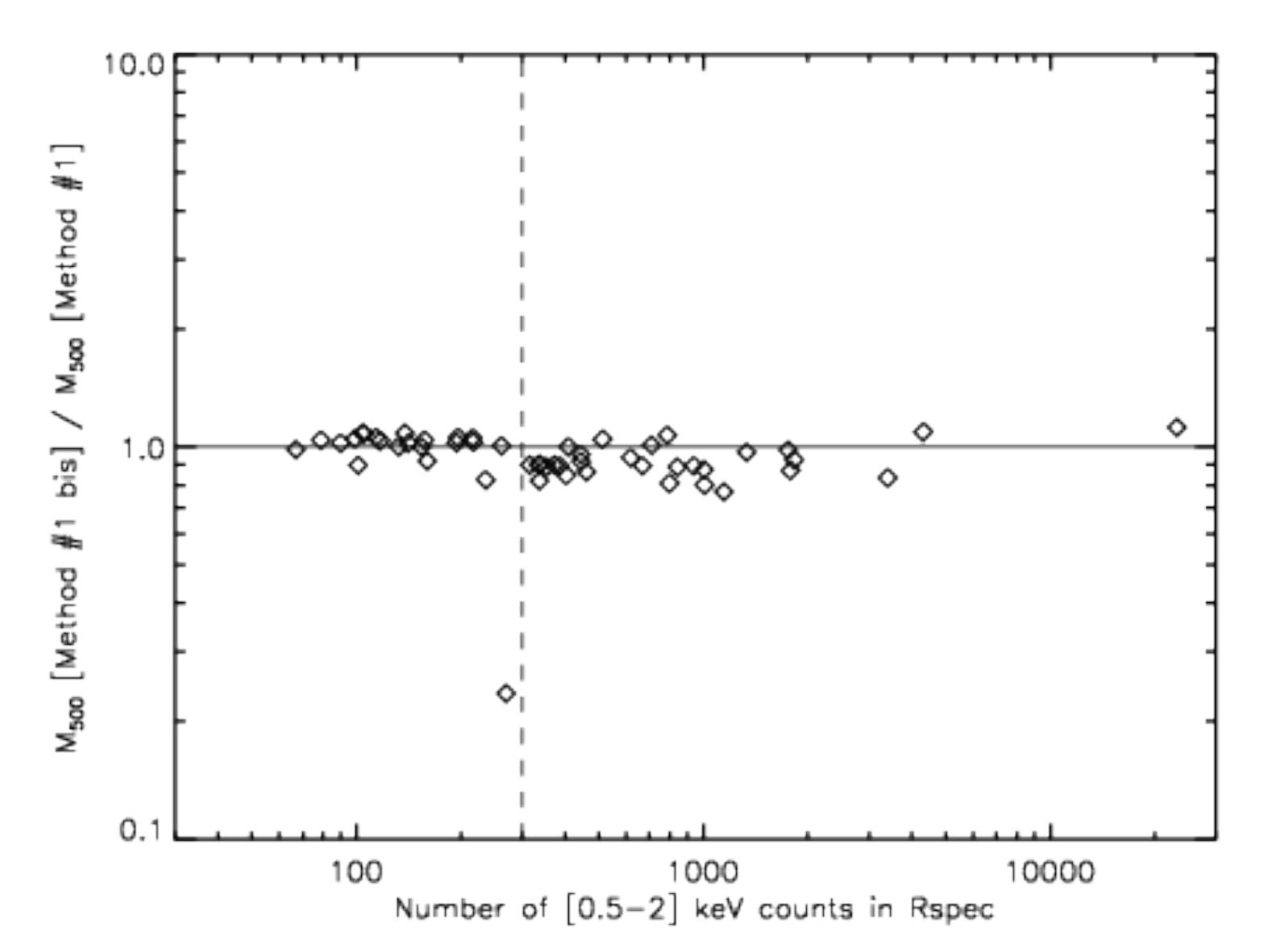}
	\caption{Comparison of mass estimates for the 52 clusters in this sample. \emph{Method~1} assumes hydrostatic equilibrium and a gas distribution following a $\beta$-model. \emph{Method~1bis} is similar but does not assume a scaling law for the value of $R_{500}$.
	Clusters located right of the dashed line have their surface brightness profile well-described by a 2-parameter $\beta$-model, while clusters on the left were imposed $\beta=2/3$ (see text).}
	\label{fig:compa_miter} 
\end{figure}

We described in Sect.~\ref{mass_method1} a method to estimate the mass of a cluster. The first step consists in estimating $R_{500}$ from a $M_{500}-T_X$ relation. However, this can be seen as an unnecessary step since the relation between $R_{500}$ and $M_{500}$ is straightforward: $M_{500} = \frac{4\pi}{3} 500 \rho_c(z) R_{500}^3$.
Combining this formula with Eq.~\ref{equ_mass_beta} provides:
\begin{equation}
R_{500} = R_c \sqrt{\frac{3.33\times 10^{14} \beta T}{2000 \pi \rho_c(z)R_c^2}-1}
\end{equation}
The $R_{500}$ values do not depend on a scaling relation and can be used in Eq.~\ref{equ_mass_beta} to provide a mass estimate. Fig.~\ref{fig:compa_miter} compares the mass estimates obtained by this method ("method 1 bis") and the method presented earlier ("method 1").
Apart from one outlier (XLSSC~080), the agreement is satisfactory, showing that our scaling used for inferring $R_{500}$ is a reasonable one.

\section[]{Notes on individual clusters}
\label{app:clusters}

Fig.~\ref{fig:c1_gallery} shows X-ray/optical overlays for 23 C1 clusters in this sample (the other 29 are shown in \citealt{pacaud07}, their Appendix~B).

\begin{itemize}
\item \textbf{XLSSC~060:} Abell cluster A0329. Our pipeline wrongly deblended multiple components, due to the exceptional extent and brightness of this source and the presence of gaps on XMM detectors. They were manually merged together for the purpose of measuring its X-ray properties. \citet{cruddace02} measured an unabsorbed, [0.1-2.4]~keV flux of $2.12 \pm 0.42 \times 10^{-12}$~ergs/s/cm$^2$ from ROSAT data. This translates into a [0.5-2]~keV, galactic-absorbed, flux of $1.2 \pm 0.2 \times 10^{-12}$~ergs/s/cm$^2$, hence entirely consistent with our XMM value.
\item \textbf{XLSSC~079:} this cluster is detected on the deepest pointing of the survey (80~ks on-axis) at an off-axis radius of $8\arcmin$. The X-ray analysis is therefore severely limited by confusion: its extended emission is contaminated by a number of point-sources. Despite our efforts to correctly mask all point sources, we recommend to consider its temperature and luminosity measurements with caution. This is reinforced by the fact that this cluster appears as a clear outlier in the $L_X-T$ diagram (lower-right point in first panel of Fig.~\ref{fig:lx-t_selfsimilar}).
\item \textbf{XLSSC~053:} this cluster is probably a group of clusters at $z\sim0.5$, as hinted by the projected distribution of galaxies and the faint, large, X-ray surface brightness. An assessment of its multiple-component nature will be possible thanks to the analysis of spectroscopic redshifts in the vicinity of this object.
\item \textbf{XLSSU J021744.1-034536:} this cluster was first presented and discussed in \citet{willis13}. We note that recent data obtained with the CARMA interferometer confirmed the presence of hot gas at the location of this object via Sunyaev-Zeldovich effect \citep{mantzsz}.
To date, its spectroscopic confirmation is awaited. A $3\times9$~min-exposure spectrum was obtained using the FORS2 instrument at ESO-VLT. A long slit was centered on the bright, blue object close to the X-ray centroid and on the bright, blue, object located 21\arcsec southwards. Both their spectra indicate these objects are probably stars. The red cD galaxy (visible on the image of \citealt{willis13}) partly falls onto the slit but is too faint to provide a spectrum. A very faint object located within the X-ray contours at $\mathrm{R.A.}=34.434$, $\delta=-3.753$ (J2000, 2\arcsec positional uncertainty) shows one emission line in all three exposures at $\lambda=5437.5$ \AA. We postulate this line is Ly-$\alpha$ emitted by an active emitter at $z=3.47$. With this data in hand, we cannot exclude that the two stars and the active object also emit X-rays contaminating the galaxy cluster emission.
\end{itemize}

\begin{figure*}
	\begin{tabular}{ m{9.5cm} m{9.5cm} }
		\includegraphics[height=90mm]{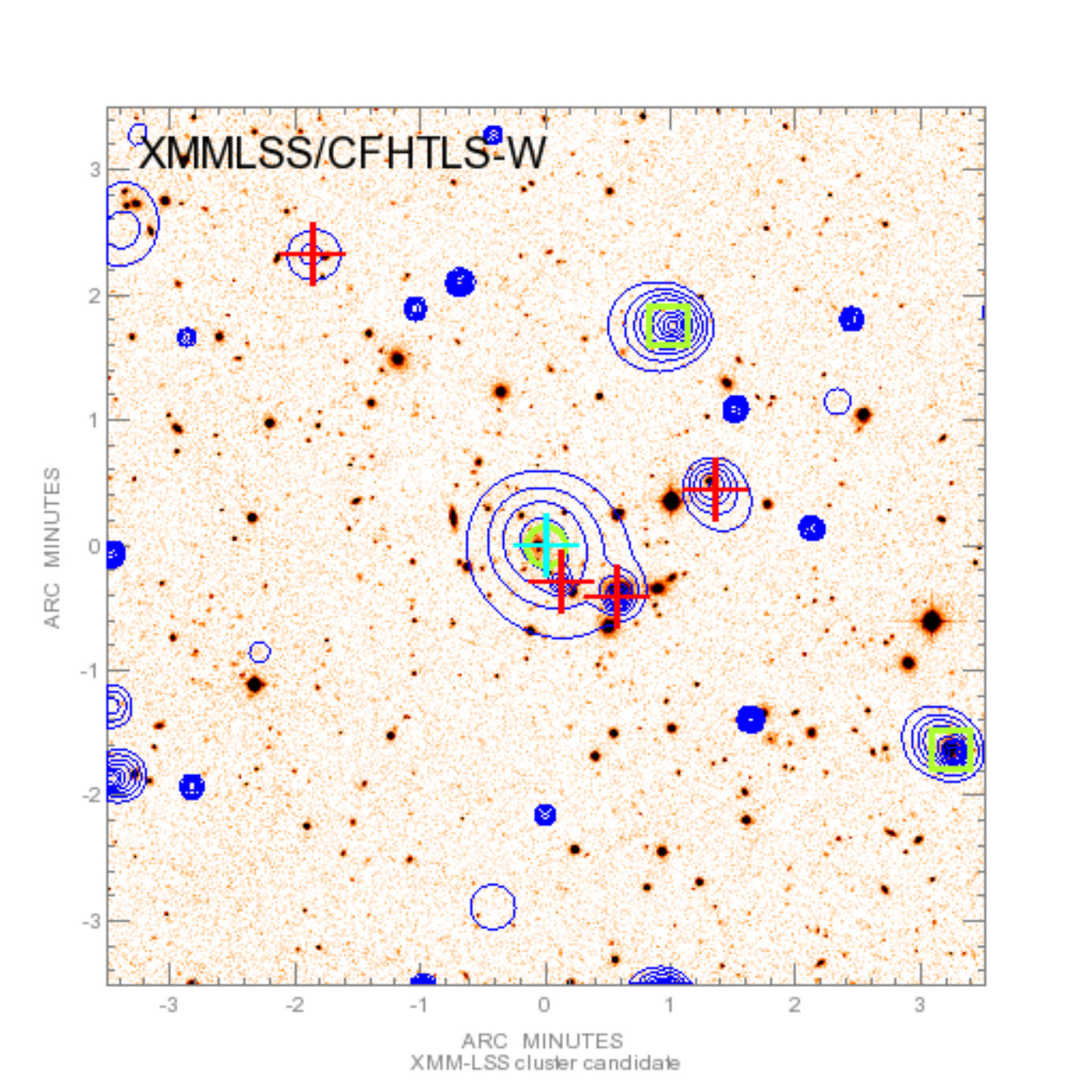} &
		\includegraphics[width=74mm]{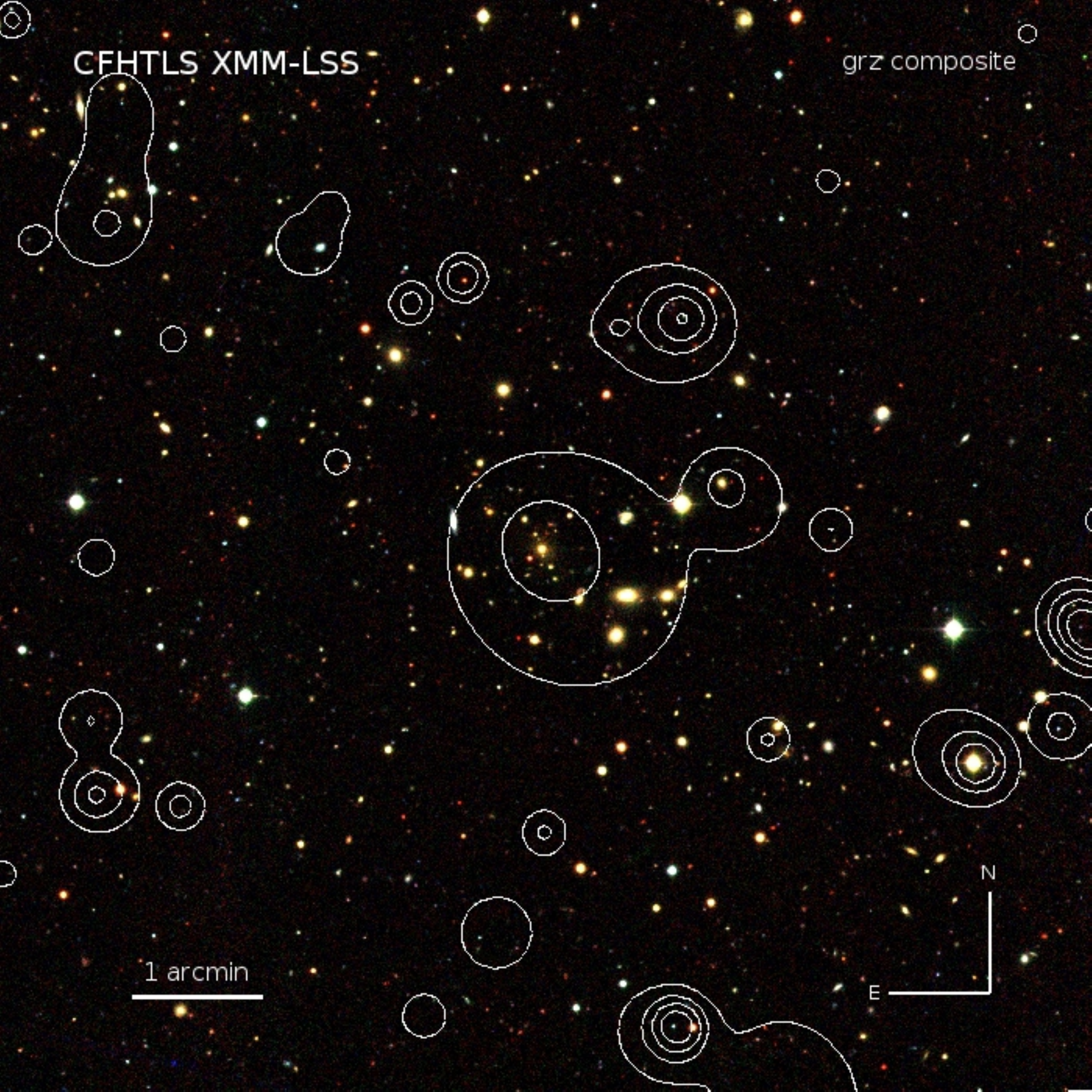} \\
		\includegraphics[height=90mm]{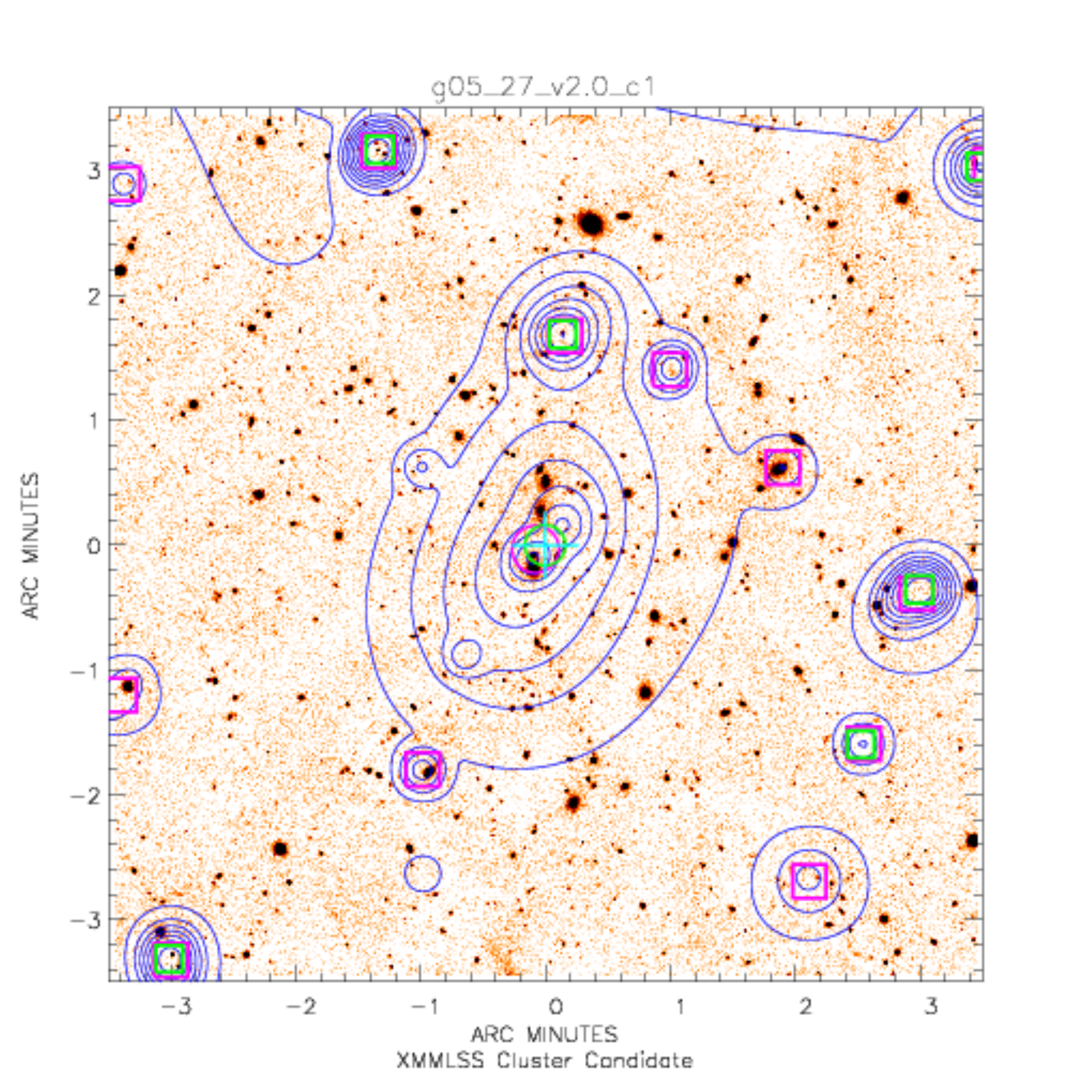} &
		\includegraphics[width=74mm]{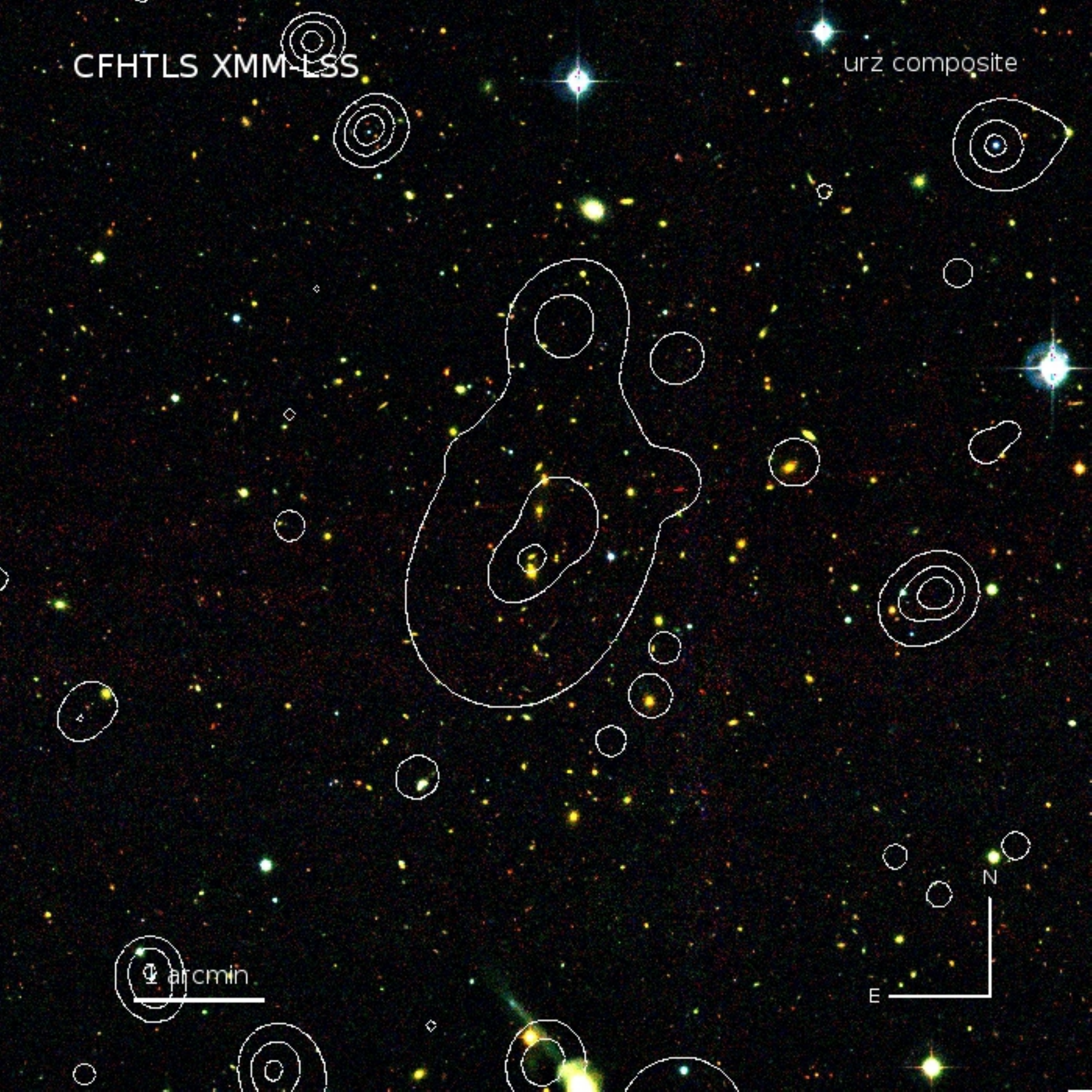} \\
	\end{tabular}
	\caption{Images of the C1 clusters not presented in Appendix~B of \citet{pacaud07} and sorted by ascending XLSSC number.
	Left: X-ray/I-band overlay (7 arcmin on a side). Squares indicate point sources (likelihood $> 15$, \citealt{chiappetti13}) and red crosses are other detections (likelihood $< 15$). Right: three-colour image with X-ray overlays. Images are centered on the pipeline best centroid, which may differ from the position used for fluxes and temperatures measurements. Contour levels are identical across all images appearing in the left column but have been chosen differently for images on the right-hand side.
	\emph{Top:} XLSSC~009 ($z=0.33$) \emph{Bottom:} XLSSC~012 ($z=0.43$).}
	\label{fig:c1_gallery}
\end{figure*}

\begin{figure*}
	\begin{tabular}{ m{9.5cm} m{9.5cm} }
		\includegraphics[height=90mm]{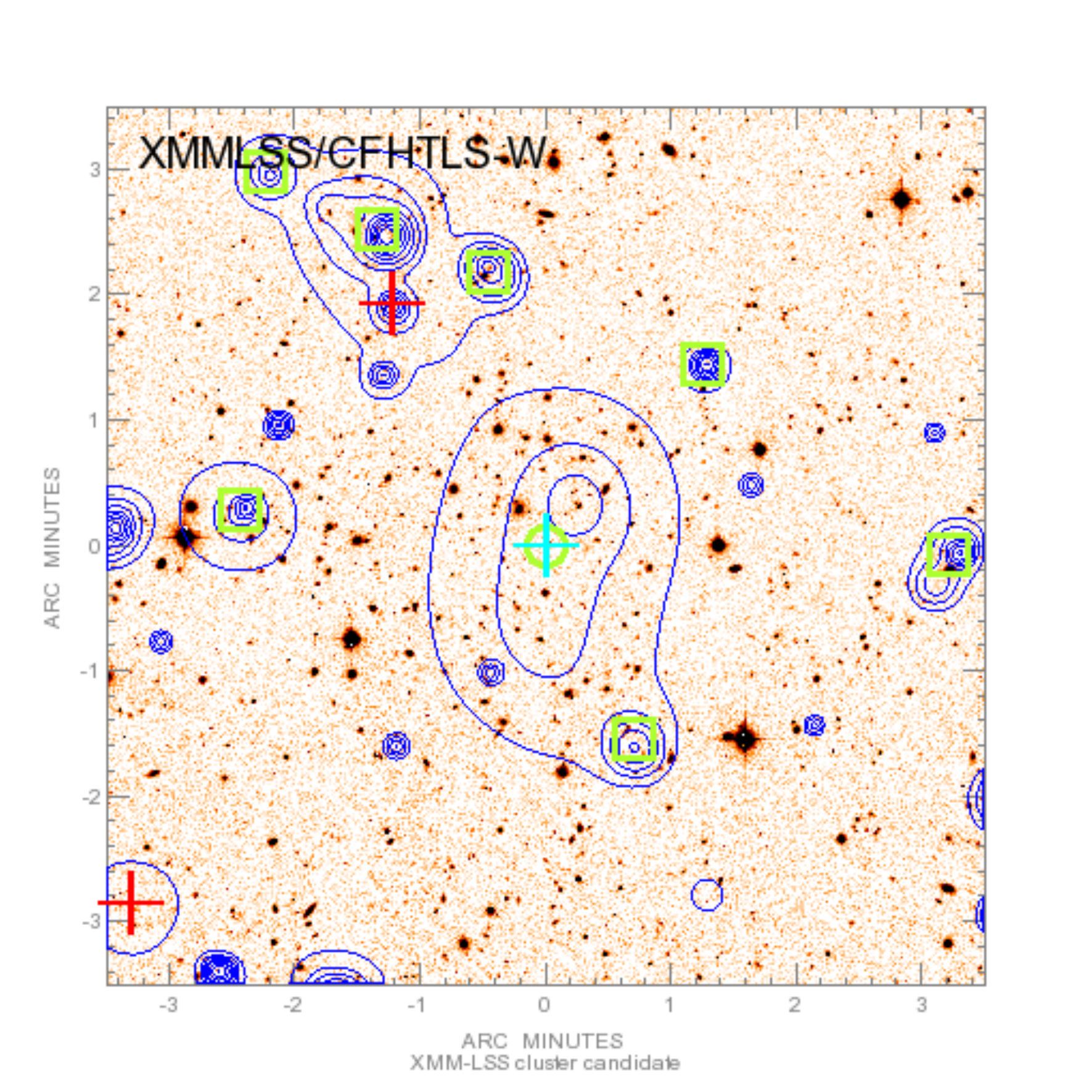} &
		\includegraphics[width=74mm]{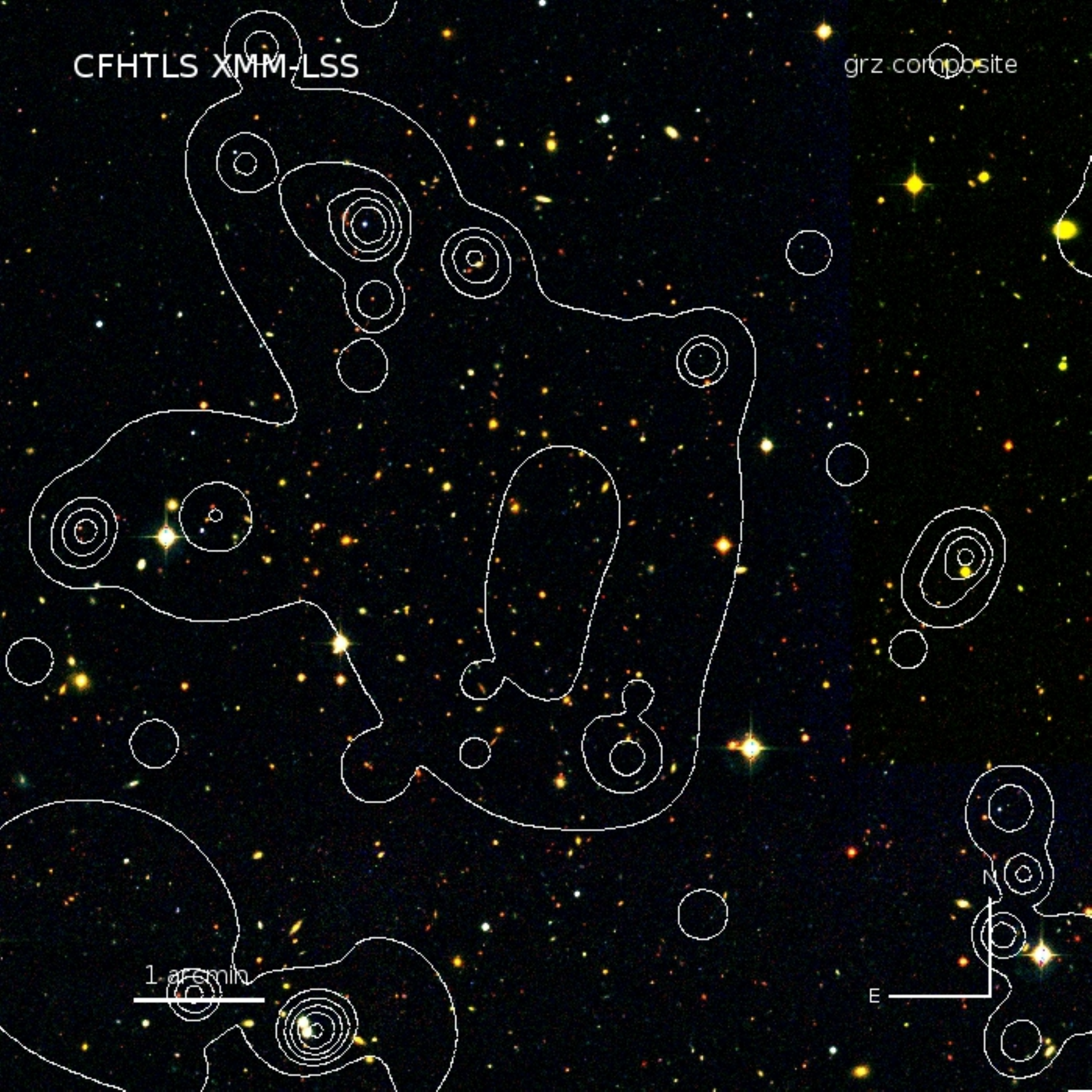} \\
		\includegraphics[height=90mm]{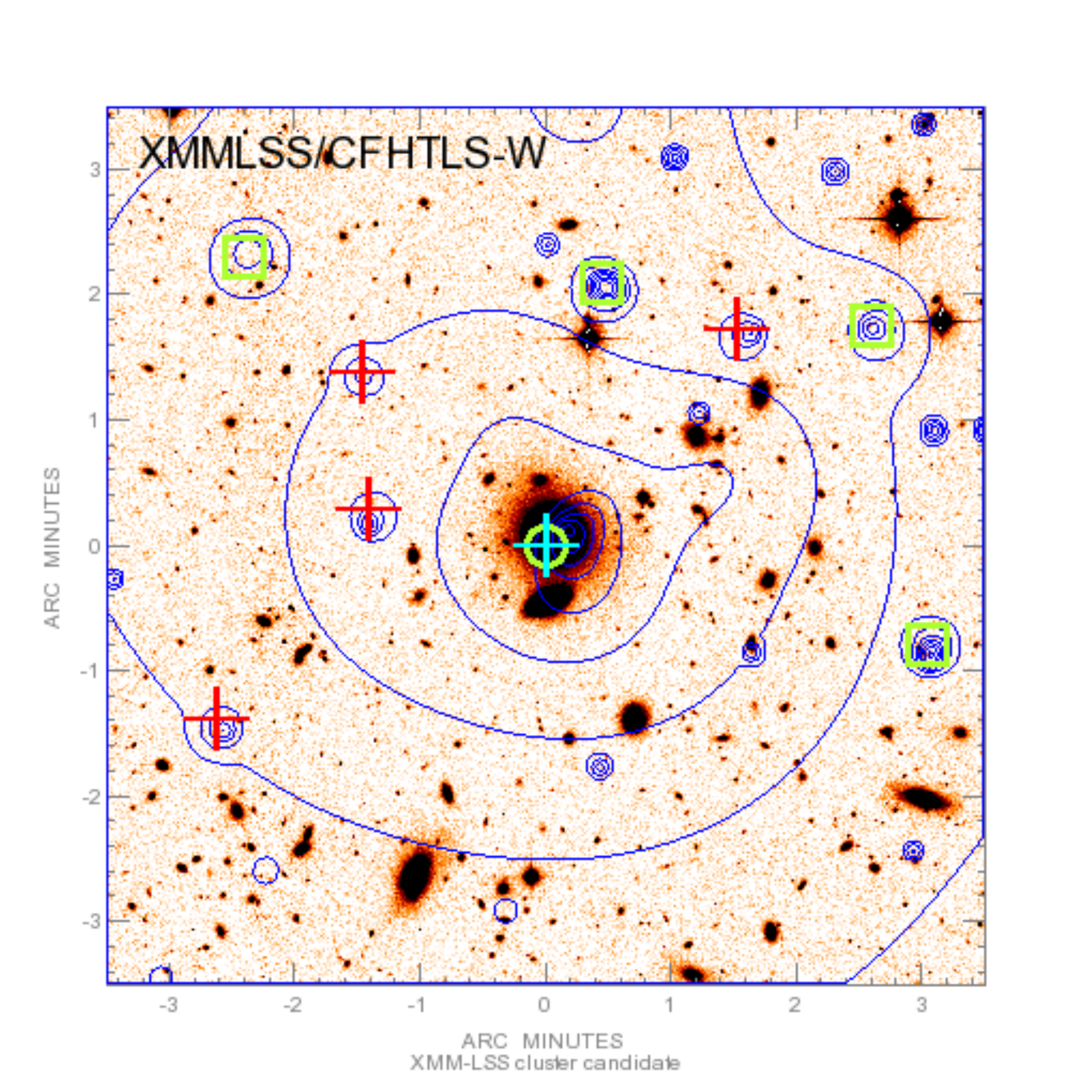} &
		\includegraphics[width=74mm]{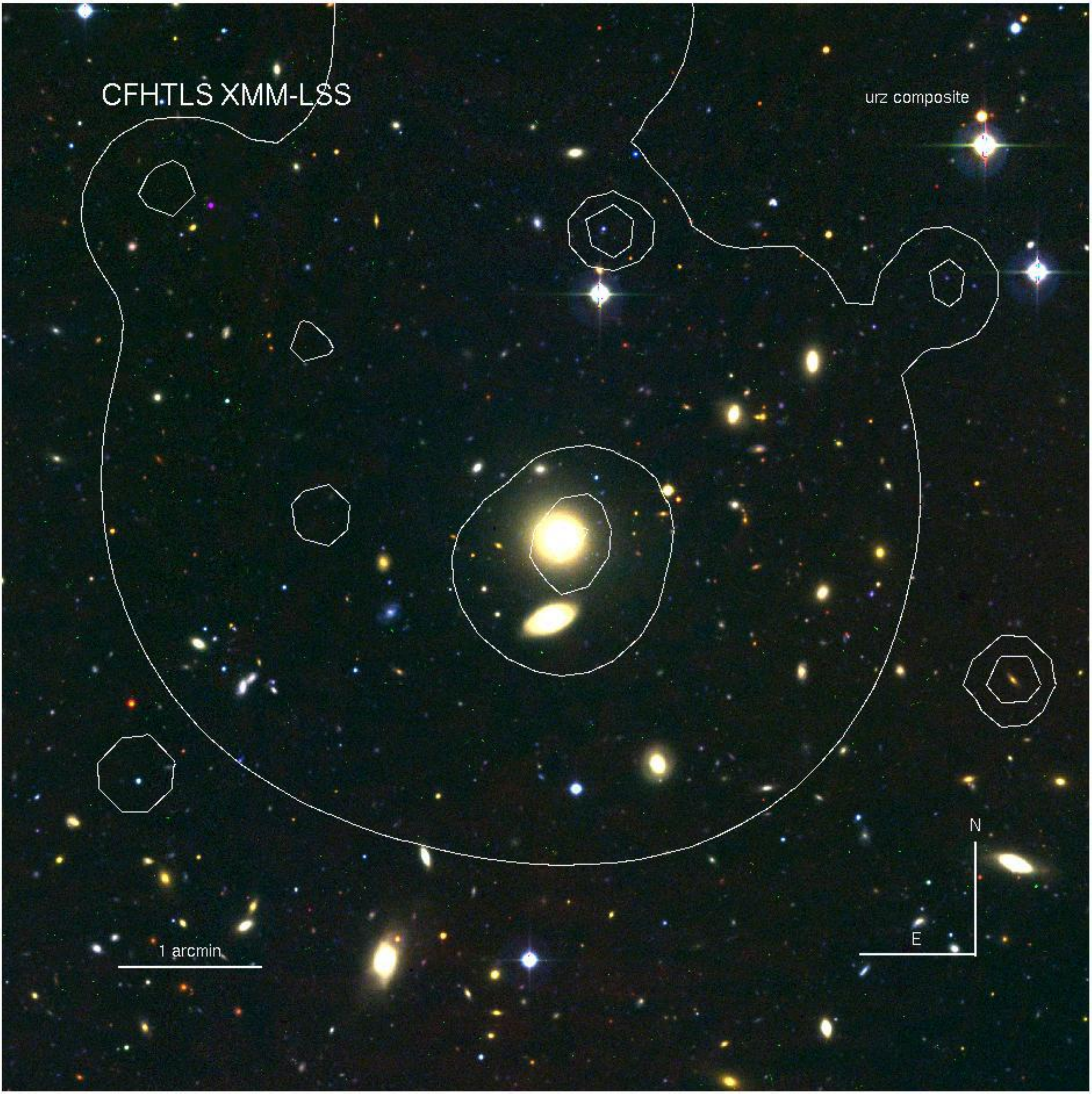} \\
	\end{tabular}
	\contcaption{Images of the C1 clusters.
	\emph{Top:} XLSSC~053 ($z=0.50$) \emph{Bottom:} XLSSC~054 ($z=0.05$).}
\end{figure*}

\begin{figure*}
	\begin{tabular}{ m{9.5cm} m{9.5cm} }
		\includegraphics[height=90mm]{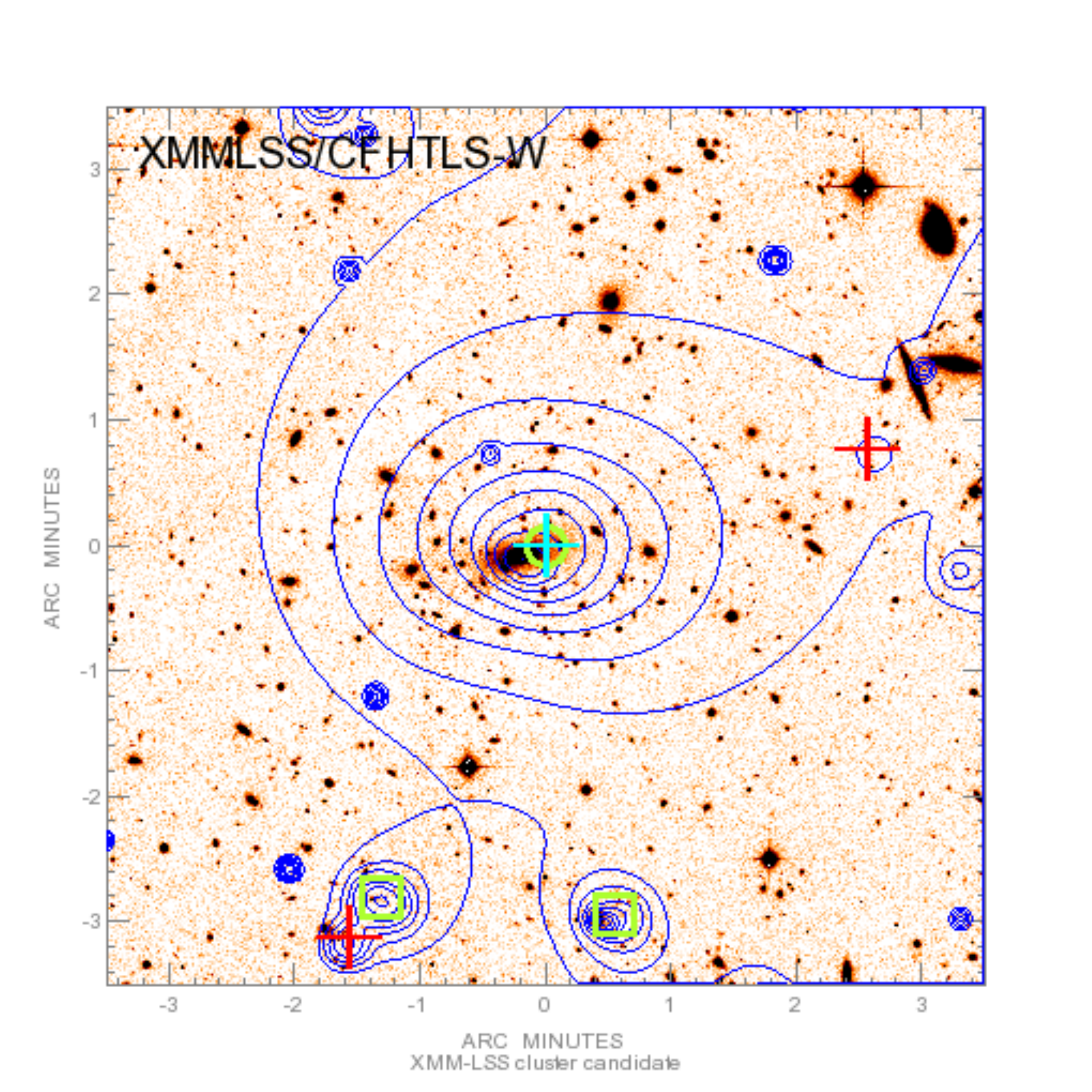} &
		\includegraphics[width=74mm]{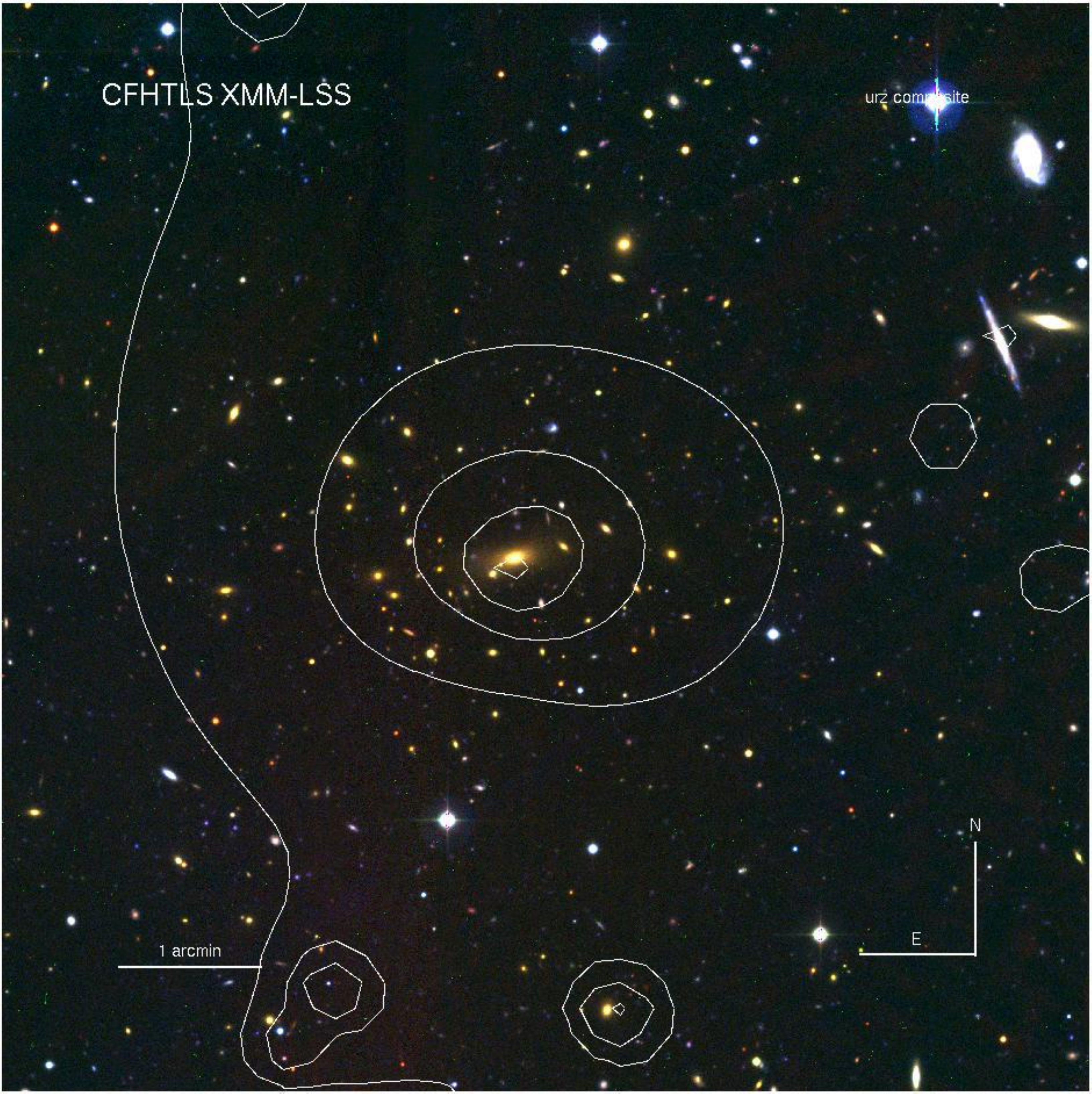} \\
		\includegraphics[height=90mm]{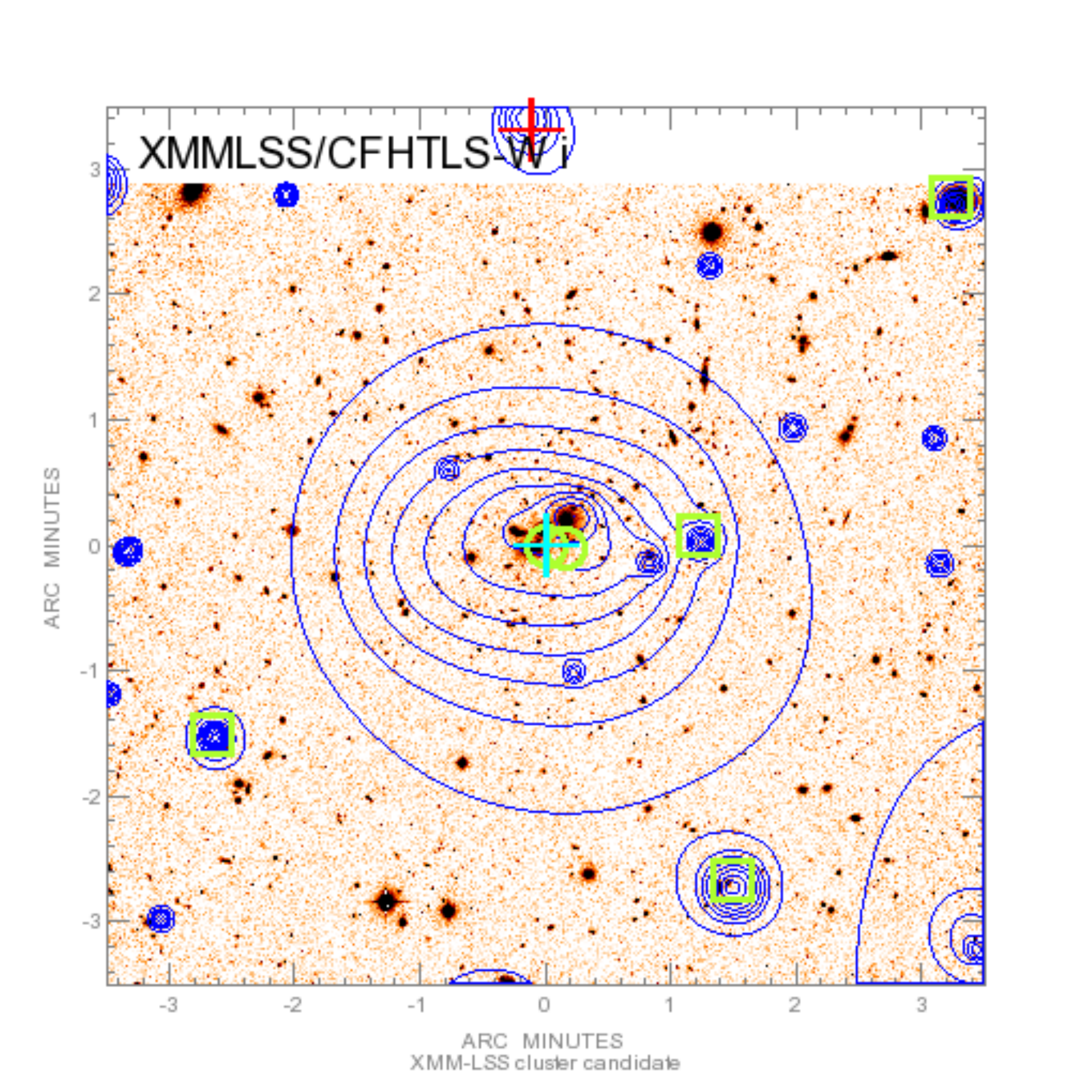} &
		\includegraphics[width=74mm]{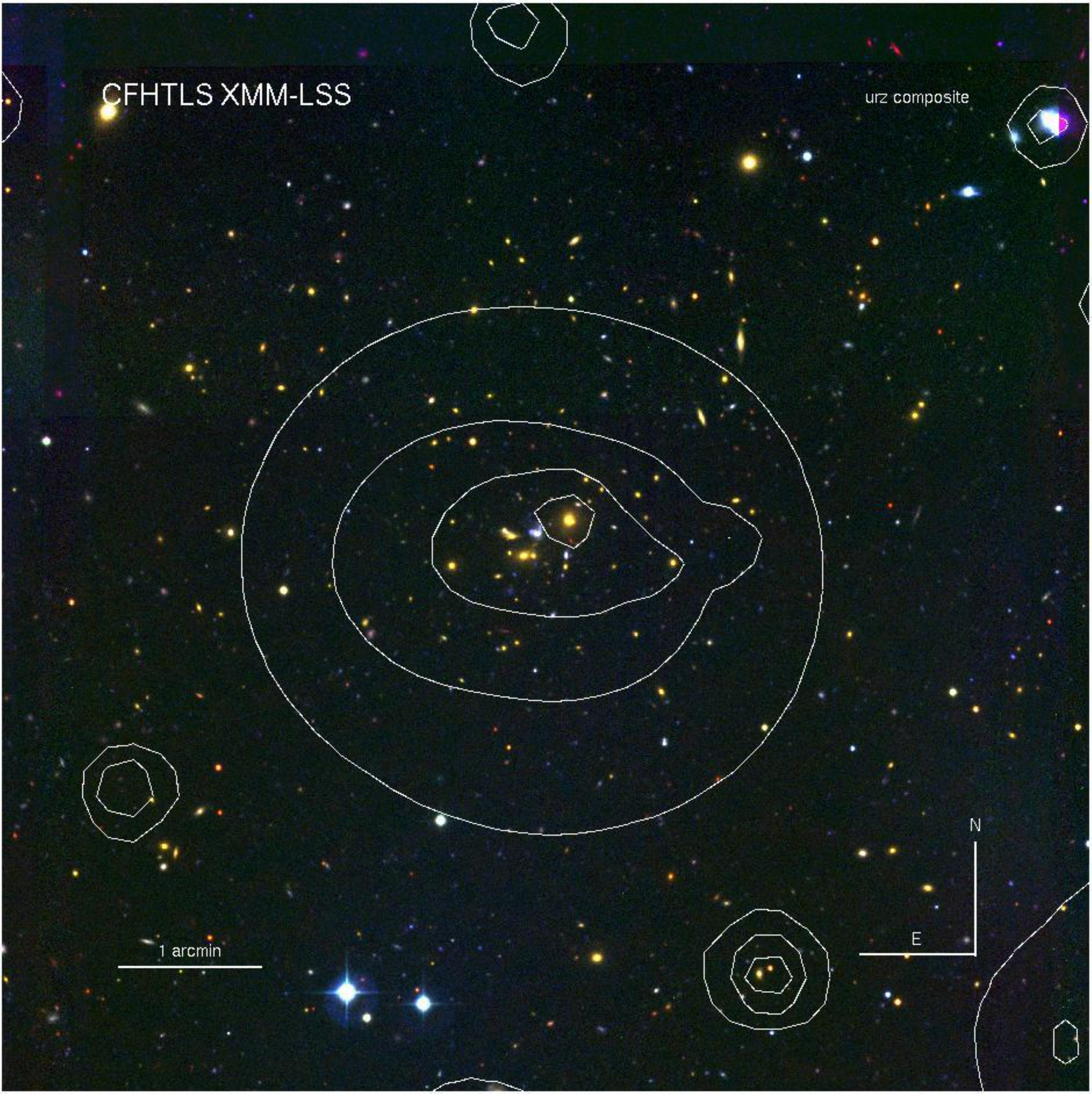} \\			
	\end{tabular}
	\contcaption{Images of the C1 clusters.
	\emph{Top:} XLSSC~055 ($z=0.23$) \emph{Bottom:} XLSSC~056 ($z=0.35$).}
\end{figure*}

\begin{figure*}
	\begin{tabular}{ m{9.5cm} m{9.5cm} }
		\includegraphics[height=90mm]{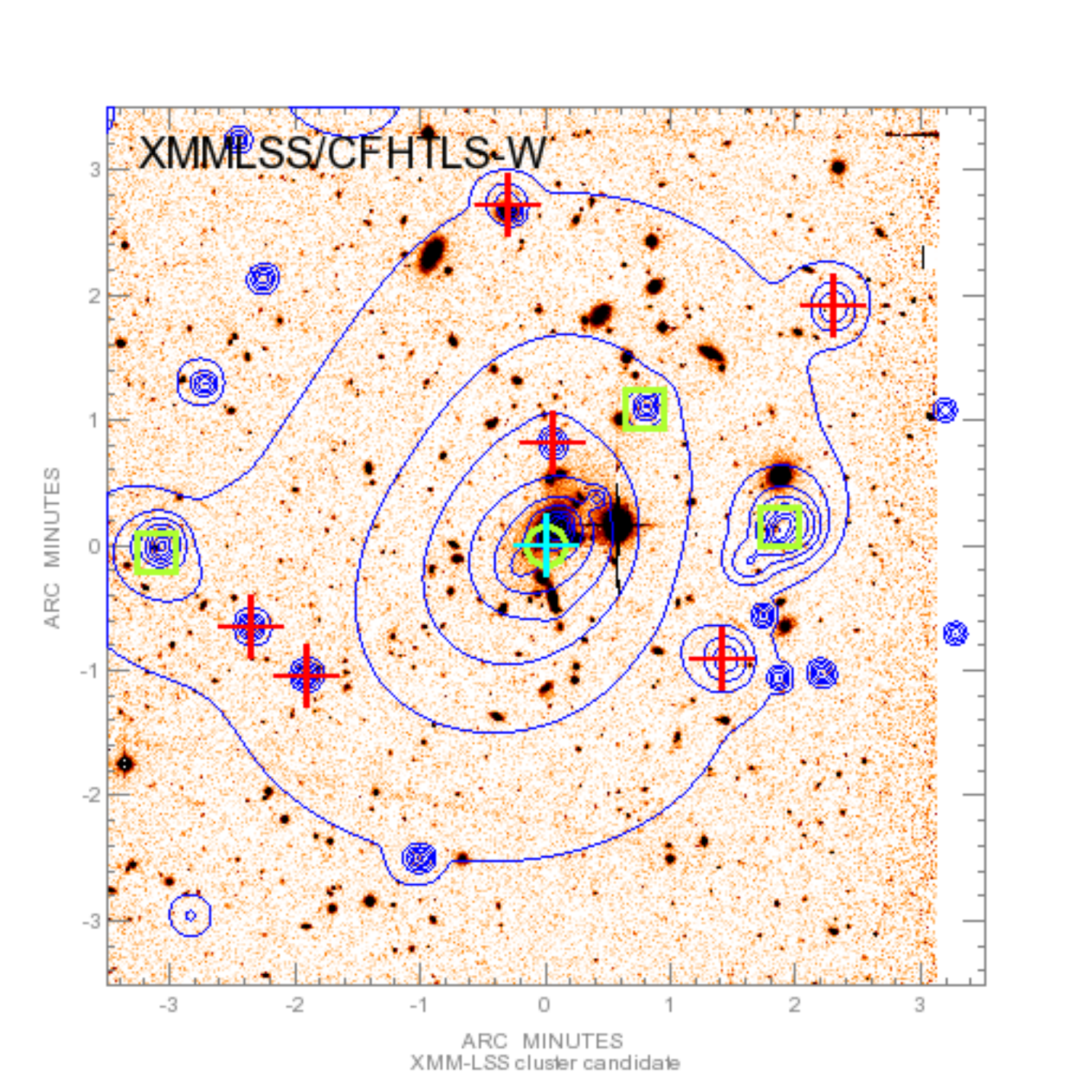} &
		\includegraphics[width=74mm]{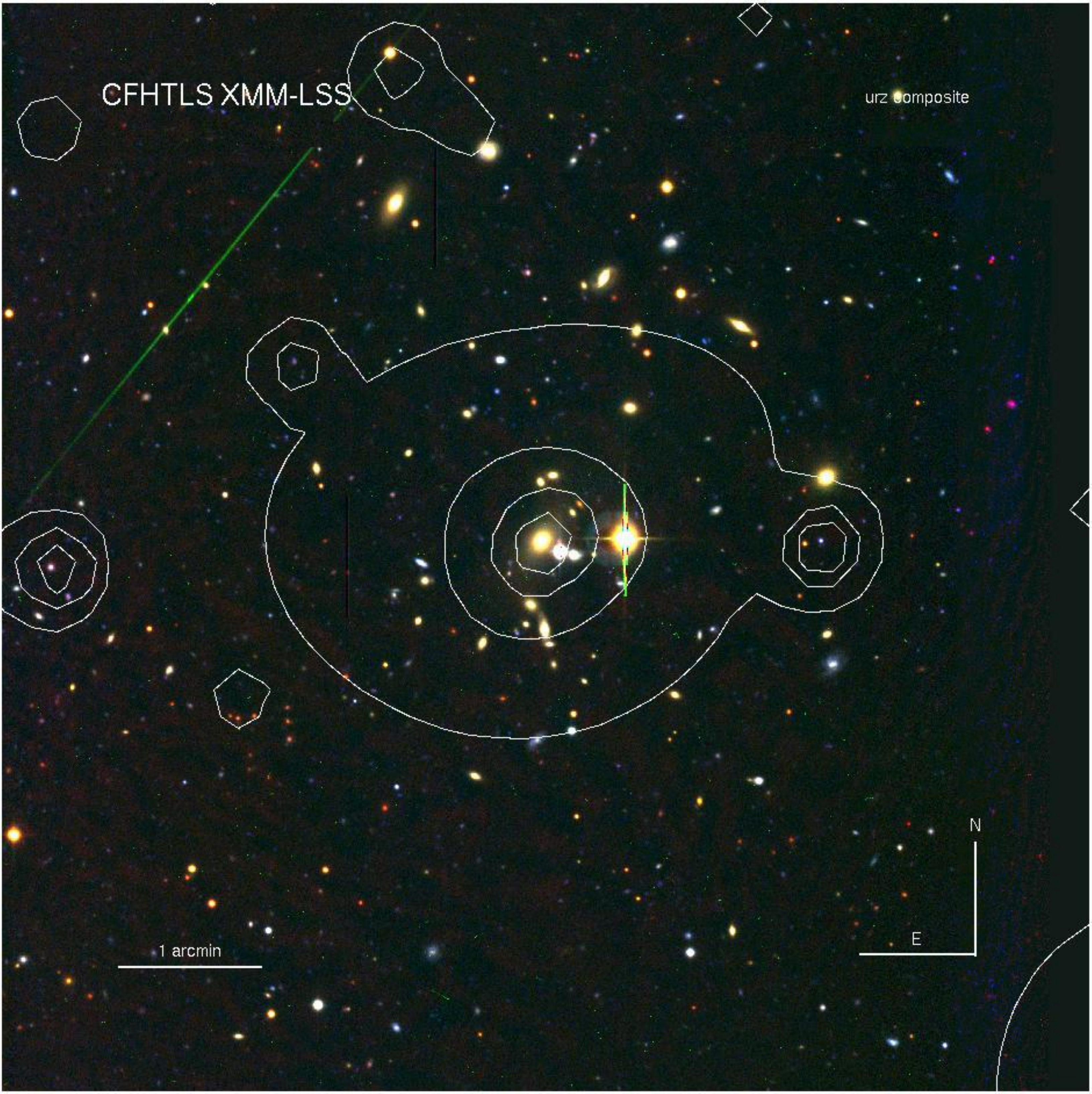} \\
		\includegraphics[height=90mm]{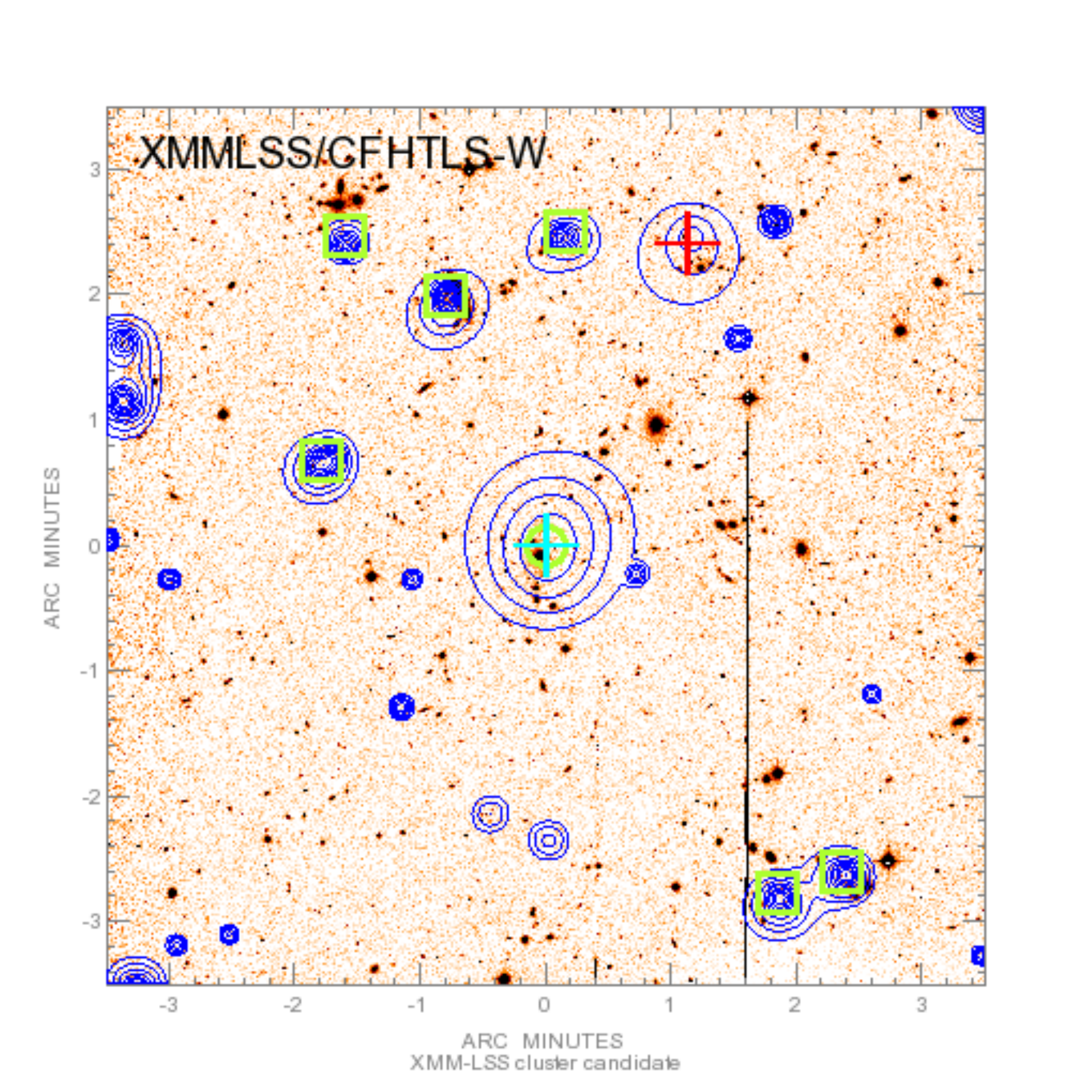} &
		\includegraphics[width=74mm]{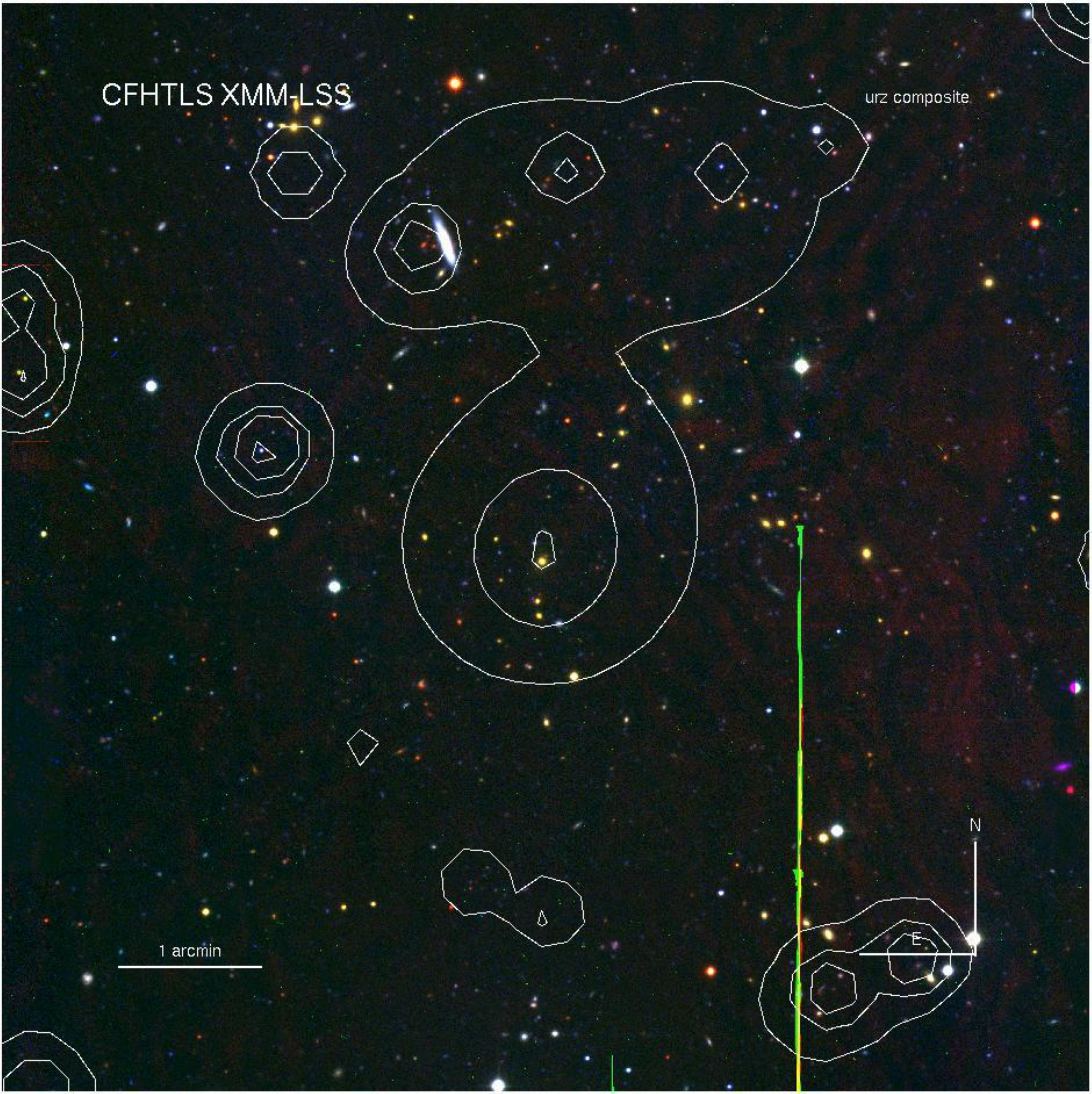} \\			
	\end{tabular}
	\contcaption{Images of the C1 clusters.
	\emph{Top:} XLSSC~057 ($z=0.15$) \emph{Bottom:} XLSSC~058 ($z=0.33$).}
\end{figure*}

\begin{figure*}
	\begin{tabular}{ m{9.5cm} m{9.5cm} }
		\includegraphics[height=90mm]{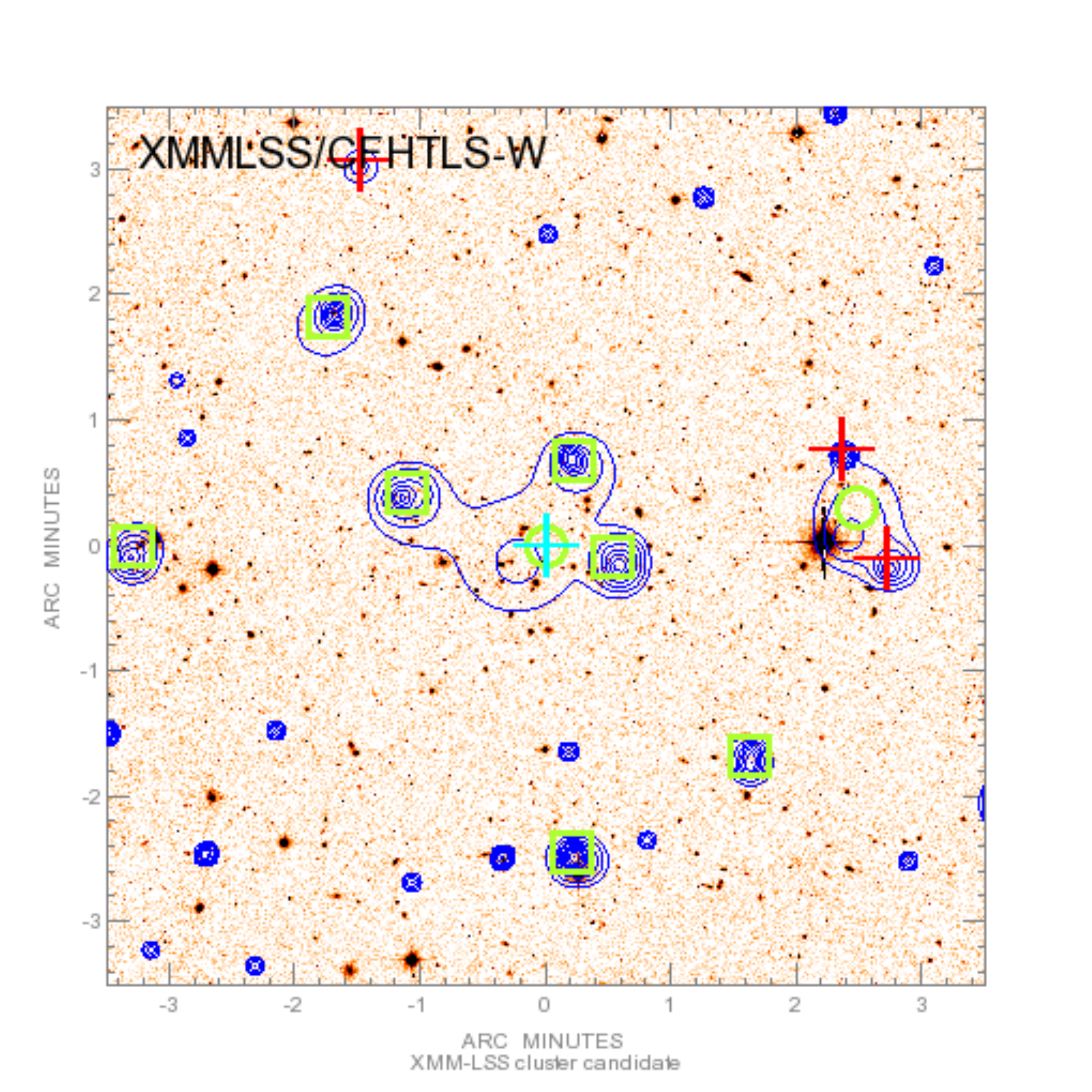} &
		\includegraphics[width=74mm]{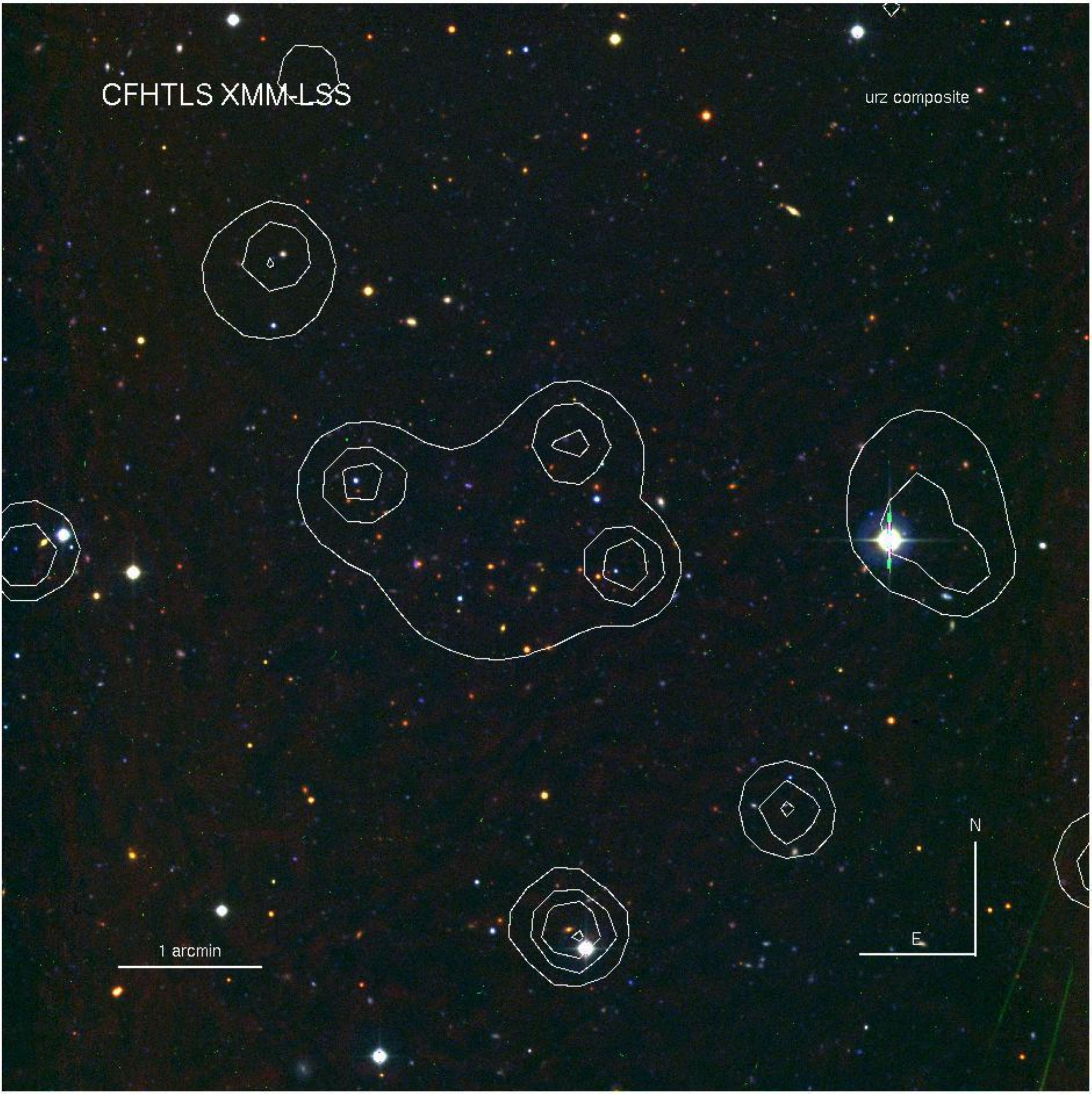} \\	
		\includegraphics[height=90mm]{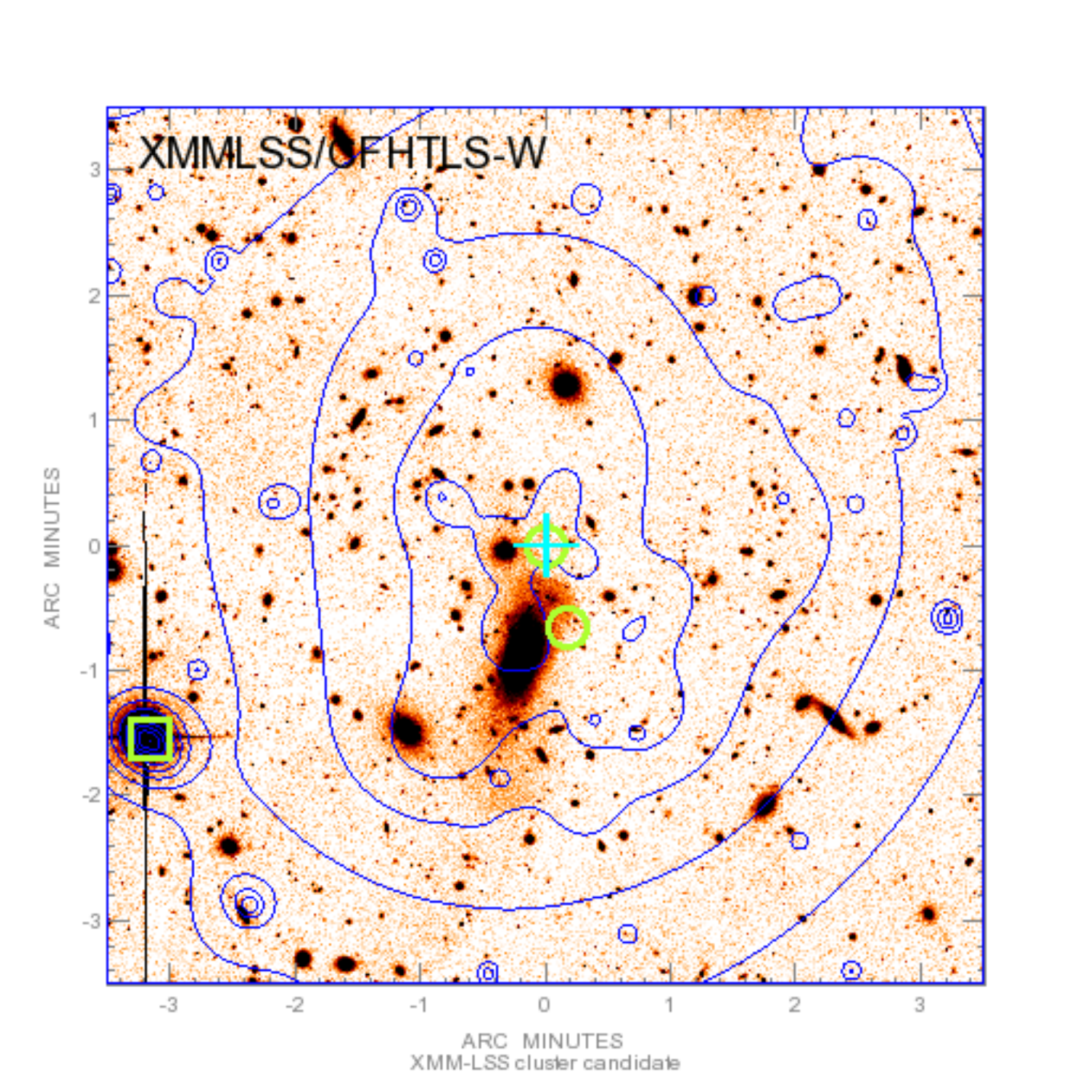} &
		\includegraphics[width=74mm]{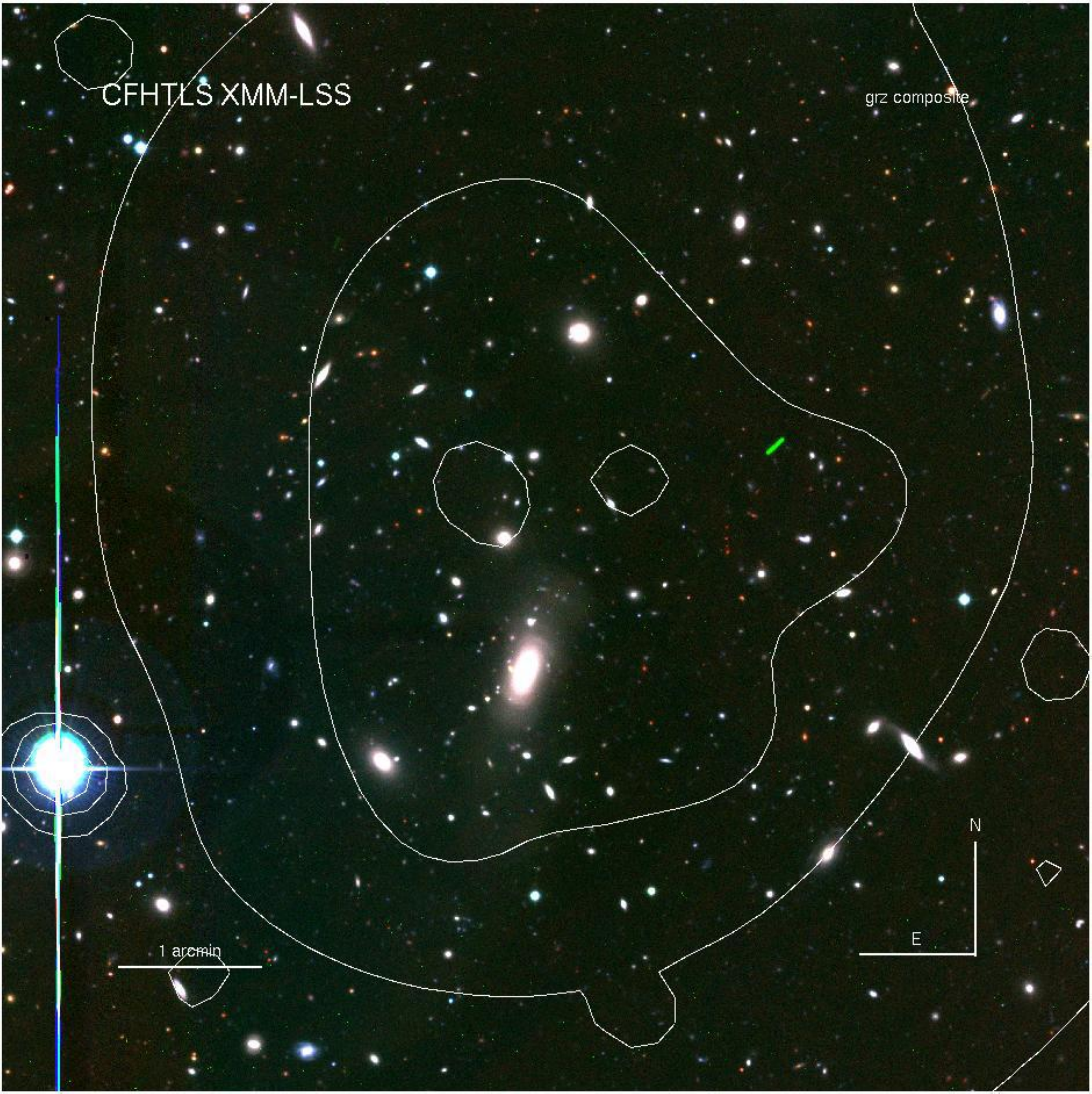} \\
	\end{tabular}
	\contcaption{Images of the C1 clusters.
	\emph{Top:} XLSSC~059 ($z=0.65$) \emph{Bottom:} XLSSC~060 ($z=0.14$).}
\end{figure*}

\begin{figure*}
	\begin{tabular}{ m{9.5cm} m{9.5cm} }
		\includegraphics[height=90mm]{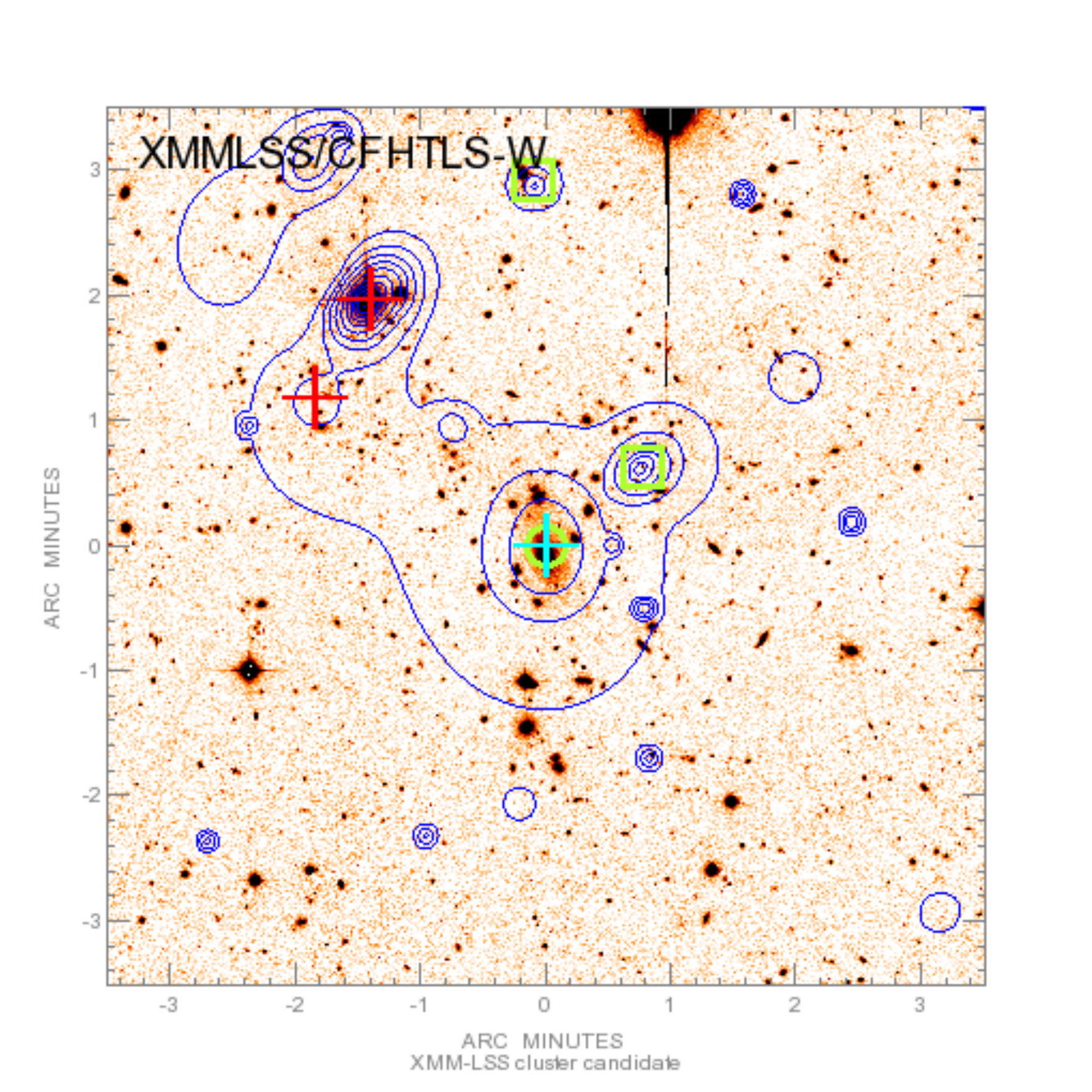} &
		\includegraphics[width=74mm]{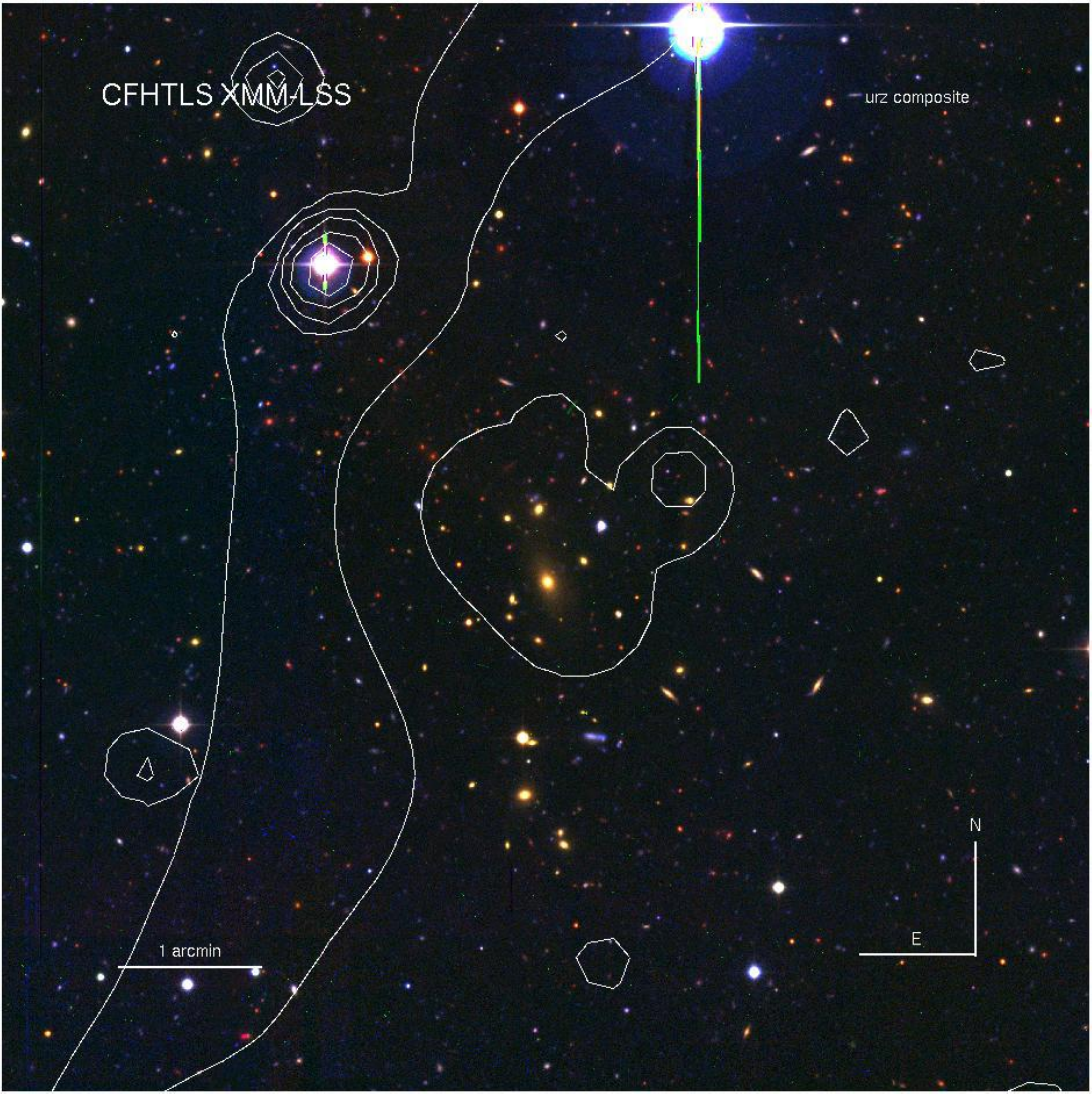} \\
		\includegraphics[height=90mm]{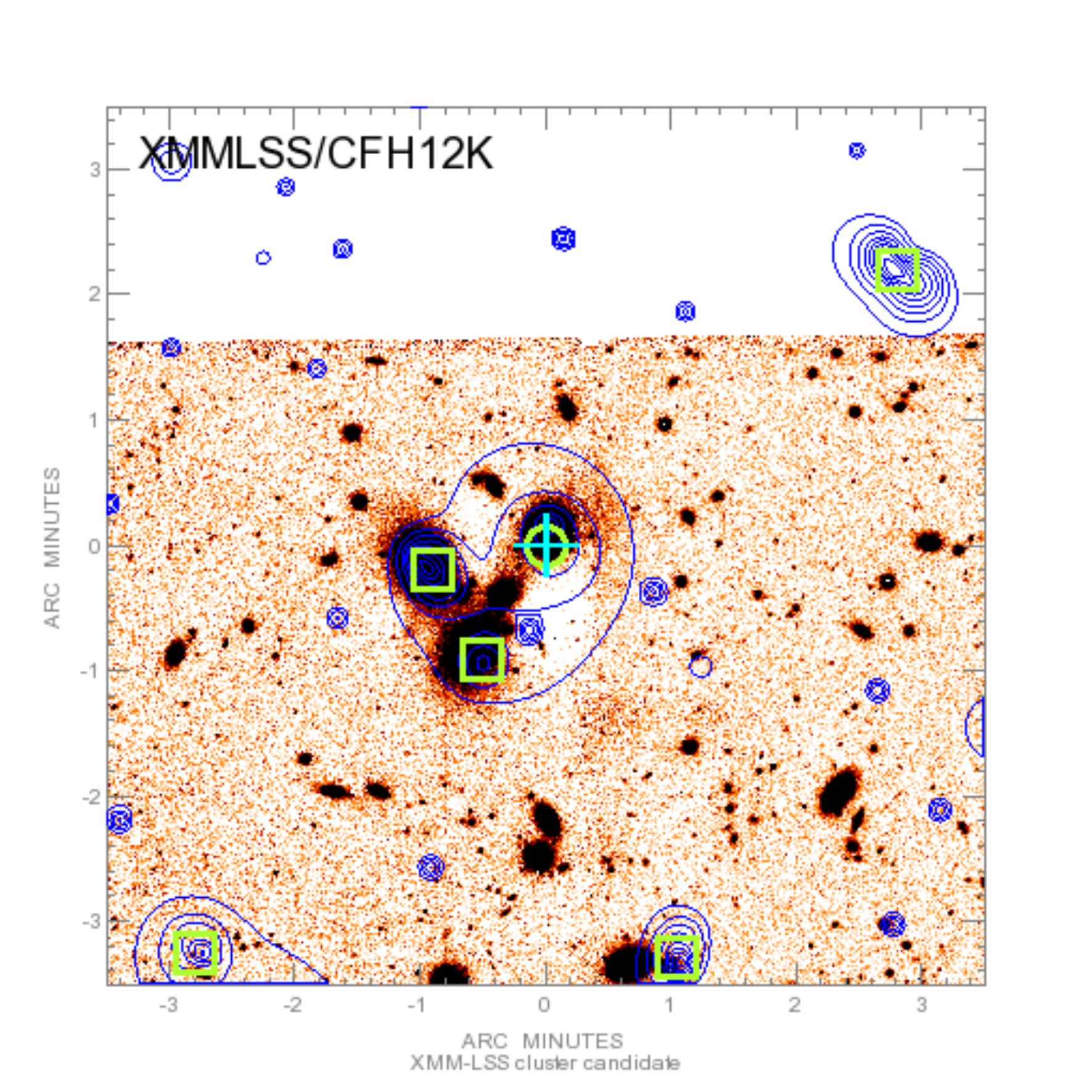} &
			\\	
	\end{tabular}
	\contcaption{Images of the C1 clusters.
	\emph{Top:} XLSSC~061 ($z=0.26$) \emph{Bottom:} XLSSC~062 ($z=0.06$, located outside of the CFHT-LS W1 footprint).}
\end{figure*}

\begin{figure*}
	\begin{tabular}{ m{9.5cm} m{9.5cm} }
		\includegraphics[height=90mm]{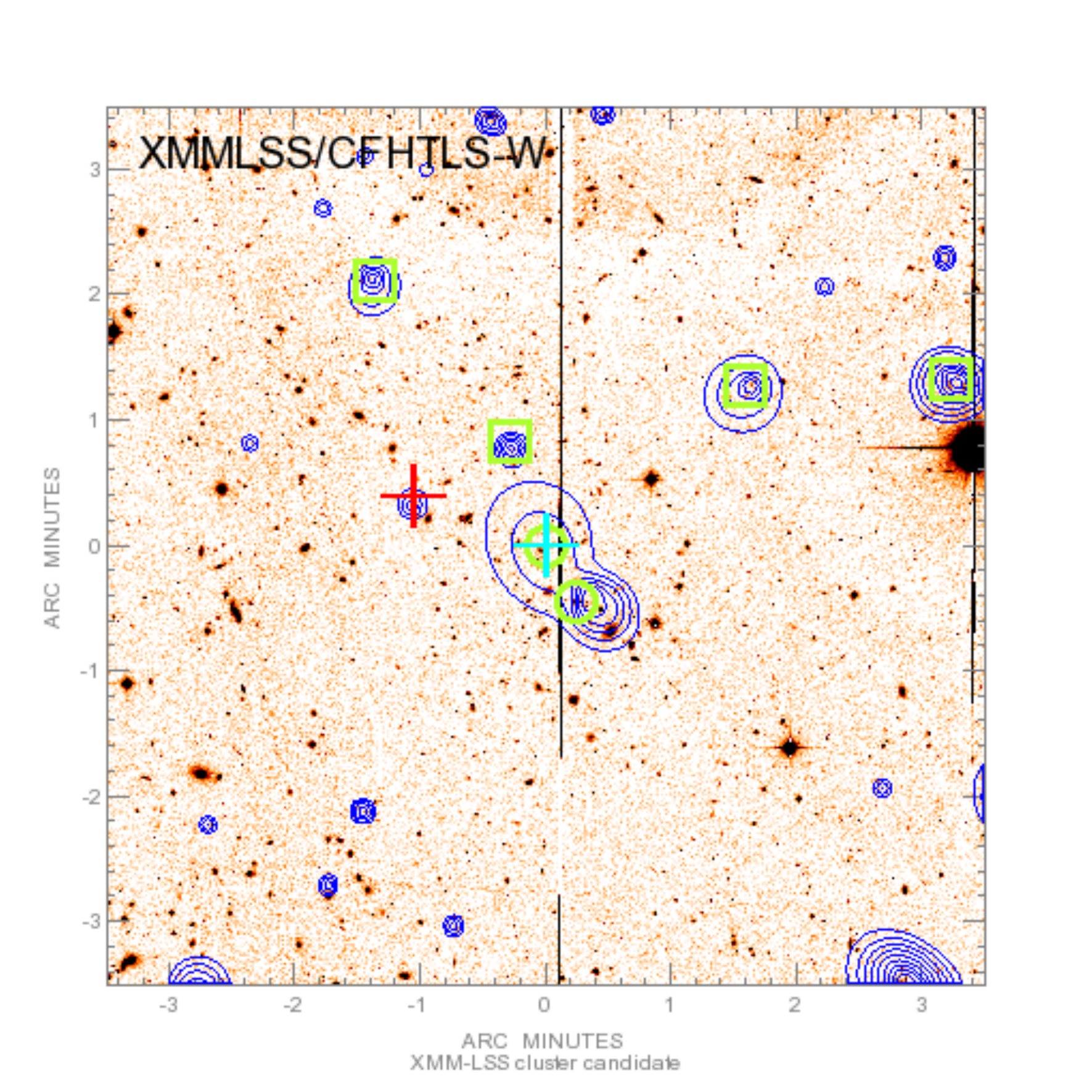} &
		\includegraphics[width=74mm]{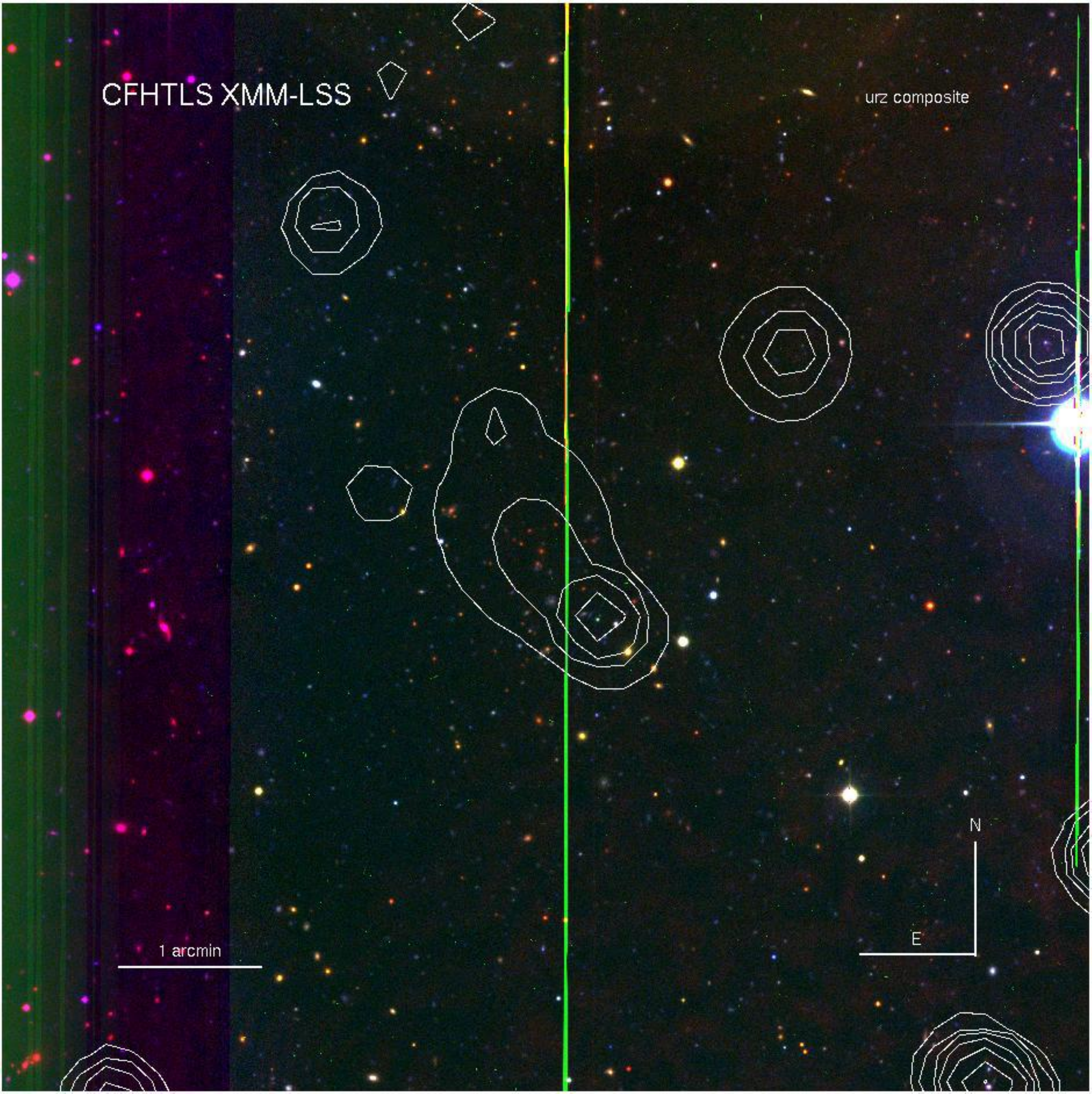} \\
		\includegraphics[height=90mm]{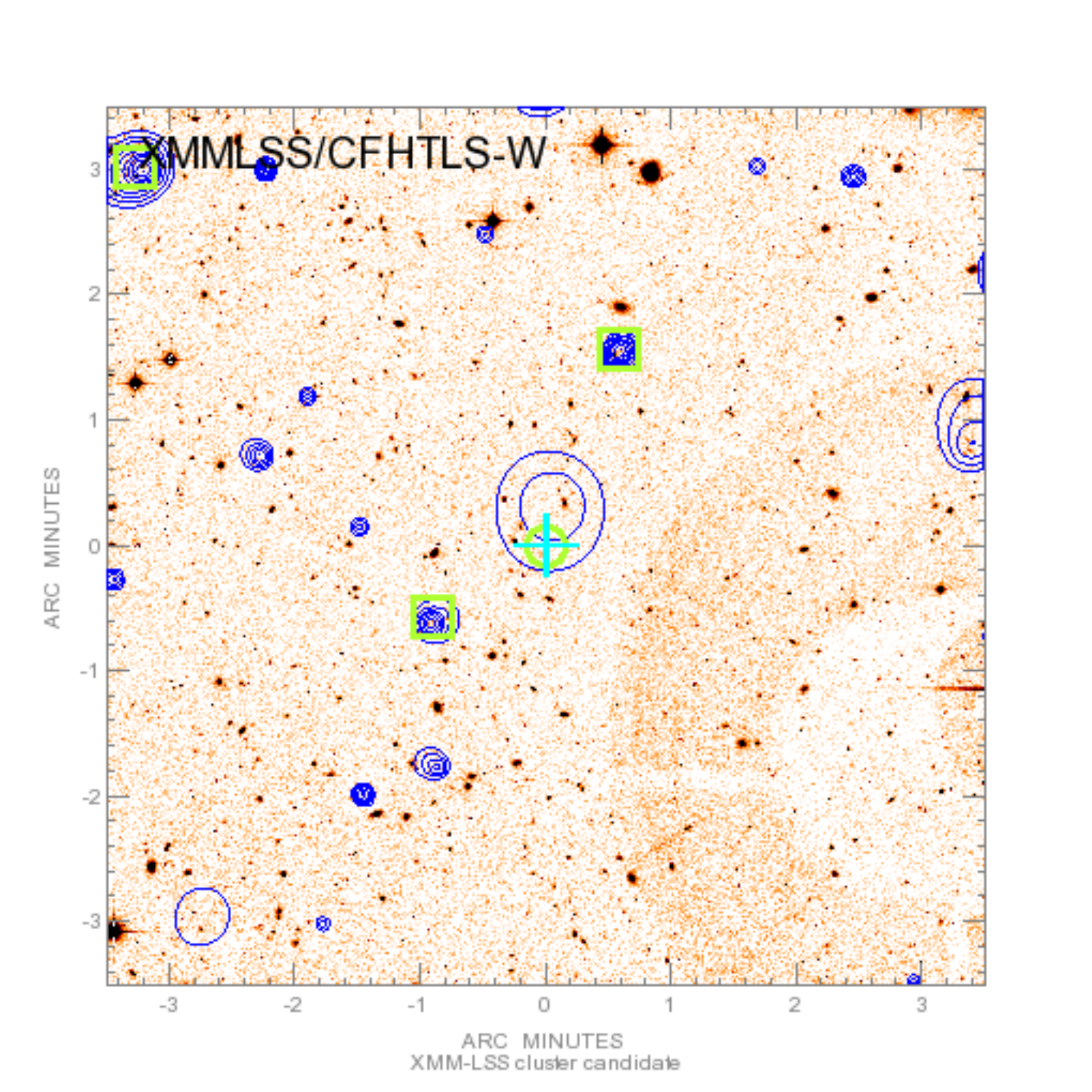} &
		\includegraphics[width=74mm]{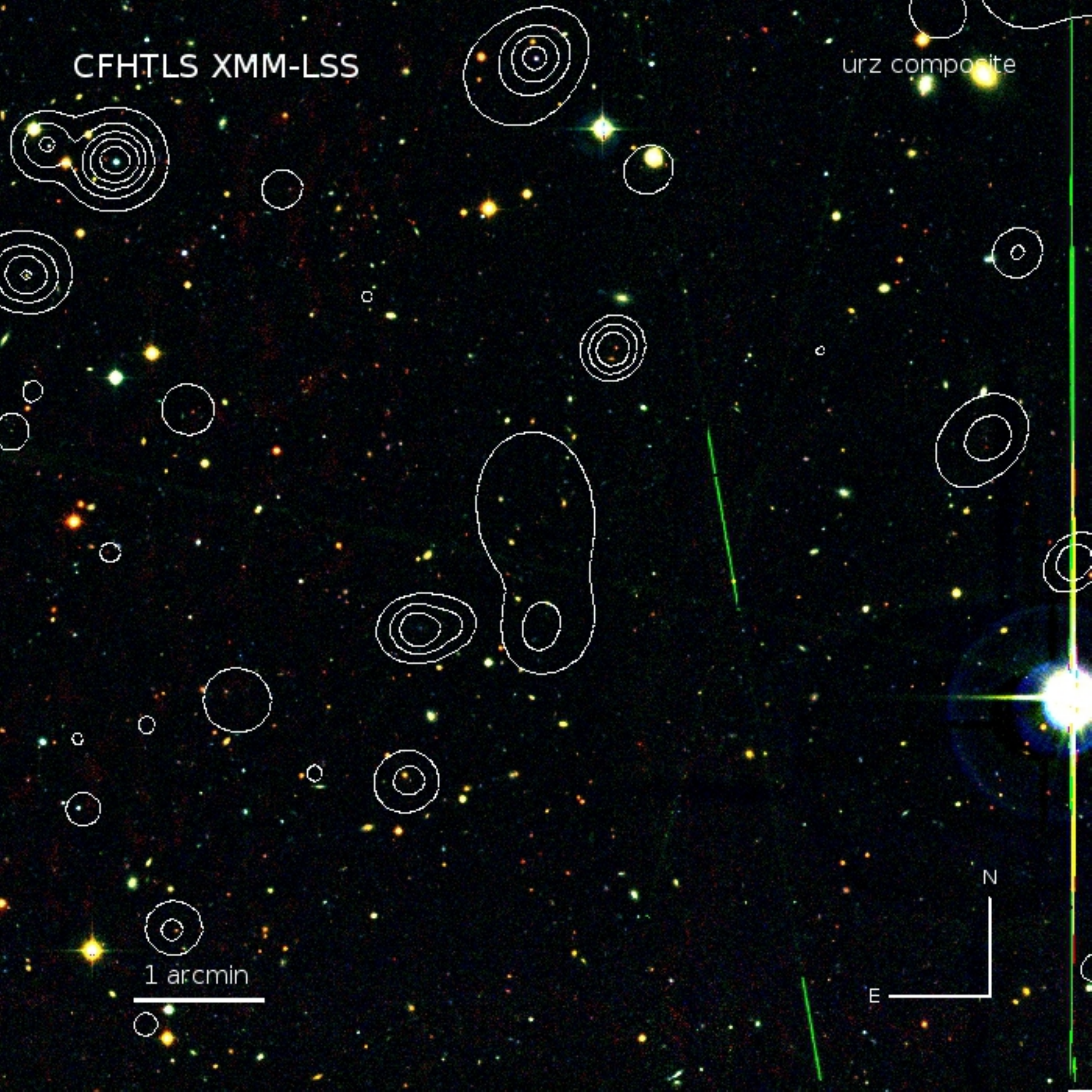} \\	
	\end{tabular}
	\contcaption{Images of the C1 clusters.
	\emph{Top:} XLSSC~064 ($z=0.88$) \emph{Bottom:} XLSSC~065 ($z=0.43$).}
\end{figure*}

\begin{figure*}
	\begin{tabular}{ m{9.5cm} m{9.5cm} }
		\includegraphics[height=90mm]{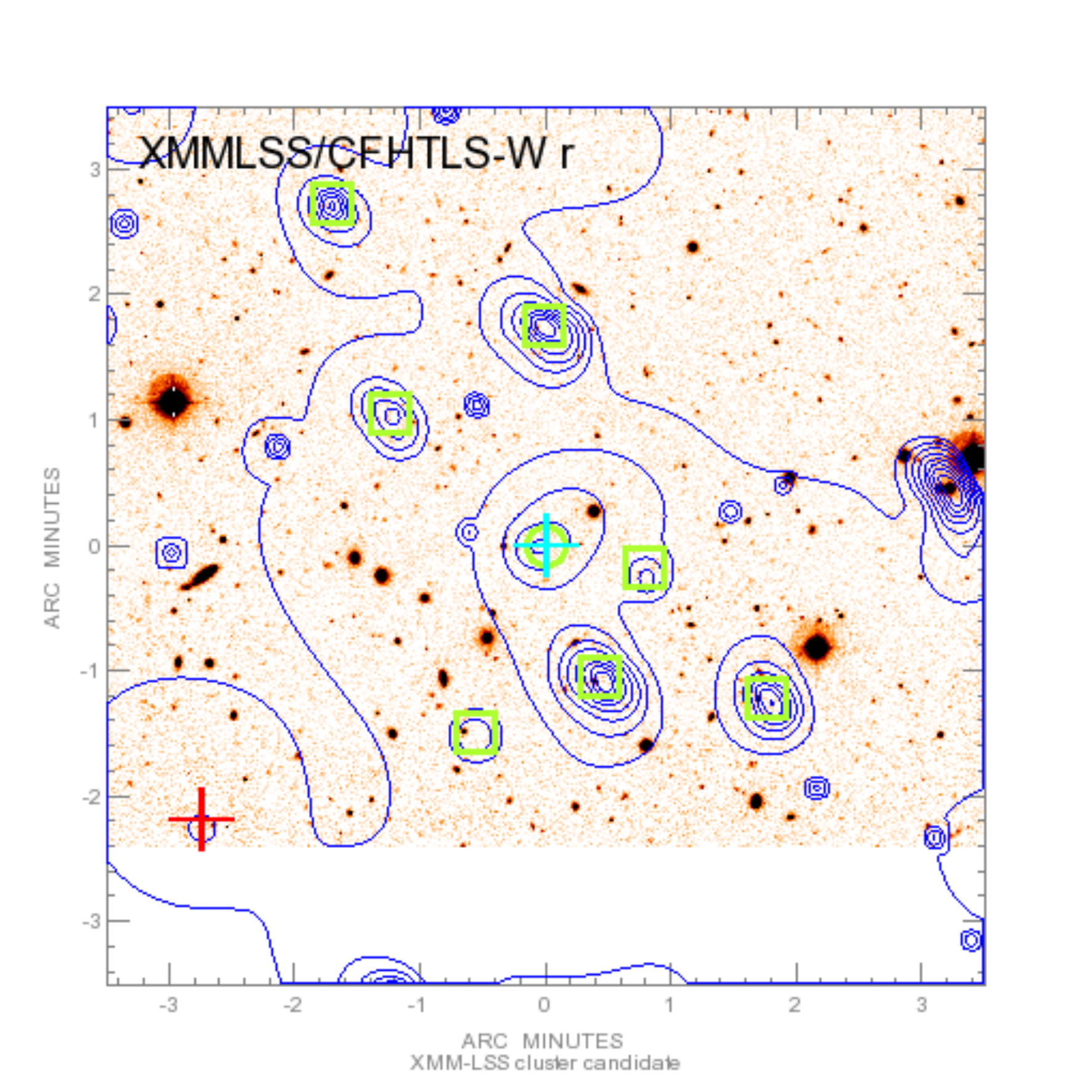} &
		\includegraphics[width=74mm]{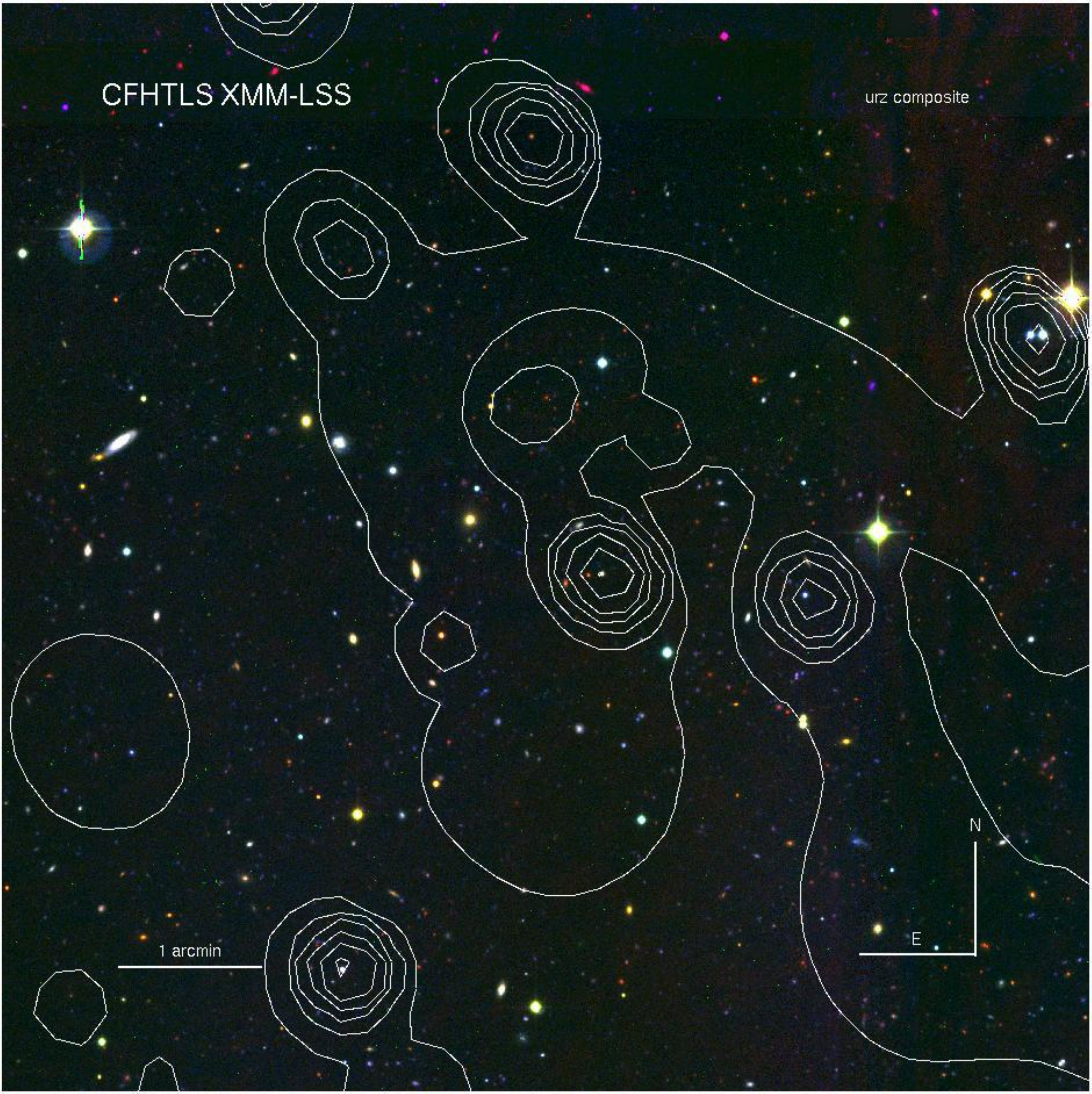} \\
		\includegraphics[height=90mm]{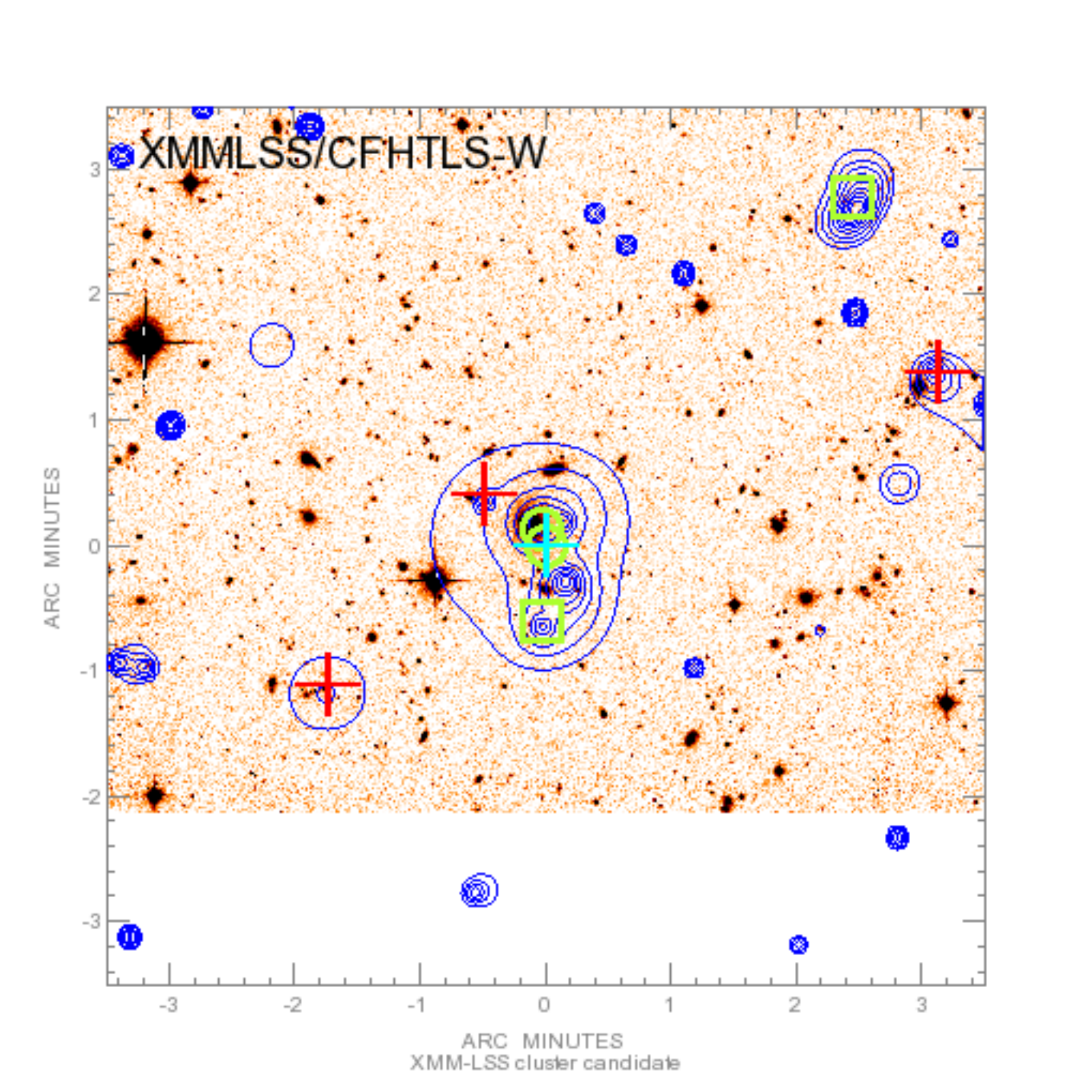} &
		\includegraphics[width=74mm]{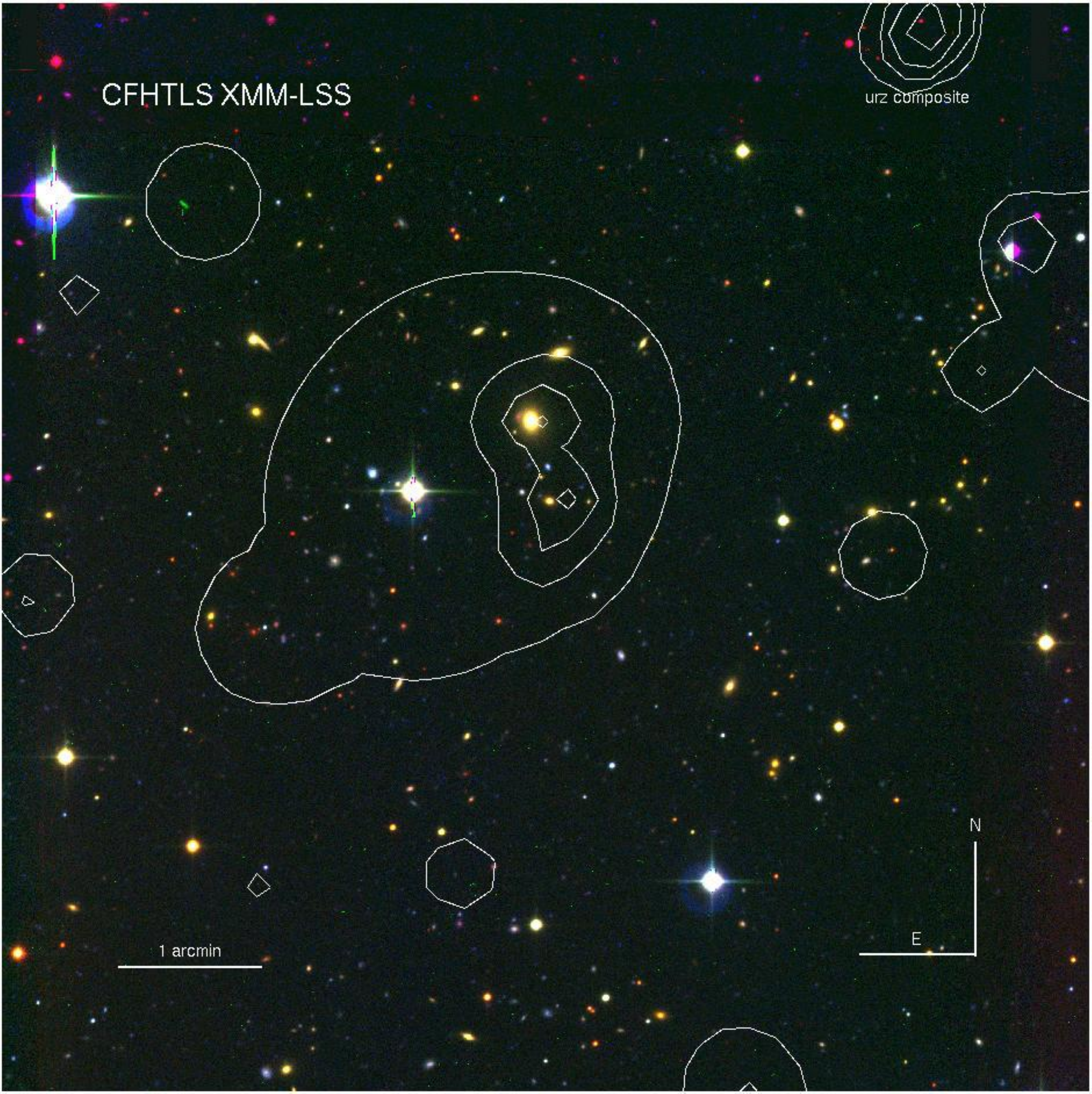} \\		
	\end{tabular}
	\contcaption{Images of the C1 clusters.
	\emph{Top:} XLSSC~072 ($z=1.00$) \emph{Bottom:} XLSSC~074 ($z=0.19$).}
\end{figure*}

\begin{figure*}
	\begin{tabular}{ m{9.5cm} m{9.5cm} }
		\includegraphics[height=90mm]{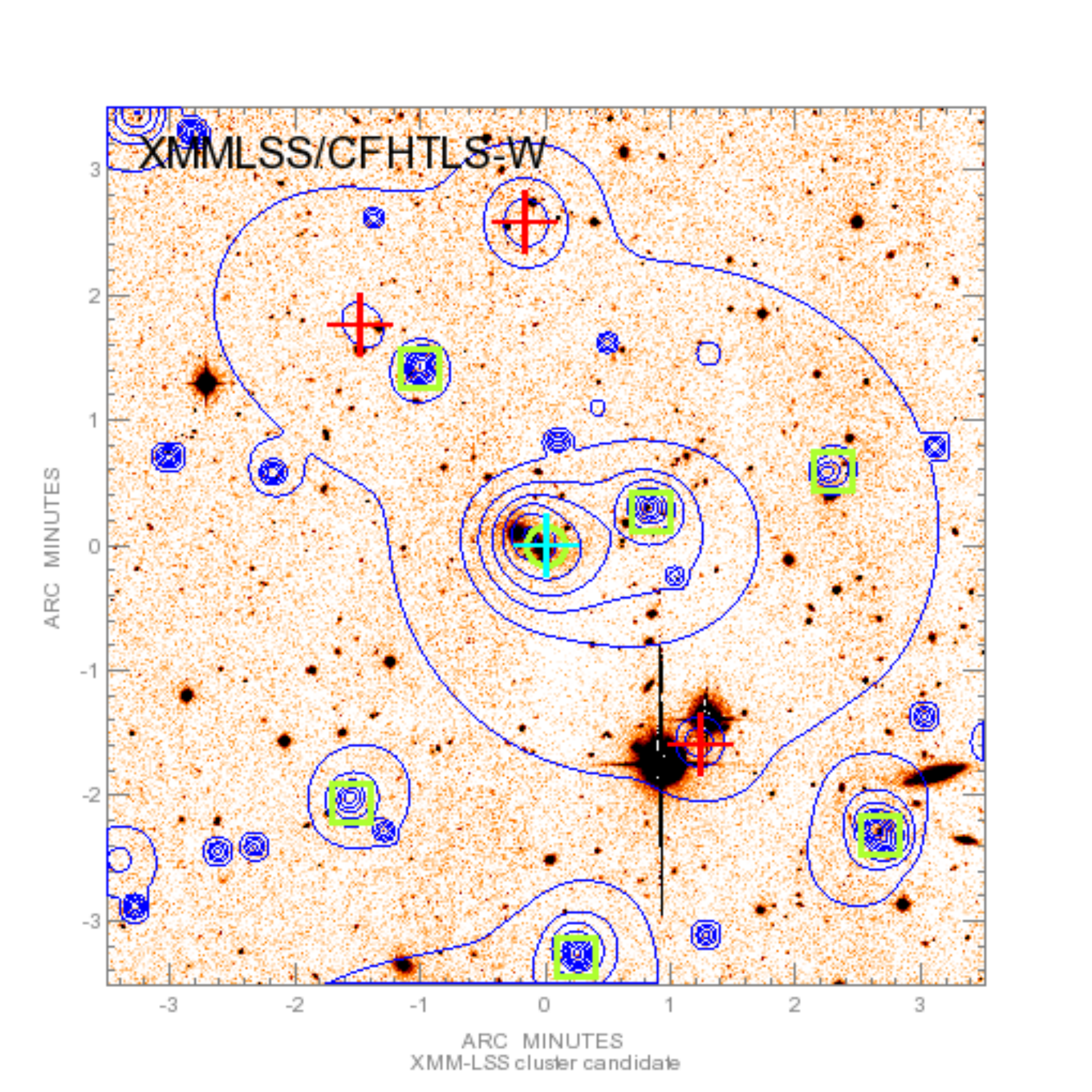} &
		\includegraphics[width=74mm]{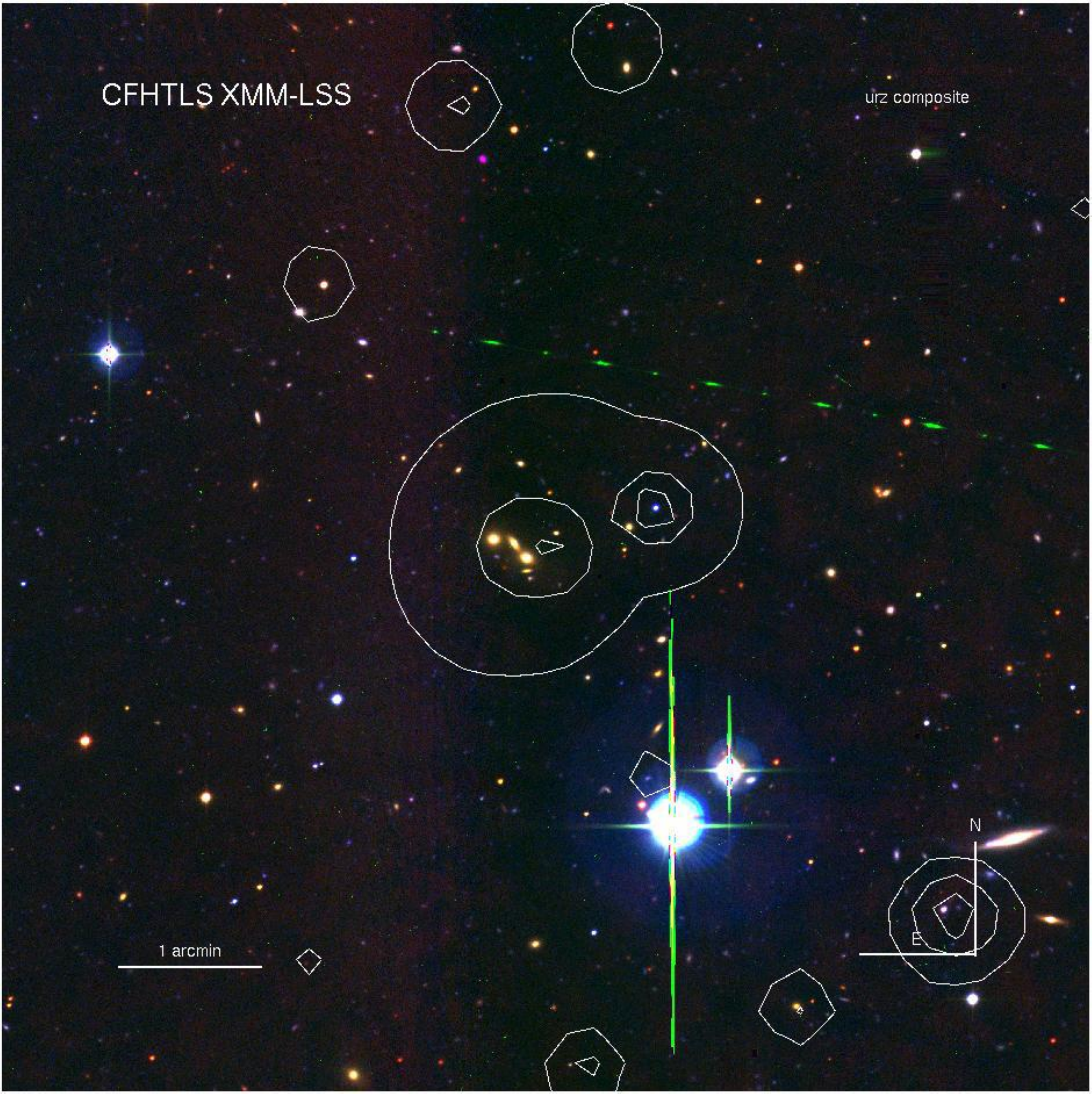} \\
		\includegraphics[height=90mm]{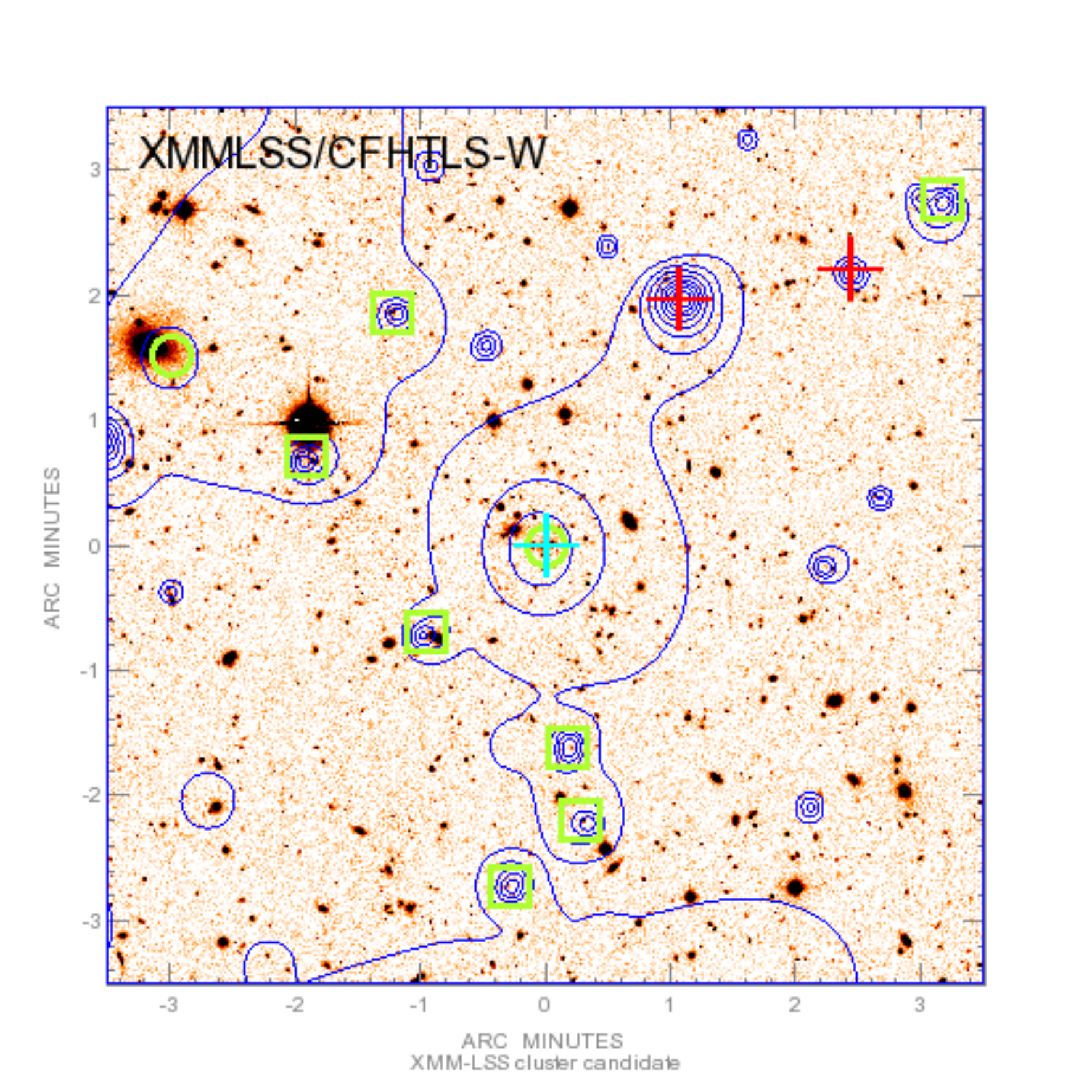} &
		\includegraphics[width=74mm]{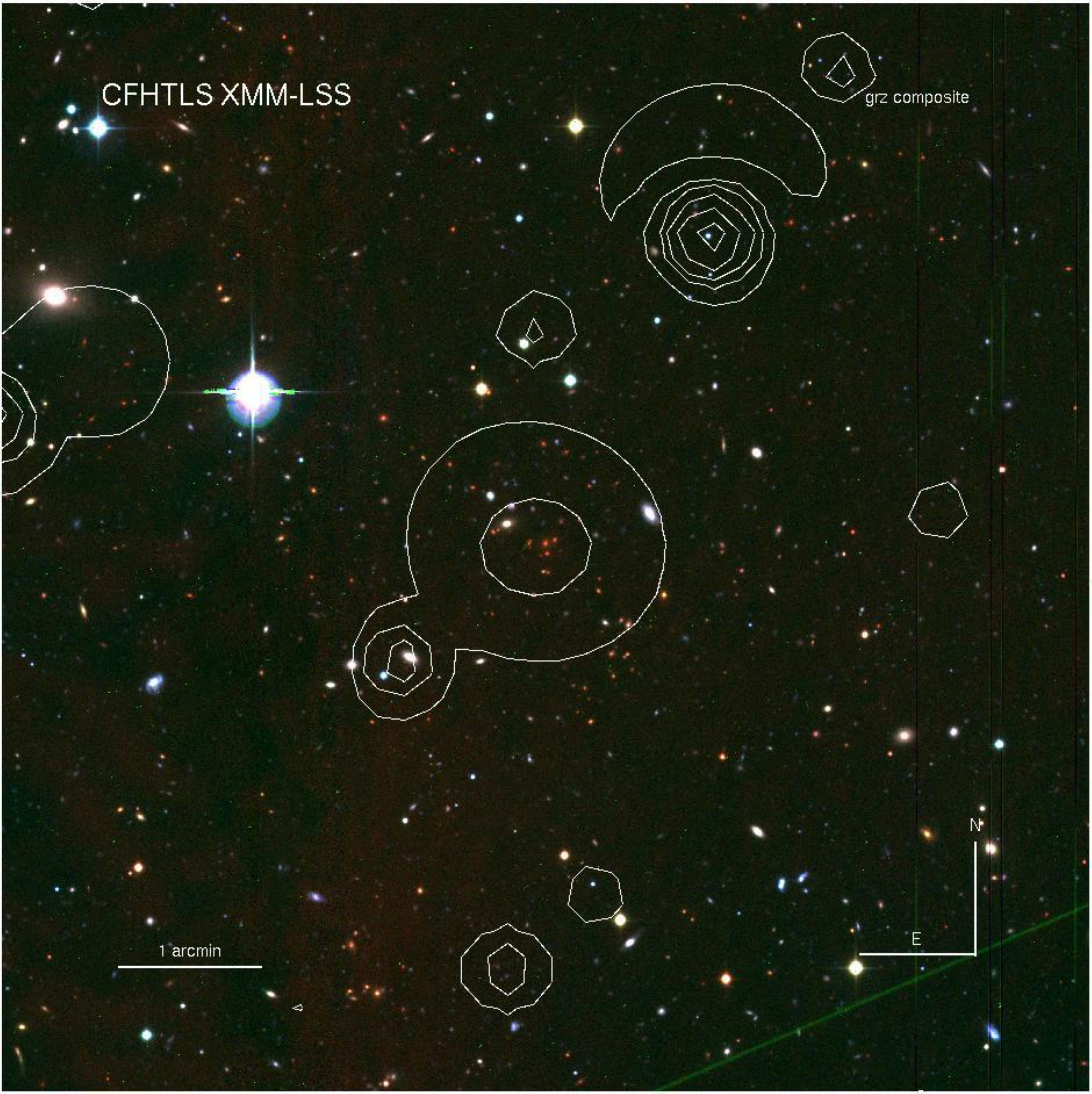} \\		
	\end{tabular}
	\contcaption{Images of the C1 clusters.
	\emph{Top:} XLSSC~075 ($z=0.21$) \emph{Bottom:} XLSSC~076 ($z=0.75$).}
\end{figure*}

\begin{figure*}
	\begin{tabular}{ m{9.5cm} m{9.5cm} }
		\includegraphics[height=90mm]{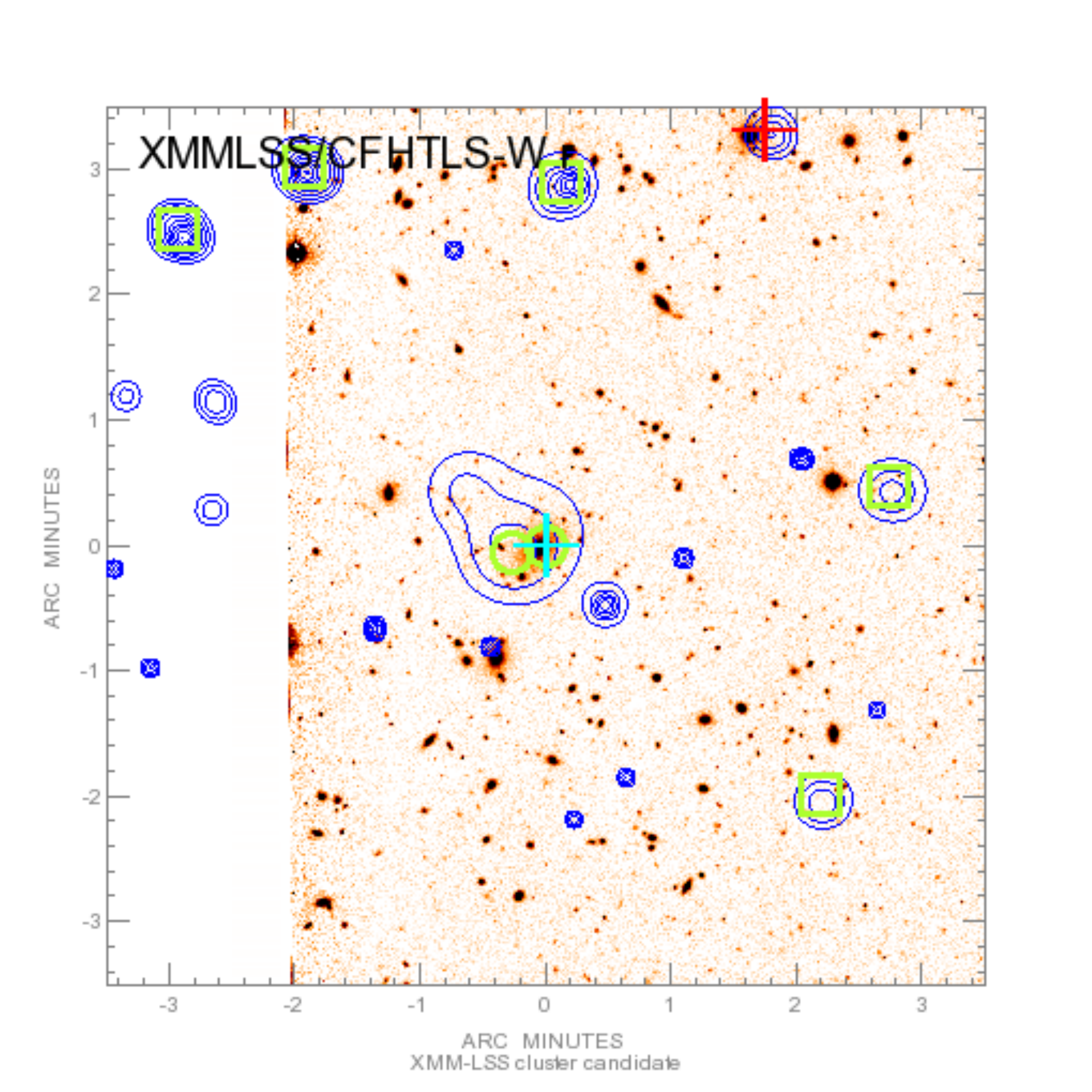} &
		\includegraphics[width=74mm]{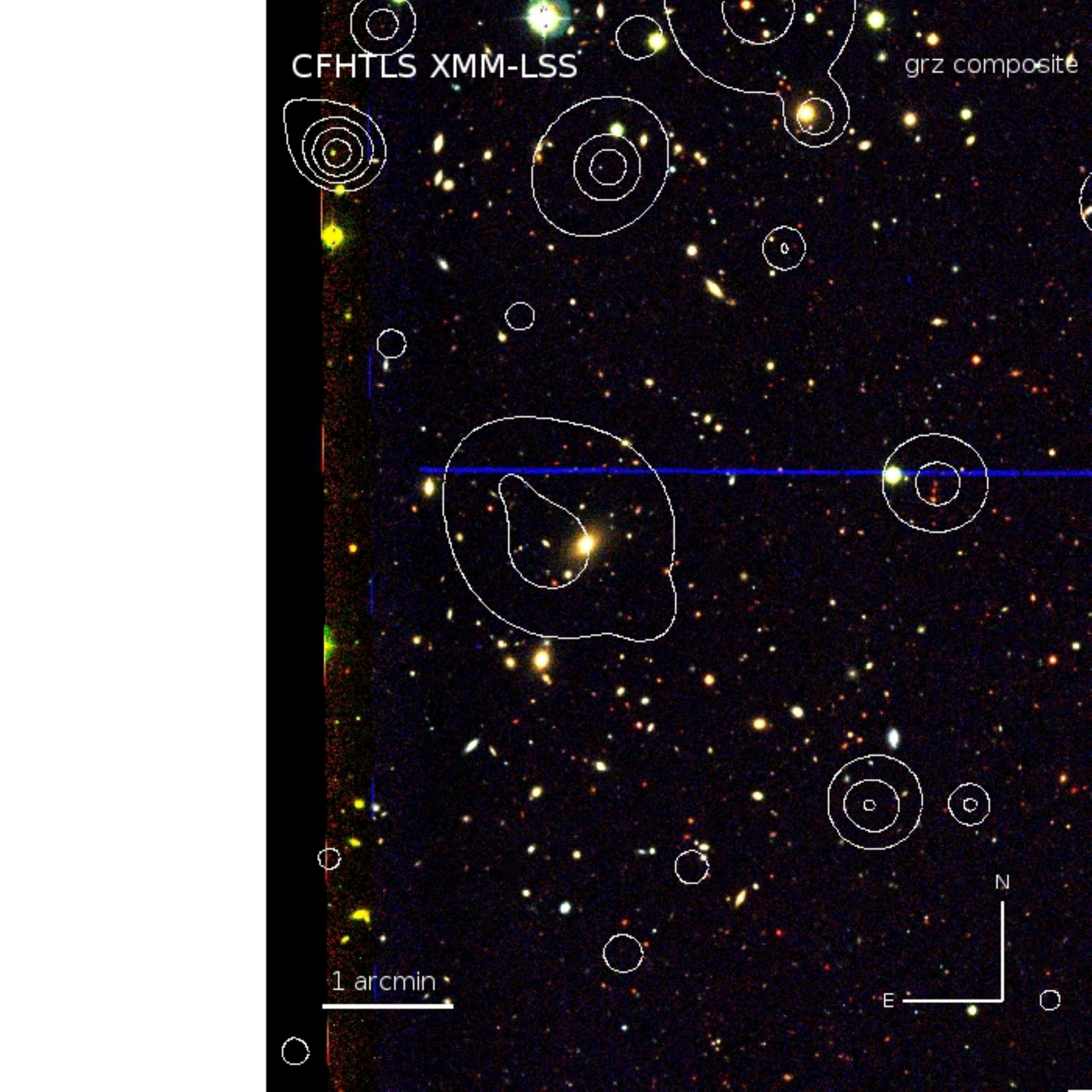} \\
		\includegraphics[height=90mm]{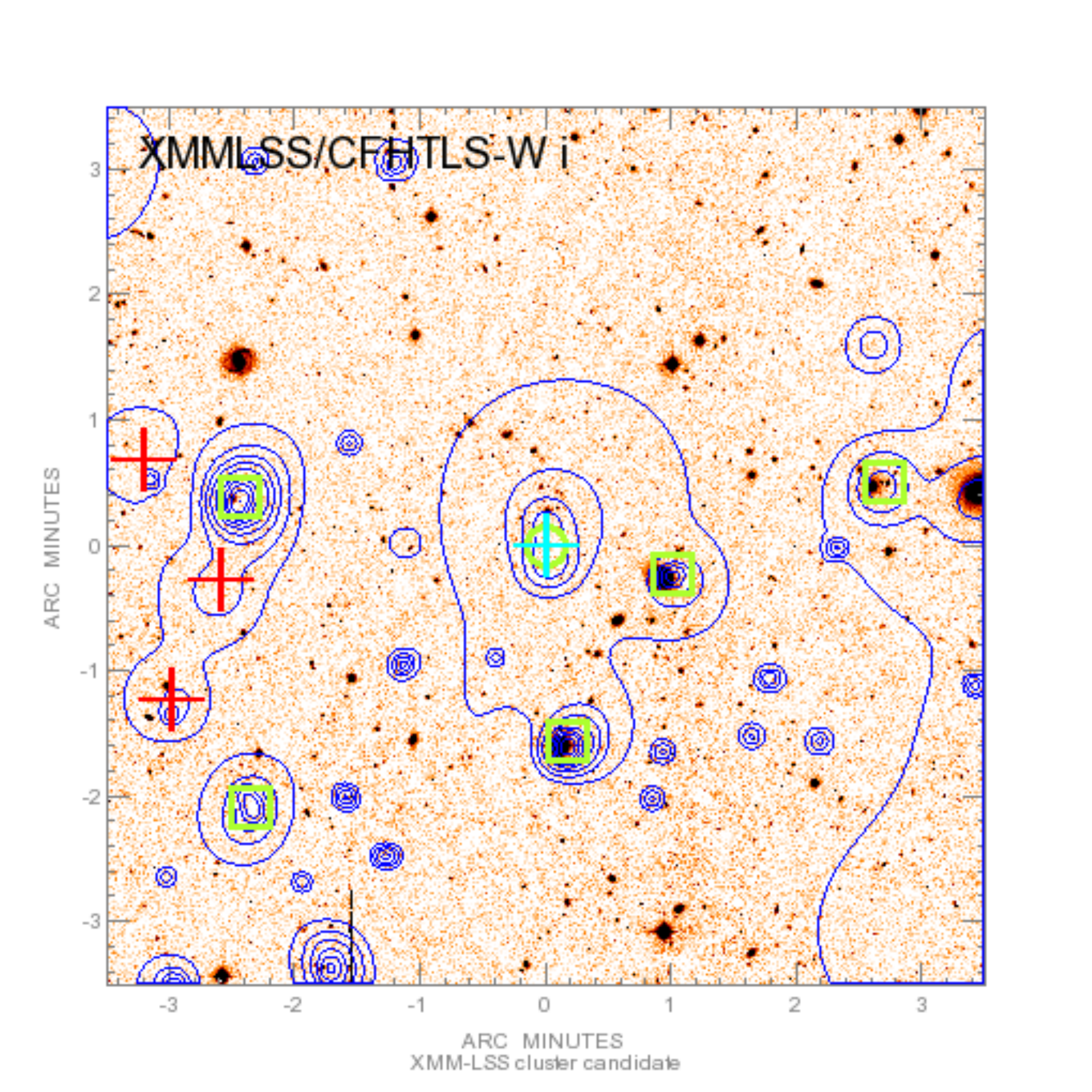} &
		\includegraphics[width=74mm]{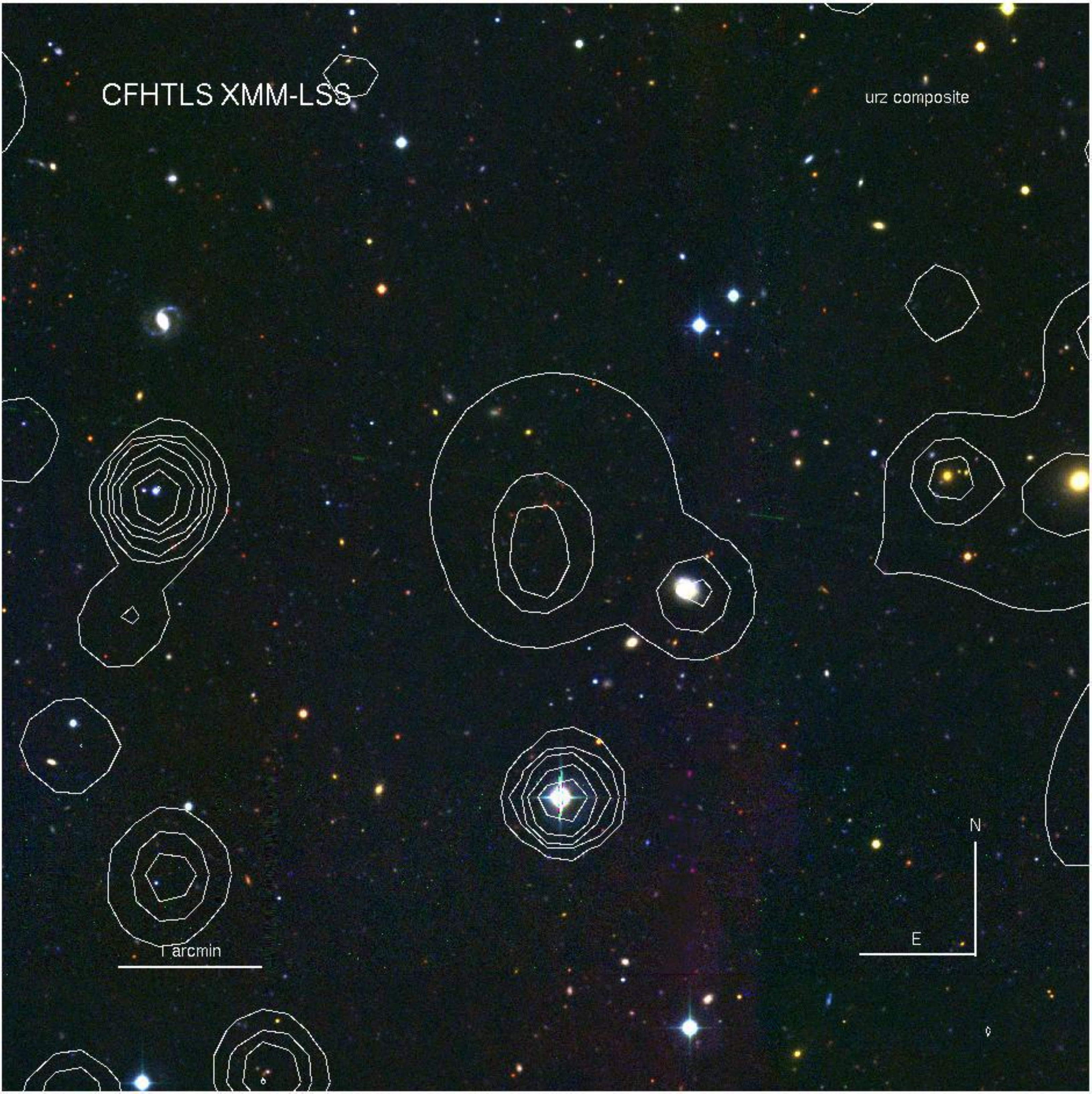} \\		
	\end{tabular}
	\contcaption{Images of the C1 clusters.
	\emph{Top:} XLSSC~077 ($z=0.20$) \emph{Bottom:} XLSSC~078 ($z=0.95$).}
\end{figure*}

\begin{figure*}
	\begin{tabular}{ m{9.5cm} m{9.5cm} }
		\includegraphics[height=90mm]{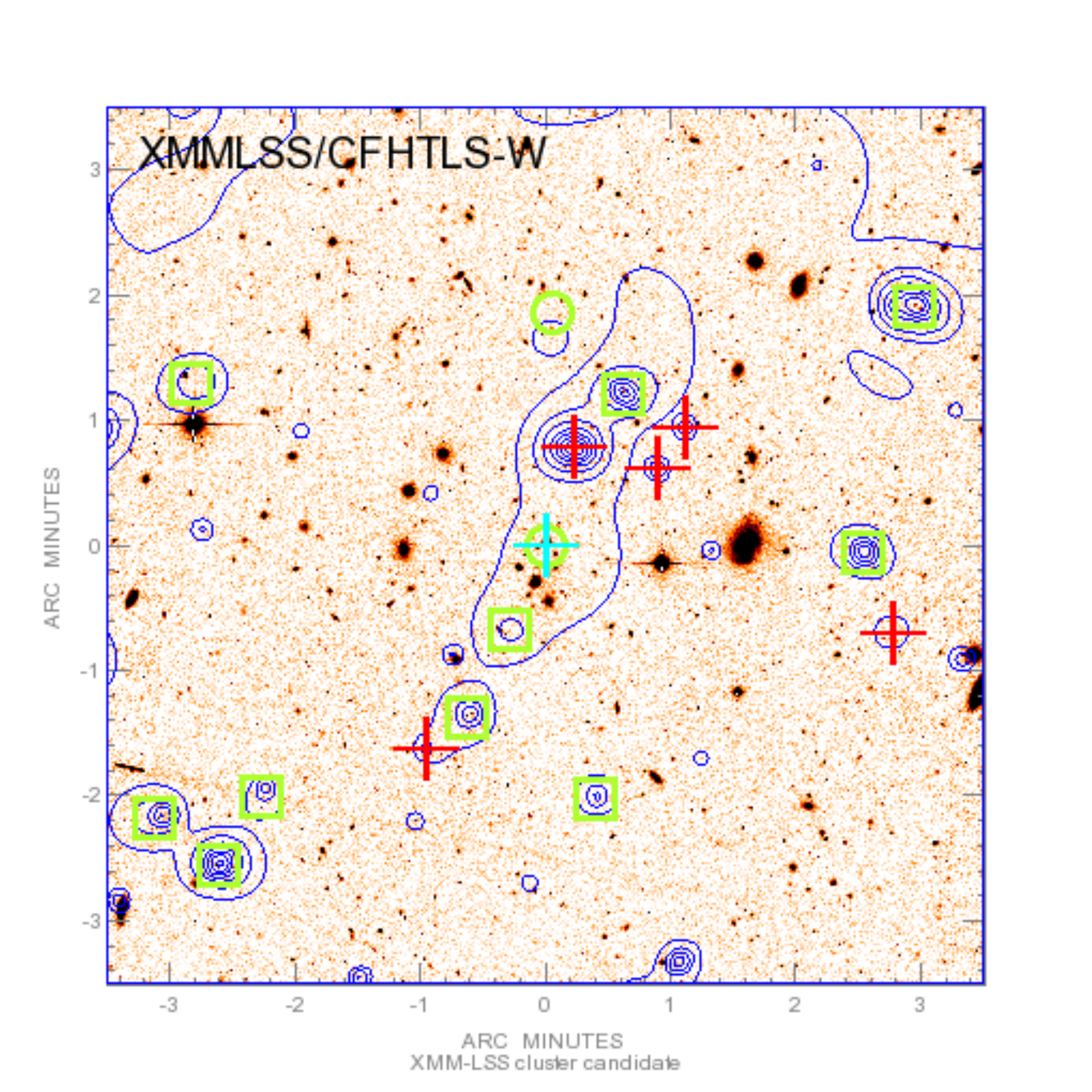} &
		\includegraphics[width=74mm]{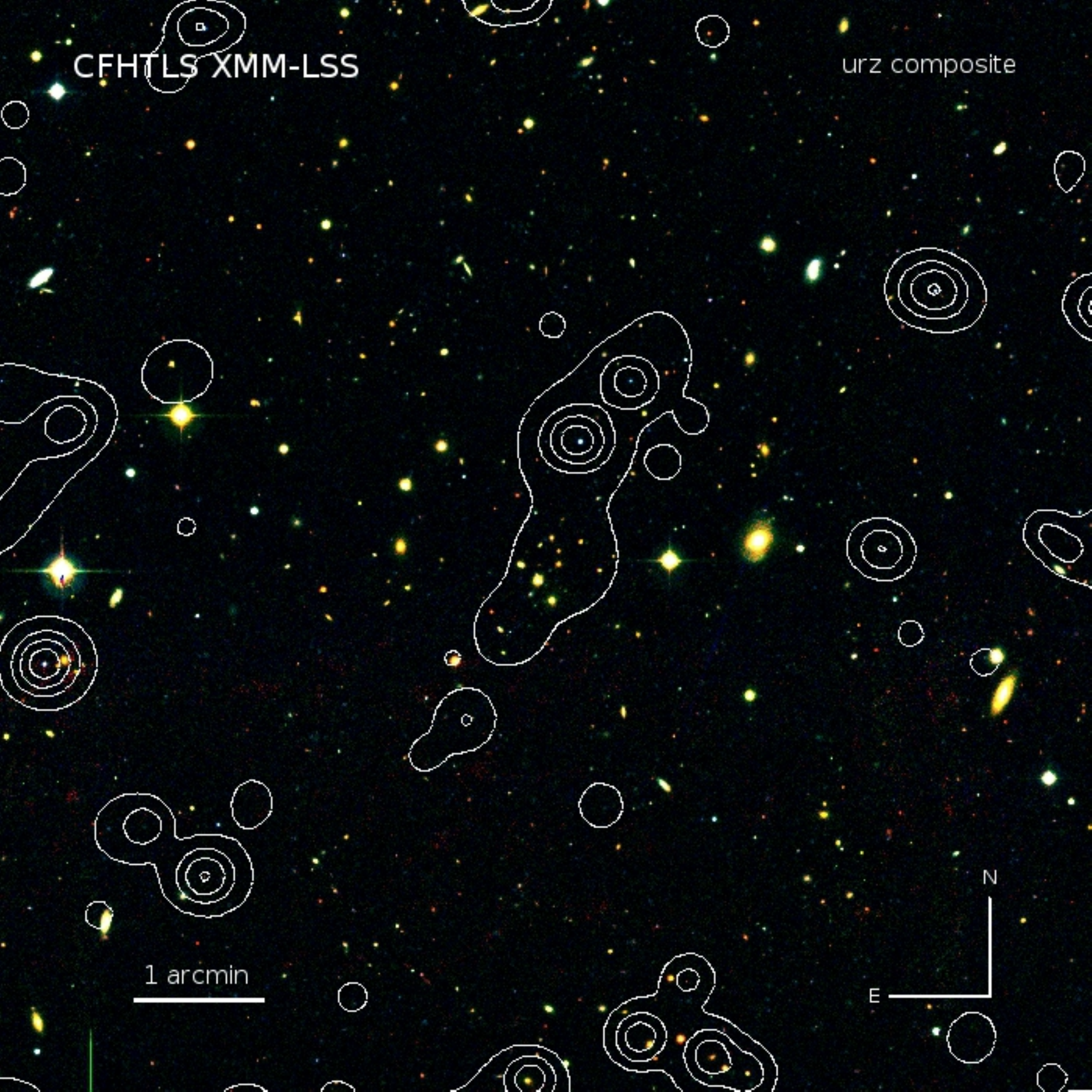} \\
		\includegraphics[height=90mm]{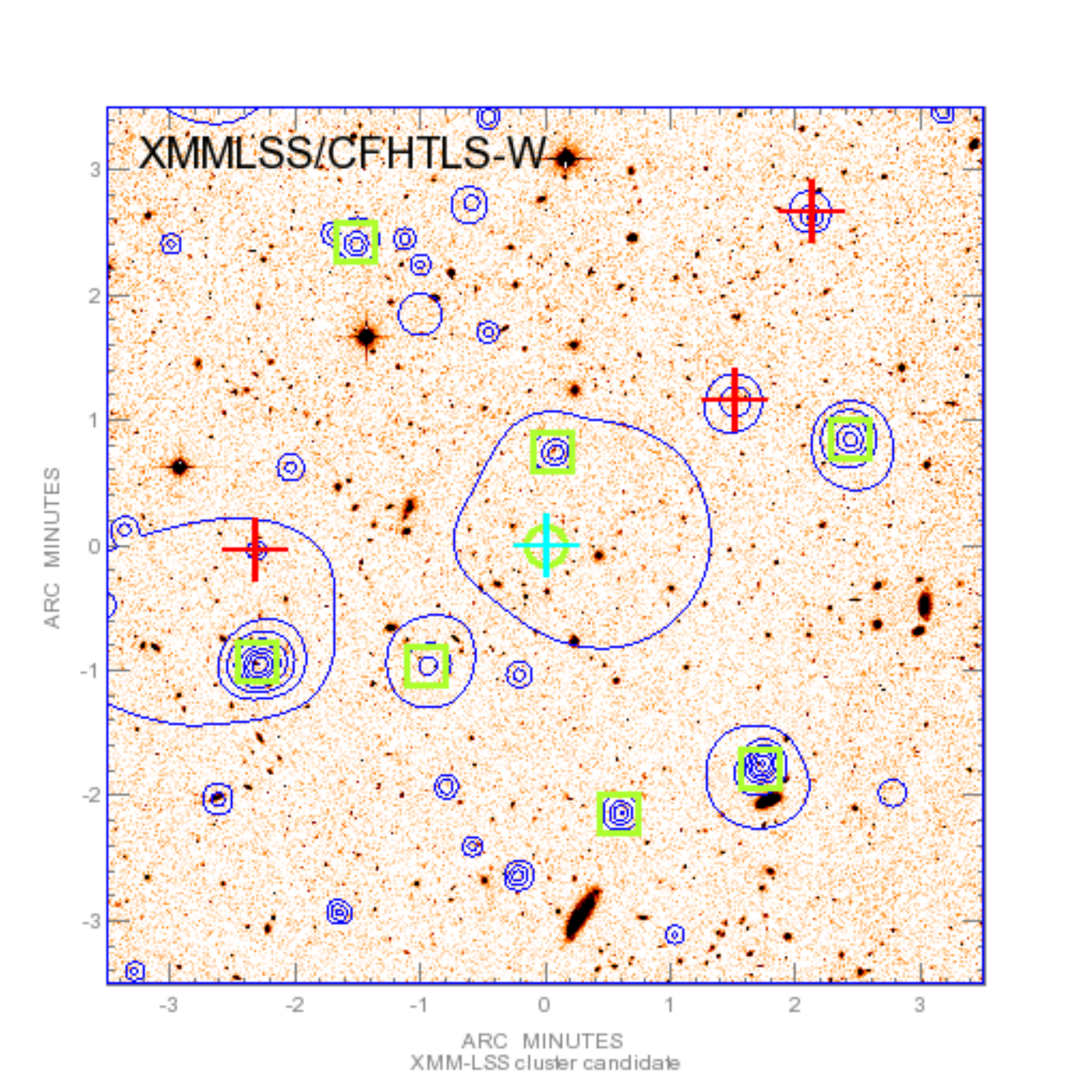} &
		\includegraphics[width=74mm]{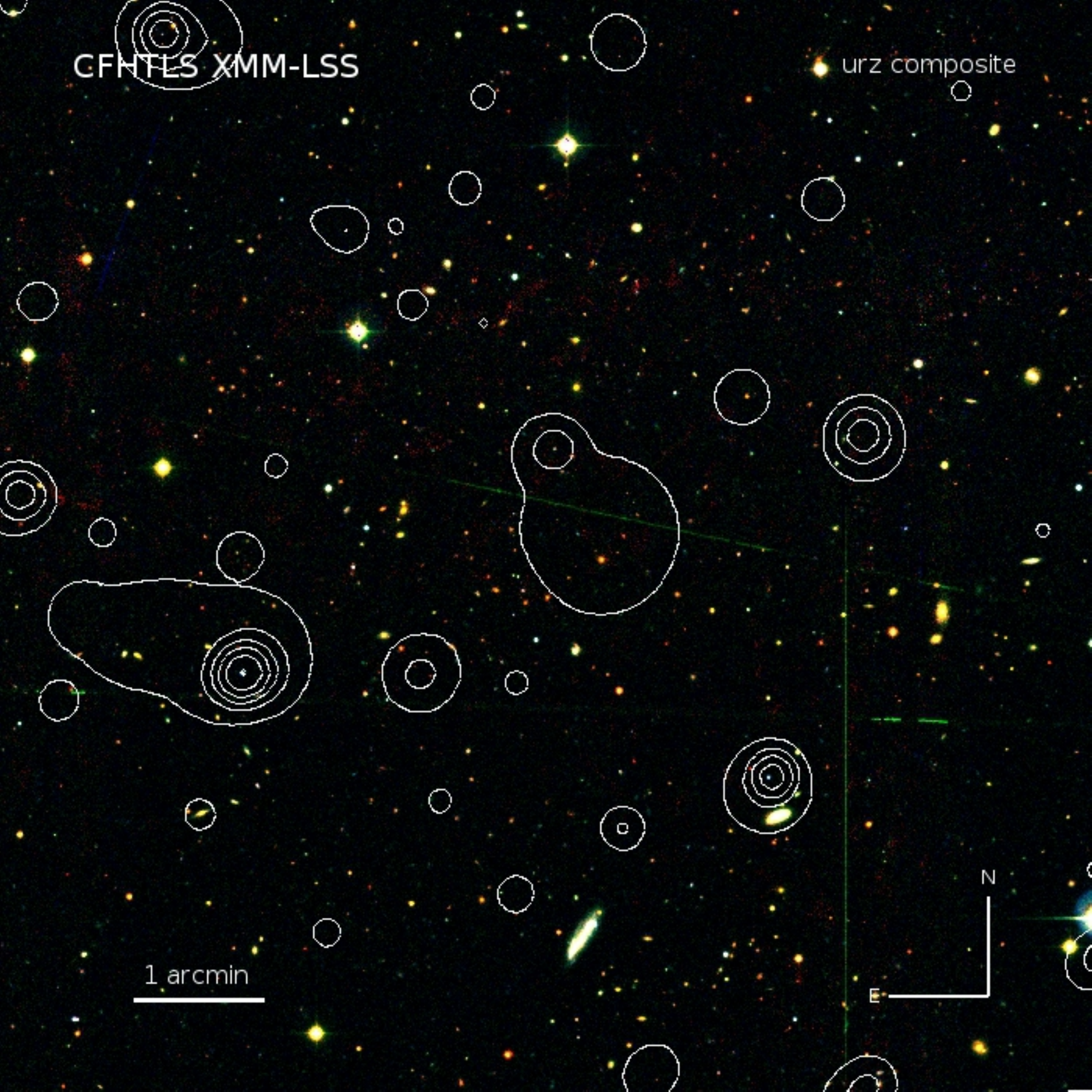} \\		
	\end{tabular}
	\contcaption{Images of the C1 clusters.
	\emph{Top:} XLSSC~079 ($z=0.19$) \emph{Bottom:} XLSSC~080 ($z=0.65$).}
\end{figure*}

\begin{figure*}
	\begin{tabular}{ m{9.5cm} m{9.5cm} }
		\includegraphics[height=90mm]{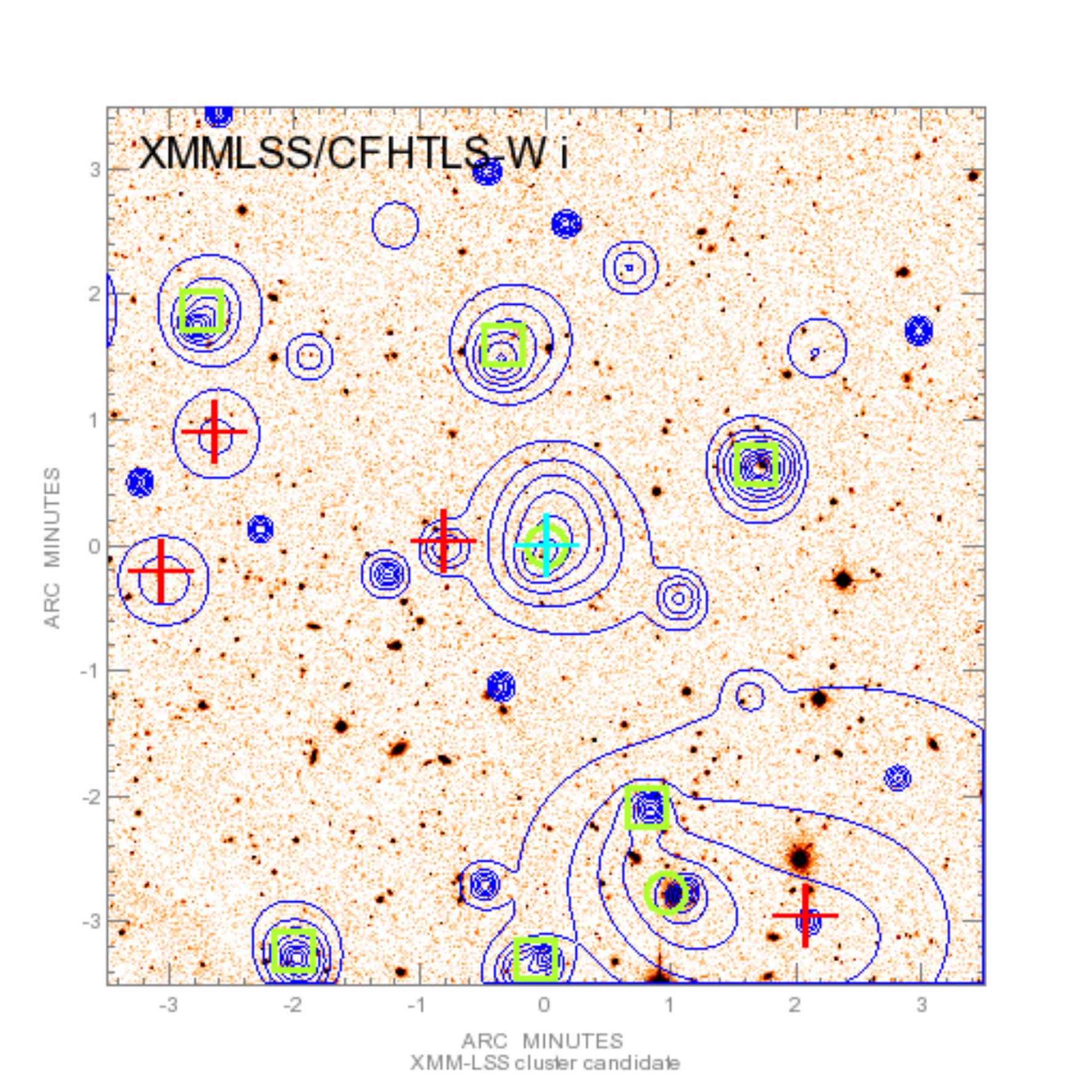} &
		\includegraphics[width=74mm]{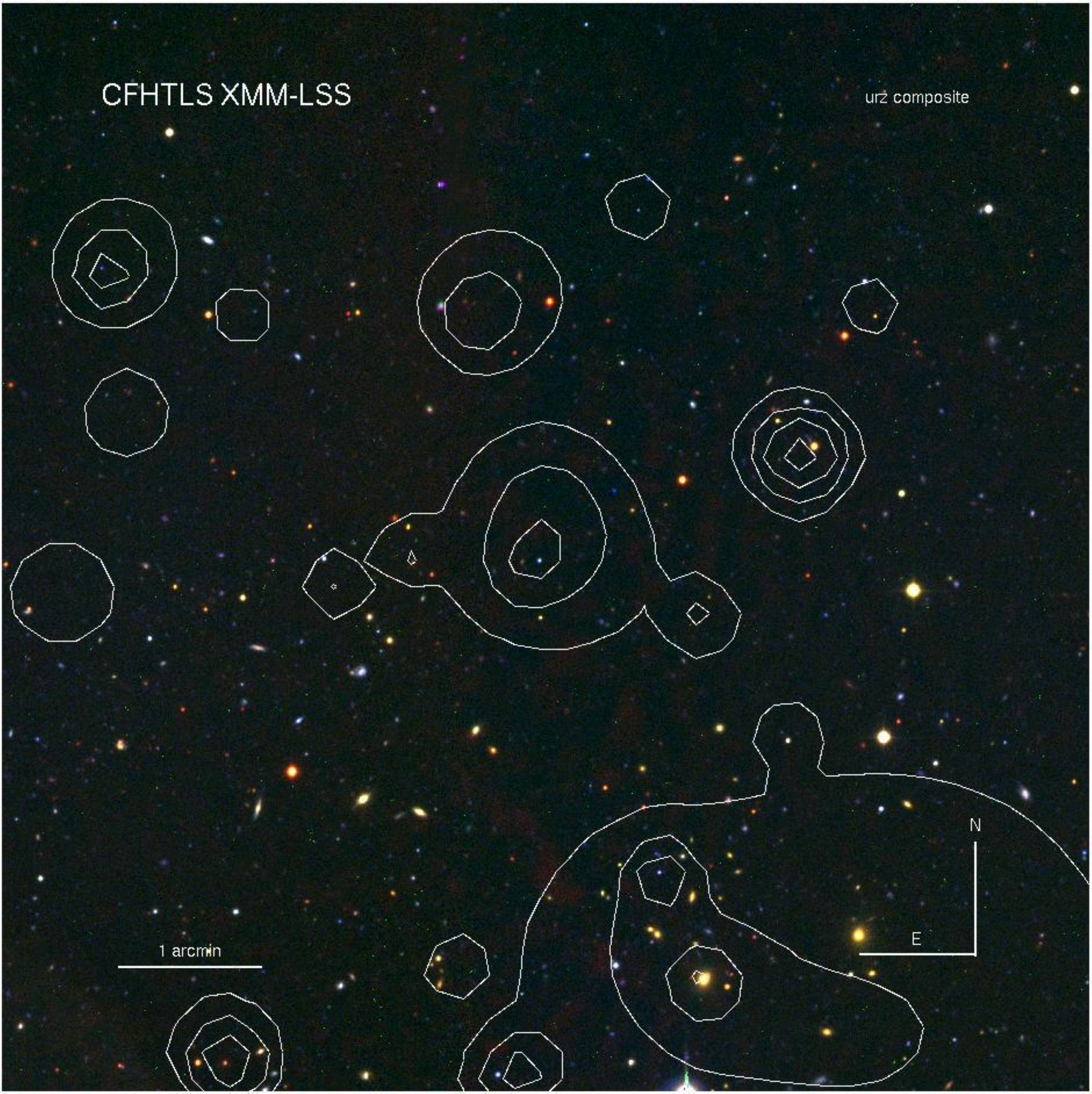} \\
	\end{tabular}
	\contcaption{Images of the C1 clusters.
	XLSSU J021744.1-034536 (see \citealt{willis13} for a near-infrared view of this cluster).}
\end{figure*}

\bsp

\label{lastpage}

\end{document}